\documentclass[12pt,a4paper, fleqn]{article}
\usepackage{etex}
\usepackage[utf8]{inputenc}
\usepackage[TS1,T1]{fontenc}
\usepackage{xspace}
\usepackage[a4paper,tmargin=3truecm,bmargin=3truecm,hmargin=1.6truecm]{geometry}

\usepackage[all]{xy}
\usepackage{xcolor}

\usepackage{relsize}

\usepackage[english]{babel}
\usepackage[bitstream-charter]{mathdesign}

\usepackage{enumitem}
\newlist{enum-hypothesis}{enumerate}{1}
\setlist[enum-hypothesis]{label=(\arabic*),itemsep=0pt, parsep=0pt}

\usepackage{amsmath,mathtools,slashed}
\usepackage[thmmarks,amsmath]{ntheorem}
\usepackage{tensor}
\usepackage{appendix}

\usepackage{hyperref}

\usepackage{pbox}

\newtheorem{theorem}{Theorem}[section]
\newtheorem{proposition}[theorem]{Proposition}
\newtheorem{lemma}[theorem]{Lemma}
\newtheorem{corollary}[theorem]{Corollary}

\newtheorem{definition}[theorem]{Definition}

\theoremsymbol{$\square$}
\theoremstyle{plain}
\theorembodyfont{\upshape}
\newtheorem{remark}[theorem]{Remark}

\theoremsymbol{}
\theoremstyle{break}
\theorembodyfont{\slshape}

\theoremheaderfont{\scshape}
\theorembodyfont{\upshape}
\theoremstyle{nonumberplain}
\theoremseparator{}
\theoremsymbol{\rule{1.3ex}{1.3ex}}
\newtheorem{proof}{Proof}

\usepackage[square,numbers,sort,compress,merge]{natbib}

\usepackage{graphicx}
\usepackage{calligra}

\hypersetup{
plainpages=false,
colorlinks=true,
linkcolor=black,
anchorcolor=black,
citecolor=black,
urlcolor=black,
menucolor=black,
filecolor=black,
bookmarksopen=true,
bookmarksnumbered=true}

\numberwithin{equation}{section}

\newcommand{\bbC}{\mathbb{C}}

\newcommand{\bbN}{\mathbb{N}}

\newcommand{\bbR}{\mathbb{R}}

\newcommand{\bbbone}{{\text{\usefont{U}{dsss}{m}{n}\char49}}}

\newcommand{\calA}{\mathcal{A}}
\newcommand{\calB}{\mathcal{B}}

\newcommand{\calD}{\mathcal{D}}

\newcommand{\calH}{\mathcal{H}}

\newcommand{\calL}{\mathcal{L}}
\newcommand{\calM}{\mathcal{M}}

\newcommand{\calP}{\mathcal{P}}
\newcommand{\calR}{\mathcal{R}}

\newcommand{\calS}{\mathcal{S}}
\newcommand{\calV}{\mathcal{V}}
\newcommand{\calZ}{\mathcal{Z}}

\newcommand{\RV}{\calR\!\calV}
\newcommand{\SR}{\calS\!\calR}

\newcommand{\hnabla}{\widehat{\nabla}}

\newcommand{\hDelta}{\widehat{\Delta}}

\newcommand{\dd}{\text{\textup{d}}}

\newcommand{\vc}{\vcentcolon =}   
\newcommand{\cv}{=\vcentcolon }   
\newcommand{\OO}{{\text{\usefont{OMS}{cmsy}{m}{n}O}}}   
\newcommand{\oo}{{\displaystyle\mathsmaller{\mathsmaller{\text{\usefont{OMS}{cmsy}{m}{n}O}}}}}  

\def\comFu_#1{[F_{#1}, u]}
\def\comFUu^#1{[F^{#1}, u]}
\def\comFN^#1_#2{[F_{#2}, N^{#1}]}
\def\anticomHLHN_#1{\{\hDelta, \hnabla_{#1}\}}
\def\Sumgigi^#1{G^{#1}}
\def\Sumgg_#1{G_{#1}}

\newcounter{termscounter}
\newcommand\EqLine[1]{
\stepcounter{termscounter}%
}

\DeclareMathOperator{\Dom}{Dom}
\DeclareMathOperator{\dom}{Dom}

\DeclareMathOperator{\Ln}{Ln}

\DeclareMathOperator{\Tr}{Tr}	   
\DeclareMathOperator{\Ker}{Ker}	 
\DeclareMathOperator{\Ran}{Ran}	 
\DeclareMathOperator{\re}{Re}  

\DeclareMathOperator*{\sumperm}{%
\mathchoice%
{\sum\kern-7pt\raise-0.5pt\hbox{\small$\mathsf{P}$}}
{\sum\kern-5.5pt\raise-0.4pt\hbox{\scriptsize$\mathsf{P}$}}
{}
{}
} 

\DeclarePairedDelimiter\abs{\lvert}{\rvert}
\DeclarePairedDelimiter\norm{\lVert}{\rVert}

\newcounter{mnotecount}[section]
\renewcommand{\themnotecount}{\thesection.\arabic{mnotecount}}
\newcommand{\mnote}[1]%
{\protect{\stepcounter{mnotecount}}${}^{\text{\footnotesize$\bullet$\themnotecount}}$%
\reversemarginpar%
\marginpar{\raggedleft\footnotesize$\bullet$\themnotecount: #1}}

\newlength{\mnotewidth}
\usepackage{calligra}

\allowdisplaybreaks

\usepackage{parskip}
\begin{document}

{
\makeatletter\def\@fnsymbol{\@arabic}\makeatother 
\title{Asymptotic behaviour of Semigroup Traces and Schatten Classes of Resolvents}

\author{Bruno Iochum\,\footnote{\,Campus de Luminy, 163 Av. de Luminy, 13009 Marseille, bruno.iochum@cpt.univ-mrs.fr} \, and Valentin A. Zagrebnov\,\footnote{\,
Site de Saint Charles, 3 place Victor Hugo, Case 19, 13331 Marseille  C\'{e}dex 3,
valentin.zagrebnov@univ-amu.fr}\\
{\small $^1$ Centre de Physique Théorique} \\
{\small Aix Marseille Univ, Université de Toulon, CNRS, CPT, Marseille, France}\\
[0.2ex]
{\small $^2$ Institut de Math\'ematiques de Marseille}
\\
{\small Universit\'e d’Aix-Marseille, CNRS, Marseille, France} \\
}
\date{}

\maketitle
}

\begin{abstract}

Motivated by examples from mathematical physics and noncommutative geometry, given a generator $A$ of a Gibbs semigroup $\{U_{A}(t) := e^{-tA}\}_{t\geq 0}$,
we re-examine the relationship between the Schatten class of its resolvents and the behaviour of the norm-trace $\norm{e^{-tA}}_1\,$ when $t$ approaches zero. Besides the applications of the Tauberian results, we specifically investigate the compatibility of asymptotic behaviours with the semigroup derivations and perturbations.
Along the course of our study, we present a novel characterisation of Gibbs semigroups.
\end{abstract}

\section{Introduction}\label{sec1}

The aim of this paper is to seek a characterisation of operators $A$ on a Hilbert space $\calH$ that generate a semigroup with a given asymptotic behaviour of its trace-norm for small $t>0$.
 When $\calH$ is infinite dimensional, the generator $A$ of such semigroup $\{e^{-tA}\}_{t\geq 0}$, called then a Gibbs semigroup (see Definition  \ref{def:GibbsSG}), is necessarily unbounded with only a discrete spectrum $\{\lambda_k(A)\}_{k \geq 1}$ such that $\abs{\lambda_k(A)} \to_{k\to \infty} \infty$.

Since the trace-norm $\norm{e^{-tA}}_1$ approaches infinity when $t$ tends to zero (see Proposition \ref{prop-behaviour at 0}), it is interesting to estimate this asymptotic divergence. To this aim
we consider, for instance, the \textit{ansatz}:
\begin{align}
\label{class of f}
\norm{e^{-tA}}_1= \OO_{t \downarrow 0}(f(t))\,, \text{ where }f(t)=t^{-p}\,\abs{\ln t}^{\,r} \text{\,\,for some }p>0,\,r\in \bbR .
\end{align}
Here, we denote by $\calL^p= \calL^p(\calH)$ the von Neumann–Schatten class of compact operators on $\calH$ equipped with the usual norm $\norm{\cdot}_p$ for $p \geq 1$ or quasi-norm for $p\in (0,1)$.

A concrete example arises from theoretical physics. Given natural numbers $n,\,m\in \bbN$, let us consider a Schr\"odinger operator on $\bbR^{n+m}$ with a scalar potential: $H\vc -\tfrac{1}{2} \Delta_{(x,y)}+ \abs{x}^{2p}\,\abs{y}^{2q}$ with parameters
$p,q>0$ and variables $x\in \bbR^n,\, y\in\bbR^m$, see \cite{Aramaki, Aramaki1} and references therein. When defined on the domain $C_0^\infty(\bbR^{n+m})\,$, this operator $H$ is essentially self-adjoint, and its self-adjoint extension $A$ serves as the generator of a Gibbs semigroup $\{e^{-tA}\}_{t\geq 0}\,$, exhibiting the following asymptotic behaviour (see definition of $ \underset{t\downarrow 0}{\sim} $ in Section \ref{Notations}):
\begin{align*}
\Tr e^{-tA} \ \underset{t\downarrow 0}{\sim} \,
\left\{ \begin{array}{ll}
c_1\, t^{-m(1+p+q)/(2q)}& \text{when } pm>qn\,,
\\  c_2\, t^{-n(1+p+q)/(2p)}& \text{when } pm<qn \,,
\\  c_3\,t^{-n(1+p+q)/(2p)}\,\ln t^{-1} & \text{when } pm=qn\,.
\end{array}
\right.
\end{align*}
We also have examples where $\Tr\,e^{-tA} \sim_{t\downarrow 0} c_q \,\abs{\ln t}^r$, with $r=2$, for the absolute value of a Dirac-type operator on the Podle\'s sphere deformed with a parameter $0<q<1$, see \cite{Eck-Ioc-Sit}, or for Casimir type operators, where $r=2,3$, also also appears in \cite{Kakehi-Masuda}. More examples can be found in \cite{Boulton} and \cite{Boulton-Dimoudis}.

Another example arises from number theory. Consider the operator $A = \text{Diag}(\{p_n\}_{n\geq 1})$ on the  Hilbert space $\ell^2(\mathbb{N})$, where $p_n$ denotes the $n$th prime number. This operator is self-adjoint  on its maximal domain and positive. By utilising the Prime Number Theorem \cite[Theorem 12, Chapter I]{Ingham} (refer also to Remark 4.12), we can deduce that:
\begin{align}
&\Tr\,e^{-tA} =\sum_{n\geq1} e^{-tp_n}  \,\underset{t \downarrow 0}{\sim}\, g(t)=\frac{1}{t\,\ln t^{-1}}\,, \label{prime semigroup}\\
&(-1)^n\Tr\,A^ne^{-tA} \,\underset{t \downarrow 0}{\sim}\, g^{(n)}(t)\,,\quad n\in \bbN\,. \label{prime semigroup1}
\end{align}
\indent
Further motivation arises from a much more precise possibility than just an asymptotic behaviour like in \eqref{class of f}.
Consider the existence of an asymptotic expansion as assumption for $F(t)=\norm{e^{-tA}}_1$
\begin{align}
\label{expansion with log 1}
F(t) \ \underset{t\downarrow 0}{\approx} \,\sum_{k=0,\,\ell=0}^\infty \, a_{k,\ell}\, t^{r_k} \,(\ln t)^{\,\ell}\,,
\end{align}
where $\{r_k\}_k$ is an unbounded non-decreasing sequence of real numbers with $r_0<0$
\footnote{\,This essentially means that in (\ref{expansion with log 1})
the difference between $F(t)$ and any truncation of the sum is small compared with next summand when $t$ goes to zero. For a precise definition of $\underset{t\downarrow 0}{\approx}\, $, see for instance \cite[Definition 1.3.3]{Estradabook} or \cite[Section 2.5]{EckIoc}.}.
We do not assume that the sum converges, as we are only interested in its asymptotics as $t \downarrow 0$. This means in particular that $F(t) \underset{t \downarrow 0}{\sim}  a_{0,\ell_{\max}} t^{r_0}\,(\ln t)^{\,\ell_{\max}}\,$.
\\
Several examples of such possibilities can be found in pseudodifferential theory.
If $A$ is an elliptic pseudodifferential operator of order $m>0$ acting on a vector bundle $E$ over a compact, boundaryless, $d$-dimensional Riemannian manifold $M$ with a constraint on its principal symbol (see \cite[Theorem 4.2.2 and Corollary 4.2.7]{Grub96} or \cite[Corollary A.7]{EckIoc} for details), then we have :
\begin{align}
\label{an expansion with log 2}
\norm{e^{-tA}}_1 \underset{t\downarrow 0}{\approx}\, \sum_{k=0}^\infty \,a_{k}\, t^{(k-d)/m} +\sum_{\ell=0}^\infty \,b_\ell \,t^\ell \,\ln t\,.
\end{align}
For instance, when $A$ is the Laplace--Beltrami operator (thus $m=2$), we get $\norm{e^{-tA}}_1 \sim_{t\downarrow 0} a_0\,t^{-d/m}\,$, and this asymptotic behaviour is equivalent to the well-known Weyl's law on the counting function: $N_A(\lambda) \sim_{\lambda \uparrow \infty} c \,\lambda^{d/2},$ (refer to Corollary \ref{thm-Karamata for semigroup}).
\\ \indent
More generally, let $CL^{m,k}(M,E)$ denote the class of pseudodifferential operators of order $m$ acting on $E$ over $M$ with polyhomogeneous symbols, and $k$ represents the highest log-power that appears in the symbol expansion. In \cite[equation (3.18)]{Lesch}, it is shown that if $A\in CL^{m,0}(M,E)$, where $m>0$, is an elliptic operator with a positive leading symbol, and $B\in CL^{a,k}(M,E)$, we have the following expansion:
\begin{align}
\label{expansion of pseudo}
\Tr\,B e^{-tA} \,\underset{t \downarrow 0}{\approx}\, \sum_{j=0}^\infty \,c_j \,  t^{(j-d-a)/m}\ln t +\sum_{j=0}^\infty \, d_j t ^j \, ,
\end{align}
where $c_j\in\bbC[x]$ are polynomials of degree less than $k+1$ and $d_j \in \bbC\,$. Thus $\Tr\, Be^{-tA} \underset{t\downarrow 0}{\sim} c_0\,t^{-(d+a)/m}\,\ln t\,$.

To conclude this set of examples (see \cite{Gof-Usa} for some others), we mention that one can easily extend this kind of asymptotic behaviours in tensor products of Hilbert spaces. That is, if $\Tr \,e^{-tA} =\OO_0 (f(t))$ and
$\Tr \,e^{-tB} =\OO_0 (g(t))\,$, then
\begin{align*}
\Tr \,e^{-t(A\otimes \bbbone +\bbbone \otimes B)}=\Tr\, e^{-tA}  \Tr\, e^{-tB} =\OO_0 (f(t)\,g(t))\,.
\end{align*}
 \indent
Furthermore, the equality $(\ln t)^r=\oo_0(t^{-\varepsilon})$ (see notations below in Section \ref{Notations} and \eqref{L and small o}), which holds for any $r\in \bbR$ and any $\varepsilon>0$, implies
$t^{-p}\,(\ln t)^r=\oo_0(t^{-(p+\varepsilon)})$. Thus, $\OO_0(t^{-p} \,(\ln t)^r)=\oo_0(t^{-(p+\varepsilon)})$.
One may wonder why we don't restrict our analysis of the asymptotic behaviour of $\norm{e^{-tA}}_1$ to the class of functions $f$ such that $f(t)=t^{-p}$ with $p>0$. However, we shall see that, due to Tauberian results, the same tools
can be applied to $\OO_0(f(t))$ or $\lim_{t\downarrow 0} f(t)$. Moreover, in the latter case, the precise values of $p$ and $r$ are crucial, as in \eqref{prime semigroup} or \eqref{prime semigroup1}.

Another motivation of this paper comes out of noncommutative geometry, where, given a spectral triple $(\calA,\,\calH,\,\calD)$, one seeks to compute the asymptotics of its spectral action \cite{ChamConnes1997}. By means of a Laplace transform, such an action relies on the asymptotic expansion of $\norm{e^{-t\abs{\calD}}}_1$ (which encompasses cases \eqref{expansion with log 1} and \eqref{an expansion with log 2}). However, proving the existence of such an asymptotic expansion is not straightforward. Essentially, one needs to establish the meromorphic extension of the zeta-function $\zeta_{\calD}(s) \vc \Tr \,\abs{\calD}^{-s}$ to the entire complex plane, having only the information that $\abs{\zeta_{\calD}(s)} <\infty$ when $\Re s >p$ for some $p$. For more details, refer to \cite[Proposition 5.1]{Grubb-seeley}, \cite{EckIoc} and \cite[Theorem 2.10]{GRU}. In the absence of a complete answer regarding this existence, one can at least examine its leading term (i.e., one works with $\sim_{t\downarrow 0}\,$ instead of $\approx_{t\downarrow 0}$) such as $\norm{e^{-tA}}_1= \OO_{0}(t^{-p}\,(\ln\,t^{-1})^r)\,$. Moreover, such an analysis is essential in the computations of Dixmier traces on noncommutative Lorentz spaces, see for instance \cite{CRSS, CGRS, Gay-Suk, LSZ, LSZ1}.

As mentioned earlier, the Laplace transform naturally emerges in this context and we consider now the following generalisation: Let $\mu$ be a non-negative $\sigma$-finite Borel measure on $[0,\infty)$, and define the Laplace--Stieltjes transform
\begin{align}
\label{def-lmu}
L_\mu(t) \vc \int_{[0,\infty)} \dd \mu(x) \, e^{-tx}, \quad t>0.
\end{align}
This expression is directly connected to our original framework, as for a positive generator $A$ with eigenvalues $\{\lambda_k\}_{k\geq 1}$, we have $\norm{e^{-tA}}_1=\sum_k e^{-t\lambda_k} =L_\mu(t)$ with the choice of the  discrete measure $\mu = \sum_k \delta_{\lambda_k}$. Since
\begin{align}
\label{Lmu and trace}
f(t^{-1})^{-1} \,L_\mu(t) = \int_0^\infty \dd x\, f(t^{-1})^{-1} \,\mu([0,t^{-1}x]) \,e^{-x}\,,
\end{align}
we are interested in functions $f$ such that $f(t^{-1})^{-1} \,L_\mu(t)$ is controlled when we commute the last integral with $\lim_{t\downarrow 0}$ or $\limsup_{t\downarrow 0}$ or $\liminf_{t\downarrow 0}$ (possibly with inequalities for the last two).
\\
Considering
\begin{align*}
\frac{\mu([0,t^{-1}x])}{f(t^{-1})} =\frac{\mu([0,t^{-1}x]) }{f(xt^{-1})}\cdot\frac{f(xt^{-1})}{f(t^{-1})}
\end{align*}
we are naturally led to the space $\RV$ of regularly varying functions $f$, see Definition \eqref{def-RV}, satisfying
\begin{align*}
\lim_{t\downarrow 0}\,\frac{f(t^{-1}x)}{f(t^{-1})} =x^p\,,\quad \text{for some $p\in \bbR\,$,}
\end{align*}
which was investigated by Kamarata in his approach to the Tauberian theorems. This is done in Theorem \ref{appendix:Thm-Karamata}, which is postponed to Appendix. Although it is considered as part of folklore, we prove it there for the sake of completeness since we were unable to find a good reference, except for one in \cite{Ambrosio} and its assertion is slightly improved in Appendix.

Note also that the space $\RV$ is the appropriate setting: When $f\in\RV$ with $p=\text{ind}_f$ (see Appendix) we still have $f(t^{-1})=\oo_0(t^{-(p+\varepsilon)})$ for any $\varepsilon>0$ as shown in  \eqref{behaviour of f(t-1)}, as already observed when $f(t^{-1})=t^{-p}\,(\ln\,t^{-1})^r$.

From the outset, the semigroup $\{e^{-t\,A}\}_{t\geq 0}$ must consist of trace-class operators, which is referred to as a Gibbs semigroup, see e.g., \cite{Zagrebnov2}. In Section 2, we provide a compilation of essential facts about these semigroups, which play a crucial role in our subsequent analysis. Additionally, the Proposition \ref{prop-behaviour at 0} and a subsequent remark present specific examples that illustrate the behaviours of $\norm{e^{-tA}}_1\,$, $\Re \Tr e^{-tA}\,$, and $\Im \Tr e^{-tA}$ (i.e. the \textit{real} and \textit{imaginary} parts of the trace) as $t$ tends to zero.

In Section \ref{Tauberian}, we deduce from Tauberian Theorem \ref{appendix:Thm-Karamata}, a characterisation of generators $A$ of Gibbs semigroups with trace-norm asymptotics $\norm{e^{-tA}}_1=\OO_0(f(t^{-1}))$ with $f \in \RV$. This characterisation covers the peculiar case $f(t^{-1})=t^{-p}\,(\ln\,t^{-1})^r$ for $p> 0$ and $r\in \bbR$ (or $p=0$ and $r>0$), as well as slightly more general cases than $A>0$. (See, for instance, Corollary \ref{thm-Karamata for semigroup}.)

Then, an easy by-product obtained in the framework of noncommutative geometry is the following: a spectral triple $(\calA,\calH,\calD)$ is called $p$-summable when $(\calD + i \bbbone)^{-1} \in \calL^{p,\infty}\,$, which implies $\norm{e^{-t\,\calD^2}}_1=\OO_0(t^{-p/2})$ (see \cite[page 450]{GracVariFigu01a}). We show in Corollary \ref{p-summability} that this last condition characterises $p$-summability within the $\theta$-summability, cf. paragraph before Corollary \ref{p-summability}.

In Section \ref{sec2}, we specifically consider the simplest case where the generators $A$ of the semigroup are positive operators. In Lemma \ref{lem-behaviour for positive operator}, we observe the appearance of {\it a deficiency} that concerns the equivalence between the asymptotics $\norm{e^{-tA}}_1 =\OO_0(t^{-p})$ and $A^{-1} \in \calL^p$.  However, this deficiency is addressed in Proposition \ref{prop-trace versus Zp}, where we establish that such asymptotics is equivalent to $A^{-1}\in \calL^{p,\infty}$. Furthermore, in Theorem \ref{singularity}, we show that such  asymptotics are stable under differentiation, when applied to functions with smooth variation at infinity, as  defined in Definition \ref{def-SR} of the Appendix.

In Section \ref{sec3}, we extend our analysis to a broader class of generators, specifically those that generate holomorphic Gibbs semigroups. Along the way, we also establish a novel characterisation of Gibbs semigroups and a complementary result concerning \textit{integrated} semigroups.

In the concluding Section~6, we investigate the \emph{stability} of trace-norm asymptotics for Gibbs semigroups under different classes of generator perturbations.

\section{Notations and preliminary material}
\label{Notations}

First we precise notations for the asymptotics. Given two complex functions $f,g$ defined on a neighborhood of $x_0\in [0,\infty) \cup \{\infty\}$, we write (with a slight abuse of notations): \\
\noindent $f (x) = \OO_{x_0} (g(x))$ if $\underset{x\to x_0}{\limsup} \,\abs{f (x)/g(x)} <\infty$, when $g$ does not vanish in a neighborhood of  $x_0$,\\
 $f(x)=\oo_{x_0}(g(x))$ if $\underset{x\to x_0}{\lim} \,\abs{f (x)/g(x)} =0$, when $g$ does not vanish in a neighborhood of $x_0$,\\
$f(x) \underset{x \to x_0}{\sim} \,g(x)$, when $f(x)-g(x)=\oo_{x_0}g(x)$,
or equivalently, when $\underset{x \to x_0}{\lim}\, \abs{f(x)/g(x)} =1 $, if the zeros of $f$ and $g$ coincide in a neighborhood of $x_0\,$, cf. \cite [page 5]{Estradabook}.

From now on, we restrict to the strongly continuous ($C_0$-)semigroups $\{U(t)\}_{t\geq0}$ on a separable complex Hilbert space $\calH$. A $C_0$-semigroup has a (infinitesimal) generator $A$ and is \textit{quasi-bounded}: $\norm{U(t)=e^{-tA}}\leq M e^{\omega t}$, where $M\geq 1$ and $\omega\in\bbR$. Taking the infimum $\omega_0$ of the numbers $\omega$ for $M\geq 1$, we denote the corresponding class of generators $A$ by $Q(M,\omega_0)$.
{If $\rho(A)$ is the resolvent set of $A$, then
$\rho(A) \supset \bbC_{-\omega_0}:= \{z \in \bbC\,\, \vert \, \Re(z) < -\omega_0 \}$,
and thus $(-\infty,-\omega_0)\subset \rho(A)\,$, see e.g., \cite[Proposition 1.12]{Zagrebnov2}.} We denote by $R_A(z)=(A-z\bbbone)^{-1}$ the resolvent operator for $z\in \rho(A)$.

Let $\calL(\calH)$ (respectively $\calL^\infty$) denote the set of bounded (respectively compact) operators on $\calH$.
For a given compact operator $K$ on $\calH$, $\{ \lambda_k(K)\}_{k\geq 1}$ denotes its set of eigenvalues. For $p > 0$, we define the von Neumann--Schatten class of operators: $\calL^p \vc \{K\in \calL^\infty \,\vert \, \,
\norm{K}_p \vc [\sum_{k=1}^\infty s_k(K)^p ]^{1/p}<\infty\}$, where the singular values $ s_k(K)\vc\sqrt{\lambda_k(K^*K)} \neq 0$ are arranged in non-increasing order counting multiplicities. The Banach space $\calL^1(\calH)$ of operators with norm $\norm{\cdot}_1$ is known as the \textit{trace-class}
(or \textit{nuclear}) operators.

Let $S_{\theta,\gamma}$ (in the complex plane $\bbC$) be an open sector of semi-angle $\theta \in (0,\pi/2]\,$ with vertex $\gamma \in \bbR\,$:
\begin{align}\label{sector}
S_{\theta,\gamma}\vc \{z \in \bbC\,\, \vert \, \Re(z-\gamma)>0 \text{ and } \abs{\arg(z-\gamma)}<\theta\}\,, \quad S_\theta \vc S_{\theta,0}\,.
\end{align}
We denote by $\overline{S}_{\theta,\gamma}$ its closure.
We also use notations $\bbN_0 \vc \bbN \cup\{0\},\,\bbR^+\vc (0,\infty)$ and $\bbR_0^+\vc [0,\infty)$.

\begin{definition} \label{def:GibbsSG} \emph{(\cite[Definition 4.1]{Zagrebnov2})}.
\emph{(a)} A $C_0$-semigroup $\{U(t)\}_{t\geq 0}$ is called an (immediately)
Gibbs semigroup if $U(t)\in \calL^1\, $ for $t \in \bbR^+ \, $.\\
\emph{(b)} A holomorphic $C_0$-semigroup $\{U(z)\}_{z\in S_{\theta}\cup\{0\}}$ is called a \textit{holomorphic} Gibbs semigroup in sector $S_{\theta}\,$
if $U(z) \in \calL^1\,$ for any $z\in S_{\theta}\,$.
\end{definition}
{Let $\{U_A(z)\}_{z\in S_{\theta} \cup \{0\}}$ be a holomorphic semigroup with generator $A$. Then it is 
a Gibbs holomorphic semigroup if and only if its restriction $\{U_A(t)\}_{t\geq 0}$ is a Gibbs semigroup.}
\begin{remark}\label{rem:hol-SG}
Because of the trace-norm continuity of multiplication on the ideals
$\calL^p$, $p \geq 1$, it follows that Gibbs semigroups are
$\norm{\cdot}_1$-continuous for $t > 0$.
Consequently, by Definition \ref{def:GibbsSG} (b), the holomorphic Gibbs semigroups are automatically holomorphic in the trace-norm topology.
\end{remark}

To establish some properties of trace-class exponentials, we first recall that
the spectrum $\sigma(A)$ of the generator $A$ of a compact semigroup satisfies the \textit{spectral mapping theorem}:
$\sigma(U_A(t)) =\{e^{-t\sigma(A)}\}\cup\{0\}$ where $\sigma(A)$ is either empty or consists of the (possibly finite) set of eigenvalues $\{\lambda_k(A)\}_{k\geq 1}$.

Next, let $A$ be a \textit{normal} operator on ${\dom}\,(A)\subset \calH\,$. Then it is known that
$\dom\,(A) = \dom\,(A^*)$ and the operator $A+A^*\,$, which is symmetric on this domain, is \textit{essentially} self-adjoint. We can then define the self-adjoint operator
${\Re A}\,$, which is called the \textit{real part} of operator $A$, by taking the closure:
\begin{align}\label{normal-oper-sum}
\Re A := \tfrac{1}{2} \ (\overline{A+A^*}) \, .
\end{align}
We warn that for a general densely defined closed operator $A$, the operator
(\ref{normal-oper-sum}) may be ill-defined, and even such that ${\dom}\,(\Re A) = \{0\}\,$. Whereas for normal operator $A$, one has
$\norm{A u} = \norm{A^*u} $ for $u \in\dom\,(A)$ and if
$A \,f = \lambda_f(A)\, f$ for some eigenvector $f \in\dom\,(A)$, then for the operator
(\ref{normal-oper-sum}) one gets: $\Re A \,f = \Re \lambda_f(A) \,f$ and $\abs{\lambda_f(A)}
= s_f(A)$.

Note that the Gibbs semigroup $\{U_A(t)\}_{t\geq 0}$ is normal (i.e. each $U_A(t)$ is a normal operator) if and only if its generator $A$ is normal.
As a consequence, we obtain: $s_k(e^{-tA})=s_k(e^{-t\Re A})=e^{-t \Re \lambda_k(A)}\,$, which yields
\begin{align}
\label{normal trace-norm}
\norm{e^{-tA}}_1=\norm{e^{-t\Re A}}_1=\sum_{k\geq 1} e^{-t \Re \lambda_k(A)}>0\,,\quad t>0\,.
\end{align}

Now, we review a few assertions that will be useful in the following sections of this paper.

\begin{lemma}
{Given a Gibbs semigroup $\{e^{-tA}\}_{t\geq 0}\,$, we have for $t>0\,$,}
 \begin{align}
\label{ineq principal}
 \abs{\Tr e^{-tA}}=\abs{\sum_k \lambda_k (e^{-t A})} \leq \sum_k \, \abs{\lambda_k (e^{-t A})}
 \leq \sum_k s_k (e^{-t A}) = \norm{e^{-tA}}_1 < \infty\,.
\end{align}
\end{lemma}
\begin{proof}
This follows from Lidski\u\i\hspace{0.05cm}'s theorem and Weyl's inequality:
$\sum_k \, \abs{\lambda_k (e^{-t A})} \leq \sum_k s_k (e^{-t A})$, with equality if the generator $A$ is normal.
\end{proof}

While our primary focus lies in the asymptotic behaviour of $\norm{e^{-tA}}_1\,$, we can also explore its connection with the asymptotics of $t$-functions: $\sum_k e^{-t \Re \lambda_k(A)}$, $\Tr e^{-tA}$,
$\norm{e^{-t\Re A}}_1\,$ where $\sigma(A) = \{\lambda_k  \mid k \in \mathbb{N}\}$ denotes the spectrum of $A\,$. To begin, we establish an estimate for the difference between the first two quantities.

\begin{lemma}
\label{lemma:gibbs_semigroup_estimate}
Let $\{e^{-tA}\}_{t \geq 0}$ be a (once-)differentiable Gibbs semigroup with normal generator $A\,$. Then, the following estimate holds:
\begin{align}
\label{estimate on d(t)}
\big\lvert \sum_{k=1}^\infty e^{-t \Re \lambda_k(A)} - \Tr e^{-tA} \big\rvert \leq t \, \norm{\,A \,e^{-tA}}_1 \,.
\end{align}
\end{lemma}
\begin{proof}
Define $x_k = \Re \lambda_k(A)$, $y_k = \Im \lambda_k(A)\,$ and $\lambda_k=x_k+iy_k\,$. By Lidskii's theorem and the spectral mapping theorem, we have: $d(t) \coloneqq \sum_{k=1}^\infty e^{-t x_k} - \Tr e^{-tA} = \sum_{k=1}^\infty e^{-t x_k} \big(1 - e^{-t i y_k}\big)\,$.
This leads to the following estimates:
\begin{align}
\abs{d(t)} &\leq \sum_k e^{-tx_k}\,\abs{1-e^{-tiy_k}} = \sum_k e^{-tx_k}\, 2 \,\abs{\sin(ty_k/2)}\leq t \sum_k e^{-tx_k}\,\abs{y_k} = t \sum_k \abs{y_ke^{-t\lambda_k}}  \label{ineq on d} \\
& \leq t\sum_k \, \abs{\lambda_k e^{-t\lambda_k}} = t \norm{\,A\,e^{-tA}}_1 \, , \label{ineq on d 1}
\end{align}
where the final equality follows from the Weyl inequality: $\sum_{k=1}^\infty \abs{\lambda_k(A e^{-t A})} \leq \sum_{k=1}^\infty s_k(A e^{-t A})\,$,
which becomes an equality for the normal operator $A\,e^{-tA}$. The differentiability of the semigroup guarantees the convergence of the trace norm.
\end{proof}
\begin{remark}
We now discuss some consequences regarding the relationships between the different asymptotic behaviours of the $t$-functions introduced earlier.

(1) Because $\big\lvert \sum_{k=1}^\infty e^{-t x_k} - \abs{\Tr e^{-tA}} \big\rvert \leq \abs{d(t)}\,$,
    we deduce from \eqref{ineq principal} and \eqref{ineq on d} the following estimates:
    \begin{align}
    \label{two-estimates}
    \sum_{k=1}^\infty e^{-t x_k} - \abs{d(t)} \leq \abs{\Tr e^{-tA}} \leq \sum_{k=1}^\infty e^{-t x_k}.
    \end{align}
    To refine previous estimate of $\abs{d(t)}$, assume that the spectrum $\sigma(A)$ is contained in the open sector $S_\alpha = \{z \in \mathbb{C} \mid \abs{\arg(z)} < \alpha\}$,  with semi-angle $\alpha < \pi/2\,$.
    Then, from \eqref{ineq on d}, we obtain:
    \begin{align}\label{d-estimate}
    \abs{d(t)} &\leq t \sum_{k=1}^\infty e^{-t x_k} \abs{y_k} \nonumber
    \leq t \sum_{k=1}^\infty e^{-t x_k} \rm{tg}(\alpha) \, x_k
    \leq {M_\alpha}, \quad t > 0.
    \end{align}   
Combining \eqref{two-estimates} and \eqref{d-estimate}, we conclude that
    \[
    \abs{\Tr e^{-tA}} \underset{t \downarrow 0}{\sim} c \sum_{k=1}^\infty e^{-t \Re \lambda_k(A)}.
    \]   
Due to \eqref{ineq principal}, this observation also applies to the asymptotics
 $ \abs{\Tr e^{-tA}} \underset{t \downarrow 0}{\sim} c \norm{e^{-tA}}_1\,$.

(2) Let $A$ be an $m$-sectorial operator with $\Re A \geq 0$ (see, e.g., \cite[Chapter 2]{Haase}).
    If $e^{-t \Re A} \in \mathcal{L}^1$ for all $t > 0$, then $\{e^{-tA}\}_{t \geq 0}$ is a Gibbs semigroup, as $\norm{e^{-tA}}_1 \leq \norm{e^{-t \Re A}}_1$ by \cite[Proposition 4.30]{Zagrebnov2}, with equality when $A$ is normal. This means that $\norm{e^{-tA}}_1= \OO_0(\norm{e^{-t\Re A}}_1)\,$.

We remark that in both cases (1) and (2), our analysis concerns \emph{holomorphic} Gibbs semigroups, which will be examined in greater detail in Section~\ref{sec3}. The challenge in establishing relationships between different asymptotic behaviours of the $t$-functions stems primarily from the limited control over the behaviour of the imaginary parts of generators and their corresponding semigroups. For another illustration of similar phenomena involving a holomorphic semigroup, we refer to the second part of Remark~\ref{rem1}.

(3) Assume $\norm{e^{-tA}}_1 \underset{t \downarrow 0}{\sim} t^{-p}$ for some $p > 0\,$. Then, from \eqref{ineq principal}, we obtain the asymptotic relations: $\abs{\Tr e^{-tA}}=\OO_0(t^{-p})\,$ and $\sum_{k=1}^\infty e^{-tx_k} \leq \norm{e^{-tA}}_1$ which  implies that $\sum_{k=1}^\infty e^{-tx_k} = \OO_0(t^{-p})\,$.
Consequently, $\abs{d(t)} = \OO_0(t^{-p})\,$. Extracting this behaviour directly from the right-hand side of \eqref{estimate on d(t)} is non-trivial: under the stronger hypotheses of Theorem~\ref{singularity-ter}, we have $\norm{\,A\, e^{-tA}}_1 \underset{t \downarrow 0}{\sim} p t^{-(p+1)},$ which through \eqref{ineq on d 1} yields again   $\abs{d(t)} \leq t \norm{\,A\, e^{-tA}}_1 \underset{t \downarrow 0}{\sim} p t^{-p}\,$.
\end{remark}

\begin{lemma} \label{lem-ReA et omega0}
Let $A\in Q(M,\omega_0)\,$ be a normal operator generating a quasi-bounded $C_0$-semigroup. \\
Then, for the self-adjoint operator (\ref{normal-oper-sum}), one gets
$\Re A \geq -\omega_0 \bbbone$.
\end{lemma}

\begin{proof}
For a normal generator $A\in Q(M,\omega_0)\,$ one has $\rho(A) \supset (-\infty,-\omega_0)$ and $\rho(A^*) \supset (-\infty,-\omega_0)$, and we obtain the same for the self-adjoint operator (\ref{normal-oper-sum}). As a consequence,  $\Re A \geq -\omega_0 \bbbone$.
\end{proof}
\begin{proposition}
\label{prop-behaviour at 0}
Let $\{U(t)\}_{t\geq 0}$ be a Gibbs semigroup with generator $A$. Then

\emph{(i)}  $\abs{\Tr U(t)}$ increases to infinity as $t$ decreases to 0. In particular,
$ \norm{U(t)}_1\, \underset{t \downarrow 0}{\longrightarrow}\, \infty\,$.

\emph{(ii)}  $U(t) \neq 0$ for any $t\geq 0\,$.

\emph{(iii)} We have $\norm{U(t)}_1 >0\,$ for $t>0$\,.

\emph{(iv)} If the semigroup is self-adjoint, then $\Tr U(t) > 0\,$ for $t>0\,$.
\end{proposition}

We say that $\{U(t)\}_{t\geq 0}$ is an {\it eventually Gibbs semigroup} if there is $t_0>0$ such that $\norm{U(t)}_1<\infty$ for $t\geq t_0$. Note that $\abs{\Tr U(t)}\leq \norm{U(t)}_1=0$ may occur for an eventually Gibbs semigroup.
In this case, the semigroup is trivially an eventually Gibbs semigroup with
$\norm{U(t>t_0)}_1=0$ and threshold $t_0=\inf_{t>0} \{t\,\vert\,U(t)=0\}\,$, see \cite[Example 4.16]{Zagrebnov2}.

\begin{proof}(i) Since $U(t)\in \calL^1$ for $t>0\,$,
$\Tr U(t)=  \sum_{k=1}^\infty \langle U(t)\,e_k,\,e_k\rangle \,$ for any orthonormal basis $\{e_k\}_{k\in \bbN}$ of $\calH$. Given $n\in \bbN$, let $S_n(t)\vc \sum_{k=1}^n \langle U(t)\,e_k,\,e_k\rangle$ and $R_n(t)\vc \sum_{k=n+1}^\infty \langle U(t)\,e_k,\,e_k\rangle\,$. \\
As $\lim_{t\downarrow 0} \norm{U(t)e_k-e_k}=0\,$,
we get
\begin{align}
\label{convergence}
\abs{S_n(t)} \underset{t\to 0}{\to} n\,, \quad \text{for any $n\in \bbN$}.
\end{align}
Since
$0\leq \big{\vert} \abs{S_n(t)}-\abs{R_n(t)} \big{\vert} \leq \abs{S_n(t)+R_n(t)}=\abs{\Tr U(t)}$ for $ t>0\,$
we have
\begin{align}
\label{inequality-remaider}
\abs{S_n(t)} \leq \abs{R_n(t)}+\abs{\Tr U(t)}
\text{ for any $t>0\,$. }
\end{align}
Let $\varepsilon>0$ be given. By \eqref{convergence}, there exists $t_\varepsilon>0$ such that if $0<t < t_\varepsilon\,$, then $n-\varepsilon \leq \abs{S_n(t)}\,$.\\
Moreover, the convergence of the series $\sum_{k=1}^\infty \langle U(t)\,e_k,\,e_k\rangle$ implies that given $t<t_\varepsilon$, there exists $N_t\in \bbN$ such that if $n>N_t$, then $\abs{R_n(t)}\leq \varepsilon\,$.
Due to \eqref{inequality-remaider}, we get $n-\varepsilon \leq \varepsilon +\abs{\Tr U(t)}$ for any $t<t_\epsilon$ and $n$ large enough.

(ii) Assume that $U(s)=0$ for some $s>0$. Then for any $n\in \bbN$, $U(s/n)$ is a quasi-nilpotent trace-class operator of order $n$. Hence, by Lidski\u\i\hspace{0.05cm}'s theorem, $\Tr U(s/n) =0$ for any
$n\in \bbN$, in contradiction with (i).

(iii) The assertion follows from (ii); in the case of a normal semigroup, it is provided by \eqref{normal trace-norm}.

(iv) This follows from the self-adjointness and the semigroup property: $U(t) = U^*({t}/{2})U({t}/{2})$.
\end{proof}

\begin{remark}\label{rem1}
The assertion in (iv) cannot be extended to \emph{normal} Gibbs semigroups while preserving the positivity of $\Re(\Tr U(t))\,$.
\\
For example, let $A$ be a normal unbounded multiplication operator in the Hilbert space $l^2(\bbN)$ with spectrum $\sigma(A) = \{(1+ic) \,k\,\vert \,k\in \bbN,\,c\in \bbR\}\,$. Then $A$ generates a Gibbs semigroup and we compute $\Tr U_A(t) = \sum_{k\geq1} e^{-t(1+ic)k}=(e^{(1+ic)t}-1)^{-1}\,$, which is a continuous $t$-oscillating function with $\Re (\Tr U_A(t))= 0$ for infinitely many $t>0$. The same assertion also holds for the imaginary part.
\\
Furthermore, $\Tr U_A(t) \sim_{t \downarrow 0}{(1-ic)}/{(1+c^2)}\,t^{-1}$ and $\norm{U_A(t)}_1= \sum_{k\geq 1} e^{-tk}=(e^{t}-1)^{-1}\sim_{t \downarrow 0} t^{-1}$. In particular, we have
$\lim_{t\to 0}\,  \abs{\Tr U_A(t)}/\norm{U_A(t)}_1 =(1+c^2)^{-1/2} \neq 1$. \\
It is worth noting  here that, $\Re (\Tr U_A(t)) \to_{t \downarrow 0} +\infty\,$
whereas $\Im (\Tr U_A(t)) \to_{t \downarrow 0}  -\text{sign}(c)\, \infty\,$, and we can change the sign of the imaginary part of $\Tr U_A(t)$ near $t=0$ by modifying the sign of $c$, without any impact on the trace-norm asymptotics of $U_A(t)$.

This observation might suggest that the imaginary part of $\Tr\, U_A(t)$ is not significant in such asymptotic behaviour. However, this is not the case. To this aim, consider
a normal operator $A$ on $l^2(\bbN)$ with spectrum $\sigma(A) = \{\lambda_k = (1+i)k \,\vert\, k \in \mathbb{N}\}$, where each eigenvalue $\lambda_k$ has \textit{multiplicity} $k$. Then for $t\downarrow 0$ the asymptotics of the trace-norm of the Gibbs semigroup generated by $A$ is determined by imaginary part of $\Tr \, e^{-tA}$.
Using the Lidski\u\i \ theorem, we obtain
\begin{align}
 &\norm{e^{-tA}}_1=\sum_{k\geq 1} k e^{-tk}=(2 \sinh \,t/2)^{-2} \underset{t\downarrow 0}{\sim} t^{-2}\,, \nonumber
\\
&
\Re (\Tr \,e^{-tA})=\Re\sum_{k\geq 1} k e^{-t (1+i)k}=\frac{(e^{3 t}+e^t) \cos \,t-2 e^{2t}}{\left(e^{2 t}-2 e^t \cos \,t+1\right)^2} \underset{t\downarrow 0}{\sim} -\frac{1}{12}\,, \label{Real part}
\\
&\Im (\Tr \,e^{-tA}) =-\frac{(e^{3 t}-e^t) \sin\,t}{\left(e^{2 t}-2 e^t \cos \,t+1\right)^2}\, \underset{t\downarrow 0}{\sim}\, -\frac{1}{2}t^{-2}\,.\nonumber
\end{align}
Furthermore, \eqref{Real part} shows that, for $t$ small enough, $\Re (\Tr\,e^{-tA})<0\,$.
\end{remark}

The following corollary is rather convenient for computing an explicit asymptotic behaviour.
\begin{corollary}
\label{cor-computation}
Let $A\in Q(M,\omega_0)$ be a normal operator generating a Gibbs semigroup and $f$:\,$(0,\infty) \mapsto \bbR$ be a measurable function.

\emph{(i)} If $\lambda_k(\Re A) \underset{k \uparrow \infty}{\sim}\,f(k)\,$, then $
\norm{e^{-tA}}_1 \,\underset{t \downarrow 0}{\sim}\, \int_0^\infty \dd x \, e^{-tf(x)}\,$.

 \emph{(ii)} If $\lambda_k(R_{\Re A}(\lambda)) =\OO_\infty((f(k))^{-1})\,$ for some $\lambda\in \rho(\Re A)\,$, then for $c=\limsup_{k\to \infty} f(k)\,\lambda_k(R_{\Re A}(\lambda)) \,$, we have  $\norm{e^{-tA}}_1 =\OO_0( \int_0^\infty \! \dd x \, e^{-t c^{-1} f(x)}\,)\,$  when the last integral converges.
 \end{corollary}

\begin{proof}
By Lemma \ref{lem-ReA et omega0}, without loss of generality, we may assume that $\omega_0 =0$,
and so $\Re A>0$.
As a consequence, $\norm{e^{-tA}}_1=\norm{e^{-t\Re A}}_1=\sum_{k\geq 1} e^{-t \Re \lambda_k(A)}$, see \eqref{normal trace-norm}, where we put $\lambda_k\vc \lambda_k(\Re A) >0$.

(i) For each $\varepsilon>0$, there exists $K\in \bbN$ such that for any $k>K$, $f(k)-\varepsilon\leq \lambda_k \leq f(k) +\varepsilon$. Thus
\begin{align}
\label{encadrement}
\sum_{k> K} e^{-t(f(k)+\varepsilon)} \leq \norm{e^{-t\Re A}}_1 -\sum_{k=1}^K e^{-t\lambda_k}\leq \sum_{k> K} e^{-t(f(k)-\varepsilon)}\,.
\end{align}
Since by Proposition \ref{prop-behaviour at 0} we have
$\norm{e^{-t\Re A}}_1 -\sum_{k=1}^K e^{-t\lambda_k} \,\underset{t \downarrow 0}{\sim}\, \norm{e^{-t\Re A}}_1 -K \underset{t \downarrow 0}{\sim}\, \norm{e^{-t\Re A}}_1=\norm{e^{-t A}}_1\,$,
the estimates of \eqref{encadrement} and the same proposition again imply
$\norm{e^{-tA}}_1 \underset{t \downarrow 0}{\sim}\,\sum_{k\geq 1} e^{-t f(k)}\,$.
\\
By hypothesis, $f$ must be positive and nondecreasing, we conclude the proof using the estimates:
\begin{align}
\label{serie/integrale}\sum_{k\geq 1} e^{-tf(k)} = \sum_{k\geq 0} e^{-tf(k+1)} \leq \sum_{k\geq 0} \int_k^{k+1} \! \dd x\,e^{-tf(x)} =\int_0^\infty \! \dd x \,e^{-tf(x)} \leq \sum_{k\geq 1} e^{-tf(k)} + 1\,.
\end{align}

(ii) By hypothesis, $(\lambda_k(\Re A))^{-1}=\lambda_k((\Re A)^{-1})=\OO_\infty(f(k)^{-1})\,$. Thus, for $k>K$ large enough, there exists $c>0$ such that $0<(\lambda_k(\Re A))^{-1} < c f(k)^{-1}\,$, and hence $e^{-t \lambda_k(\Re A)} < e^{-t c^{-1} f(k)}$. From this inequality, we deduce the claim since $\sum_{k\geq 1}e^{-t c^{-1} f(k)}$ and $\int_0^\infty \!\!\dd x\,e^{-tc^{-1} f(x)}$ have the same asymptotics for $t \downarrow 0$ by virtue of
\eqref{serie/integrale}.
\end{proof}

\section{Tauberian results for generators of Gibbs semigroups}
\label{Tauberian}

The following result adapts Theorem~A.2 from the Appendix to our context. This is significant because it connects the asymptotic behaviour of the trace-norm of Gibbs semigroups for $t \downarrow 0$
with that of the \textit{counting function} $N(\lambda)$ of their generators.

\begin{theorem}
\label{thm-Kamarata for series}
Let $f \in \RV$ (so $f(x) = x^p\, L(x)$ with $p=\textup{ind}_f \geq 0$, and $L(x)$ a positive measurable function slowly varying at infinity) and let $\{\lambda_k\}_{k\geq 1}$ be a non-decreasing unbounded family of real numbers such that
\begin{align*}
N: \lambda \in [0,\infty) \mapsto N(\lambda)\vc {\rm{card}}\{k \in \bbN \, \vert \,\lambda_k
\leq \lambda\} \quad  \text{is finite for each $\lambda$}.
\end{align*}
Then

\emph{(a)}  The following two assertions are equivalent:

\hspace{0.3cm} \emph{(i)} $N(\lambda)=\OO_\infty(f(\lambda))\,$.

\hspace{0.3cm} \emph{(ii)} $\sum_{k=1}^\infty \,e^{-t\lambda_k} = \OO_0(f(t^{-1}))\,$.

\emph{(b)}  The following implications hold:

\hspace{0.3cm} \!\emph{(iii)} If  $\underset{\lambda \uparrow \infty} {\liminf} \,\,f(\lambda)^{-1}\,N(\lambda) \geq c\,$,\,\, then \,\,$\underset{t \downarrow 0} {\liminf} \,\,f(t^{-1})^{-1} \sum_{k=1}^\infty e^{-t\lambda_k}\geq c \,\Gamma(1+p)\,$.

\hspace{0.3cm} \emph{(iv)}
 In the opposite direction, we have
\begin{align*}
\text{If }\,\underset{t \downarrow 0} {\liminf} \,\,f(t^{-1})^{-1} \sum_{k=1}^\infty e^{-t\lambda_k}>0 \,\,\ {\text{and}}\ \,\,  \sum_{k=1}^\infty e^{-t\lambda_k} =\OO_0(f(t^{-1}))\,,\,
\text{ then, } \underset{\lambda \uparrow \infty} {\liminf} \,\,f(\lambda)^{-1}\,N(\lambda) >0\,.
\end{align*}

\emph{(c)} Assume that $\textup{ind}_f>0\,$. Then, for any $C \in [0,\infty)\,$, the following two assertions are equivalent

\hspace{0.3cm} \emph{(v)} $N(\lambda) \underset{\lambda \uparrow \infty}{\sim}\, C\, f(\lambda)\,$.

\hspace{0.3cm} \emph{(vi)} $\sum_{k=1}^\infty e^{-t\lambda_k}  \underset{t \downarrow 0}{\sim}\, \Gamma(1+p)\, C\,f(t^{-1})\,$.

\emph{(d)} Moreover, in this theorem, if $f(x) = x^p  \, \Ln_k(x)^r$ (see definition in \eqref{def-Lnk}) with $r\in \bbR\,$,
then $\Ln_k(\cdot)$ can be replaced by $\ln^{\circ k}(\cdot)$ (see definition in \eqref{def-lnk}).
\end{theorem}

\begin{proof}
Assume first that $\lambda_k \geq 0$ for all $k\geq 1$. If $\mu\vc \sum_{k\geq 1} \delta_{\lambda_k}$, then $\mu$ is a positive measure which is $\sigma$-finite since $\mu([0,\lambda])=N(\lambda)$.
As a consequence, the theorem is a direct application of Theorem \ref{appendix:Thm-Karamata} since we have
$L_\mu(t)=\int_{[0,\infty)} \dd \mu(x)\, e^{-tx}= \sum_{k\geq 1} e^{-t\lambda_k}$.
\\
If there exists some $\lambda_k<0$, we can define $\lambda'_k \vc \lambda_k-\lambda_1$. Then $\lambda'_k \geq 0$ and
\begin{align*}
\sum_{k\geq 1} e^{-t\lambda_k} =e^{-t\lambda_1}\sum_{k\geq 1} e^{-t\lambda'_k} \underset{t \downarrow 0}{\sim}\,
\sum_{k\geq 1} e^{-t\lambda'_k}\,.
\end{align*}
which brings us back to the previous assumption.
\end{proof}
In (c), the case ind$_f=0$ is excluded.
Otherwise, as discussed in Remark \ref{mu finite}, the implication (vi) $\implies$ (v) gives a finite $\mu$, which is incompatible with the hypothesis in Theorem \ref{thm-Kamarata for series} that $\lambda_k\to \infty$.

Since our analysis primarily focuses on the behaviour of $\norm{e^{-tA}}_1$ as $t\downarrow 0$, we now present the following key consequence of previous theorem. To this aim, we introduce for an operator $A$ with only discrete spectrum $\{\lambda_k(A)\}_{k\geq 1}$ the counting function:
\begin{align}
\label{def-NA}
N_A(\lambda)\vc  {\rm{card}}\{k\,\vert \,\lambda_k(A) \leq \lambda \}\,.
\end{align}

\begin{corollary} \label{thm-Karamata for semigroup}
Let $A$ be a normal operator with compact resolvent generating a $C_0$-semigroup.

\emph{(a)} Given $p>0$ with $\,r\in \bbR$, or $p=0$ with $r>0$, and $k\geq1$, the following behaviours are equivalent:

\hspace{0.3cm}\emph{(i)} $A$ is the generator of a Gibbs semigroup with asymptotics $\norm{e^{-tA}}_1 = \OO_0(t^{-p}\,(\Ln_k t^{-1})^{\,r}) \,$.

\hspace{0.3cm}\emph{(ii)} $\Re A \neq 0$ and $N_{\Re A}(\lambda)= \OO_\infty(\lambda^{p} \,(\Ln_k \lambda)^r)\,$.

\hspace{0.3cm}\emph{(ii')} $\Re A \neq 0$ and $N_{\Re A}(e^{W_0(\lambda)})=\OO_\infty(\lambda^p \,W_0(\lambda)^{r-p})\,$ when $k=1$\,.

\hspace{0.3cm} Here, $W_0$ is the Lambert function which is defined for $xe^x=y$ with $x\geq 0$ by $x=W_0(y)\,$.

\hspace{0.3cm}\emph{(iii)}  When $r=0$, this is equivalent to $\Re A \neq 0$ and $\lambda_k(\Re A)^{-1} = \OO_\infty(k^{-1/p})\,$.

\emph{(b)}
 \!\emph{(iv)} If  $\underset{\lambda \uparrow \infty} {\liminf} \,\,f(\lambda)^{-1}\,N_{\Re A}(\lambda) \geq c\,$,\,\, then, \,\,$\underset{t \downarrow 0} {\liminf} \,\,f(t^{-1})^{-1}\,\norm{e^{-tA}}_1\geq c \,\Gamma(1+p)\,$.

\hspace{0.3cm} \emph{(v)}
In the opposite direction, given $f\in \RV$,
\begin{align*}
\text{If }\,\underset{t \downarrow 0} {\liminf} \,\,f(t^{-1})^{-1} \norm{e^{-tA}}_1>0 \,\,\,{\text{and}}\,\,\, \norm{e^{-tA}}_1 =\OO_0(f(t^{-1}))\,,\,
\text{ then, } \underset{\lambda \uparrow \infty} {\liminf} \,\,f(\lambda)^{-1}\,N(\lambda) >0\,.
\end{align*}
\indent
\emph{(c)} Assume $p>0$ with $\,r\in \bbR$ or $p=0$ with $r>0$ and $k\in \bbN$. For any $C \in [0,\infty)\,$, the following are equivalent

\hspace{0.3cm} \emph{(vi)} $N_{\Re A}(\lambda) \underset{\lambda \uparrow \infty}{\sim}\, C\, \lambda^p \,(\ln^{\circ k} \lambda)^r\,$.

\hspace{0.3cm} \emph{(vii)} $\norm{e^{-tA}}_1  \underset{t \downarrow 0}{\sim}\, \Gamma(1+p) C\,t^{-p}\,(\ln^{\circ k} t^{-1})^r$.
\end{corollary}

\begin{remark}
\label{rem-exclusion}

In the hypothesis of Corollary \ref{thm-Karamata for semigroup},
we require $\{e^{-tA}\}_{t\geq 0}$ in (a) to be a Gibbs semigroup and thus, we may assume that $A\in Q(M, \omega_0)$.
Note that the case $\Re A=0$ is excluded since otherwise $\{e^{-tA}\}_{t\geq 0}$ becomes a unitary (semi-)group.

Then the case: \(p=0,\; r\leq 0\) in (i) is also excluded since for a Gibbs semigroup, one has \(\Re A \neq 0\). Indeed, if we were to obtain $\norm{e^{-tA}}_1 \sim_{t\downarrow 0} \norm{e^{-t(\Re A+\omega_0\bbbone)}}_1 = \OO_0(t^0\,(\ln t^{-1})^{\,r})\,$,
this would contradict Proposition \ref{prop-behaviour at 0}. See also Remark \ref{mu finite} which states that if
$\norm{e^{-tA}}_1 = L_\mu(t) \sim_{t\downarrow 0}\, t^0$ for \(A>0\), then the measure \(\mu=\sum_k\delta_{\lambda_k}\) is finite.

The case $p=0$ and $r>0$ is allowed: For instance, $N_{\Re A}(\lambda)= \OO_\infty((\ln\lambda)^r)$ in (ii) means that for
$\lambda_k=\lambda_k(\Re A)$, when $K\in \bbN$ is large enough, one obtains
\begin{align}
\label{equi}
k < c_1\, \ln(\lambda_k)^r \iff e^{c_2\,k^{1/r}} < \lambda_k \iff \lambda_k^{-1} < e^{-c_2\,k^{1/r}} \quad \text{for } k>K \, ,
\end{align}
and in particular, $N_{\Re A}(\lambda)= \OO_\infty((\ln \lambda)^r)  \iff \lambda_k^{-1} = \OO_\infty (e^{-c \,k^{1/r}})\,$.
Thus $\{e^{-tA}\}_{t\geq 0}$ with a normal generator $A$ is, in turn, a normal semigroup, and consequently a Gibbs semigroup since \eqref{equi} implies
\begin{align*}
\norm{e^{-tA}}_1 =\norm{e^{-t\Re A}}_1 = \sum_{k=1}^K e^{-t\lambda_k} +\sum_{k=K+1}^\infty e^{-t\lambda_k}
< \sum_{k=1}^K e^{-t\lambda_k}+ \sum_{k=1}^\infty e^{-t c_2 \exp{(k^{1/r})}}
<\infty\,.
\end{align*}
\end{remark}

\begin{proof}\! \!\!of Corollary \ref{thm-Karamata for semigroup}:\\
Assume $A\in Q(M,\omega_0)$. The compactness of the resolvent yields that a normal generator $A$ has only a discrete spectrum. Thus, the normal operator $B = A+\omega_0 \bbbone$, as well as  its real part $\Re B$, have discrete spectra. Since $B$ is normal, we have $s_k(e^{-tB)})=s_k(e^{-t\Re B})$ for any $k\geq 1$ and $\lambda_k=\lambda_k(\Re B)\geq 0$ by virtue of Lemma \ref{lem-ReA et omega0}.
For this reason, $\norm{e^{-tB}}_1=\norm{e^{-t\Re B}}_1$ is bounded when $A$ is a Gibbs semigroup generator.

In the latter case, the boundedness of
$\norm{e^{-tA}}_1=\norm{e^{-t\Re A}}_1$ implies that
$\lambda_k(\Re A) \to \infty$, when $k\to \infty$ together with  $N_{\Re A}(\lambda)<\infty$ for any $\lambda$ and the same for $\lbrace\lambda_k(\Re B)\rbrace_{k\geq 1}$.
Thus the hypothesis of Theorem \ref{thm-Kamarata for series} is fulfilled since the function $f:x>0 \mapsto x^p\,(\Ln_k x)^r$ is in $\RV$ proving points (b) and (c) and the equivalence in (a) of (i) and (ii).

To show the equivalence between (ii) and (ii'), we first remark that for $x(\lambda)\vc e^{W_0(\lambda)}$ with $\lambda >e$, we obtain $W_0(\lambda)>1$ and $x(\lambda)\to \infty$ when $\lambda \to \infty$. Hence,
\begin{align*}
x(\lambda)^p\,(\Ln x(\lambda))^r = e^{pW_0(\lambda)} (W_0(\lambda))^r = (W_0(\lambda) \,e^{W_0(\lambda)})^p\,(W_0(\lambda))^{r-p} =\lambda^p \,(W_0(\lambda))^{r-p}\,.
\end{align*}
The quoted equivalence follows from the asymptotic correspondence between $\lambda$ and $x(\lambda)$ through this bijective transformation.

(i) $\implies$ (iii): As in the proof of Theorem \ref{thm-Kamarata for series}, we may assume that $\lambda_k(\Re A)\cv \lambda_k>0$ for any $k\geq 1$. By hypothesis, for any $ k\in \bbN$, $ke^{-t\lambda_k} <\sum_{n=1}^k e^{-t\lambda_n} \leq \sum_{n\geq 1} e^{-t\lambda_n} =\OO_0(t^{-p})$. Thus there exists $c>0$ and $t_0>0$ such that $ke^{-t\lambda_k} < ct^{-p}$ for any $t<t_0$. Choosing $k$ large enough to get $t=\lambda_k^{-1}<t_0\,$, we obtain, still for $k$ large enough,
$k\,e<c\lambda_k^p\,$. Thus $\lambda_k^{-1}=\OO_\infty(k^{-1/p})\,$.

 (iii) $\implies$ (i):  We still assume $\lambda_k(\Re A)\cv \lambda_k>0$ for any $k\geq 1$. By hypothesis, there exists $k_0\in \bbN$,  $c>0$ such that for $k>k_0\,$, $\lambda_k >ck^{1/p}$ and
$\sum_{k\geq 1} e^{-t\lambda_k} =\sum_{k= 1}^{k_0} e^{-t\lambda_k} +\sum_{k> k_0} e^{-t\lambda_k} \leq k_0 + \sum_{k\geq 1} e^{-ctk^{1/p} }$.
By Corollary \ref{cor-computation} (i), we have $\sum_{k\geq 1} e^{-ctk^{1/p} } \sim_{t\downarrow 0}\, c't^{-p}$ implying  $\,\sum_{k\geq 1} e^{-t\lambda_k} =\OO_0(t^{-p})\,$.
\end{proof}

If one relaxes the condition of normality for generator $A$ in Corollary \ref{thm-Karamata for semigroup}, then
the corresponding $C_0$-semigroup also loses this property. As a result, we have difficultiess with the
definition of the real part $\Re A$ (see \eqref{normal-oper-sum}) and consequently with estimate of the trace-norm $\norm{e^{-t{A}}}_1$. However, if the operator $A$ is $m$-sectorial, this issue can becan be circumvented since,
as discussed in Remark \ref{rem:hol-SG}, in that case, $A$ generates a holomorphic Gibbs semigroup.

\begin{proposition}
\label{prop-Karamata-hol}
Let $A$ be an $m$-{sectorial} operator with spectrum $\sigma(A)$ contained in sector $S_{\pi/2 - \theta, \, -\omega_0} \,$  which has vertex at $- \omega_0 \in \bbR\,$ and semi-angle $\pi/2 - \theta \leq \pi/2$, cf. \emph{(\ref{sector})}.

Then the following assertions hold\emph{:}

\emph{(a)} There exists a self-adjoint operator
${\re }\, A := (A \dot{+} A^*)/2 \, > - \omega_0 \bbbone \, $,
where the real part of $A$ is defined on ${\dom}\,({\re }\, A)$ as the \textit{form-sum}
of operators $A$ and $A^*$.

\emph{(b)} If $e^{- t {\re }A} \in \calL^1$ for $t > 0\,$, then
$\{e^{- z A}\}_{z\in S_{\theta}}$ is a quasi-bounded holomorphic Gibbs semigroup \emph{(}cf. Remark \ref{rem:hol-SG}), and for $z\in S_{\theta-\epsilon}$ with $0 < \epsilon < \theta$, we have
$\norm{e^{-t A}}_1 \leq \norm{e^{-t {\re }A}}_1 \, $ for $t : = {\re} \, z > 0\,$.

\emph{(c)} If $\dim(\Ker({A})) <\infty\,$, then for given $c> 0$, $k\in \bbN$, $p>0$ and $r\in \bbR$, or for $p=0$ and $r>0\,$, the asymptotics
$\norm{e^{-t{{\re }A}}}_1  \underset{t\downarrow  0}{\sim} c \,t^{-p}\, (\ln^{\circ k} t^{-1})^r$ and
$N_{{\re }A}(\lambda) \underset{\lambda \uparrow  \infty}{\sim}
\tfrac{c}{\Gamma(p+1)}\,\lambda^{p} \,(\ln^{\circ k} \lambda)^r\,$
are equivalent.
\end{proposition}

We recall that the inequality between norms in
Proposition \ref{prop-Karamata-hol} (b) becomes an equality if, in addition, the operator $A$ is normal, see \cite[Corollary 4.31]{Zagrebnov2}. Therefore, Proposition \ref{prop-Karamata-hol} provides an assertion, which is an \textit{alternative} to Corollary \ref{thm-Karamata for semigroup} for the real part ${\re\, A}\,$, instead of ${\Re\, A}\,$.

\begin{proof}(a) The proof is standard, see for example \cite[Chapter VI \S2]{Kato}.

(b) By Definition \ref{def:GibbsSG} and Remark \ref{rem:hol-SG},
the extension of the $C_0$-semigroup $\{e^{- t A}\}_{t \geq 0}$ to a quasi-bounded holomorphic semigroup for $z\in S_{\theta-\epsilon}$ and $0 < \epsilon < \theta$ is well-established
and the inequality $\norm{e^{-t A}}_1 \leq \norm{e^{-t {\re }A}}_1 $ for $t > 0$, follows from assertion \cite[Proposition 4.30]{Zagrebnov2} since it is independent of $\omega_0$. Then
$\{U_{A}(z)\}_{z\in S_\theta \, \cup \, \{0\}} \,$
is a quasi-bounded holomorphic Gibbs semigroup.

(c)  The condition $\dim(\Ker(A)) < \infty$ prevents the operator $A$ from having an infinite-dimensional null space, which would be incompatible with its role as the generator of a Gibbs semigroup. Furthermore, since $\Re A > -\omega_0 \bbbone$, the remainder of the proof proceeds analogously to Corollary \ref{thm-Karamata for semigroup}.
\end{proof}

\section{Case of positive generators}\label{sec2}

Let $A$ be a densely defined closed operator (possibly unbounded and non-invertible) on a separable Hilbert space $\calH$. According to a von Neumann theorem, the operator $A^*A$ is positive and self-adjoint, and its domain $\Dom(A^*A)$ serves as a core for operator $A$.
So, the operator $(\bbbone +A^*A)^{-1}$ is a bounded self-adjoint operator satisfying $0 \leq (\bbbone +A^*A)^{-1} \leq \bbbone$.

\begin{lemma}
\label{lem-behaviour for positive operator}
Let $A$ be a densely defined closed operator on $\calH$. We introduce the operators
$B:=A^*A$ and $X:=\bbbone +B$, as well as a parameter $p >0$.

\emph{(i)}  If $R_{A^*A}(z = -1)=X^{-1} \in \calL^{p}$, then $\{e^{-t\,B}\}_{t\geq 0}$ is a Gibbs semigroup,
and $\norm{e^{-t\,B}}_1=\Tr\,e^{-t\,B} = \OO_0(t^{-p})\,$.

\emph{(ii)} Conversely, suppose $\{e^{-t\,B}\}_{t\geq 0}$ is a Gibbs semigroup and
$\norm{e^{-t\,B}}_1 = \OO_0(t^{-p})\,$. Then $X^{-1} \in \calL^q$ for any  $q>p\,$.
\end{lemma}
\begin{proof}
Note that $X^{-1}\in \calL^p$ implies that $X^{-p}\in \calL^1$ and that for a positive operator $X$, the converse also holds true.

(i) (See also \cite{Connes-Mosco}, \cite[Proposition 2.3]{EckIoc}.) Note that a non-negative self-adjoint operator $B$ generates a strongly continuous contractive semigroup $\{e^{-t\,B}\}_{t\geq 0}\,$.
Now, similarly to \cite[Lemma 10.8]{GracVariFigu01a}, we have $e^{-tB}=X^{\,p} \,e^{-tB} \,X^{-p} \in \calL^1$
since $X^{\,p} \,e^{-tB} $ is bounded and  $X^{-p}\in \calL^1$. More precisely, the mapping
$x>0 \mapsto (1+x)^{\,p}\,e^{-tx}$ has a supremum equals to
$p^p e^{-p} e^t t^{-p}$ for $t>0$. Thus, we obtain a Gibbs semigroup such that
$\norm{e^{-tB}}_1  \leq p^p e^{-p}\norm{ X^{-p} }_1 \,\,e^t t^{-p}<\infty$ and $\norm{e^{-tB}}_1  =\OO_0(t^{-p})\,$.

(ii) Since $B$ generates a Gibbs semigroup, its spectrum consists entirely of discrete positive eigenvalues: $\{\lambda_k = \lambda_k (B)\}_{k\geq1} \,$, cf. \cite[Proposition 4.21]{Zagrebnov2}.\\
If $q> p > 0 $, then the mapping $t>0 \mapsto \Tr\,(e^{-tX})\,t^{q-1}$ is continuous (see \cite[Proposition 4.2]{Zagrebnov2}) and integrable on $[0, \infty)$ because its behaviour at infinity is $\norm{e^{-tX}}_1\,t^{q-1} =\OO_\infty(e^{-t}) $ and at zero, by hypothesis, $\norm{e^{-tX}}_1\,t^{q-1}=\OO_0(t^{-p})\,t^{q-1}=\OO_0(t^{q-p-1})\,$, where $q> p > 0 \,$.
\\
Therefore, using the monotone convergence theorem for positive functions,
we can interchange the sum and the integral in the second equality:
\begin{align}
\label{main equality2}
\Tr\,X^{-q} &= \sum_{k\geq 1} \,\,(1+\lambda_k)^{-q}= \sum_{k\geq 1} \, \Gamma(q)^{-1} \int_0^\infty \,\dd t \, t^{q-1} \ \!e^{-t(1+\lambda_k)}
=\Gamma(q)^{-1}\! \int_0^\infty \!\dd t \, t^{q-1} \ \sum_{k\geq 1} \, e^{-t(1+\lambda_k)}  \nonumber
\\ &= \Gamma(q)^{-1} \int_0^\infty \! \dd t \, t^{q-1} \, \Tr e^{-tX}  < \infty\,,
\end{align}
and consequently $X^{-1}\in \calL^q\,$ for any $q>p\,$.
\end{proof}

\begin{remark}
In Lemma \ref{lem-behaviour for positive operator} the assertion \rm{(ii)} can be improved as follows: \\
\rm{(ii')} Conversely, suppose $\{e^{-t\,B}\}_{t\geq 0}$ is a Gibbs semigroup, and
$\norm{e^{-t\,B}}_1 = \OO_0(f(t^{-1}))$ with $f\in \RV$. Then, $X^{-1} \in \calL^q$ for any $q>p=\text{ind}_f$.
The proof follows the same steps as before, starting with $q=p+\varepsilon$ instead of $p$, since $f(t^{-1})=\oo_0(t^{-(p+\varepsilon)})$ for any $\varepsilon>0$ thanks to the behaviour in
\eqref{behaviour of f(t-1)}.
\end{remark}

\begin{remark}
\label{p<q}
(a)  The first part of assertion (i) can be derived from \cite[Definition 4.26 and Proposition 4.27]{Zagrebnov2}. In fact, under the conditions of the lemma, the operator $B$ is $m$-sectorial with semi-angle $\pi/2$ and its resolvent $R_{B}(z = -1) \in \mathcal{L}^{p}$ for $p\geq 1$. According to \cite[Definition 4.26]{Zagrebnov2}, this implies that $B$ is a $p$-generator. Then, using \cite[Proposition 4.27]{Zagrebnov2}, we conclude that $\{e^{-tB}\}_{t\geq 0}$ is a \textit{holomorphic} Gibbs semigroup.

(b)  In part (ii), the inequality $q>p>0$ cannot be relaxed to $q = p$. To see this, consider on the space $\calH=l^2(\bbN)$ a diagonal multiplication operator $X$ with spectrum  $\sigma(X) = \{k^{1/p} : \, k\in \bbN\}$. Thus, $X$ is an unbounded, self-adjoint on its domain, positive, invertible operator and
\begin{align}
\label{sum and integral}
\sum_{k= 1}^\infty e^{-tk^{1/p}} =\sum_{k=0}^\infty e^{-t(k+1)^{1/p}} \leq \sum_{k=0}^\infty \int_k^{k+1}\!\!\!\dd x\, e^{-t x^{1/p}}=\int_0^\infty \!\! \!\dd x\,e^{-tx^{1/p}} =p! t^{-p} \leq \sum_{k=1}^\infty e^{-tk^{1/p}}+1\,.
\end{align}
Therefore, $\norm{e^{-tX}}_1=\OO_0(t^{-p})$ since the sums are finite. However,  $X^{-1}\notin \calL^p$ because $\norm{X^{-p}}_1= \sum_{k\geq 1}\,k^{-1}$ is not finite.
\end{remark}

\begin{remark}
\label{rem:2.5}
If $A$ is \textit{not} closed but $B:=A^*A$ is densely defined, then the positive operator $B$ is symmetric with equal deficiency indices. Consequently, $B$ has positive self-adjoint extensions. Among these extensions, there are two distinguished ones: $B_K$ and $B_F$, known as the \textit{Krein--von Neumann} (or "soft") and \textit{Friedrichs} (or "hard") extensions, respectively. By the Krein theorem, for any positive self-adjoint extension $\tilde{B}$ of $B$, the following inequalities hold:
\begin{align}
\label{B_F-B_K}
(B_F+\lambda\bbbone)^{-1} \leq (\tilde{B}+\lambda\bbbone)^{-1} \leq (B_K + \lambda \bbbone)^{-1}, \quad \forall \lambda>0 \,.
\end{align}
These extensions are discussed in details in \cite[Chapter 10, Section 3]{Birman}, or  \cite[Sections 13.3 and 14.8]{Schmudgen}.
\\
Recall that for a self-adjoint extension $\tilde{B}$, one can use the \textit{minimax principle} (see, e.g., \cite[Proposition 4.23]{Zagrebnov2}) to define a sequence of numbers $\{\mu_n (\tilde{B}) \}_{n\geq 1}$. If $\lim_{n \to \infty} \mu_n (\tilde{B}) = \infty$, then the operator $\tilde{B}$ has a \textit{point spectrum} (with finite multiplicity), and the sequence $\{\mu_n (\tilde{B})\}_{n\geq 1}$ coincides with the eigenvalues of $\tilde{B}$, denoted by $\{\lambda_n (\tilde{B})\}_{n\geq 1}$. In light of \eqref{B_F-B_K}, we can establish the following estimates:
\begin{align}
\label{B_F>B_K}
\lambda_n (B_K) \leq \lambda_n (\tilde{B}) \leq \lambda_n (B_F), \quad n\geq 1.
\end{align}
Using the minimax principle and \eqref{B_F>B_K}, we can conclude that the extension $B_F$ has a \textit{discrete spectrum} and the kernel $\Ker\,(B_F) = \{0\}$, while the extension $B_K$ may have a \textit{point spectrum}, as the kernel $\Ker\,(B_K) = \Ker\,(B^*)$ can be infinite-dimensional. If the extension $B_K$ has a \textit{discrete spectrum}, then the same holds for all extensions $\tilde{B}$.

As a consequence, if $A$ is \textit{not} closed, then a self-adjoint extension $\tilde{B}$ of the operator $B=A^*A$ may fail to generate a Gibbs semigroup, as the kernel $\Ker\,(\tilde{B})$ could be infinite-dimensional. Nevertheless, Lemma \ref{lem-behaviour for positive operator} remains valid for $\tilde{B}$ instead of $B$, because the hypotheses in (i): $(\bbbone+\tilde{B})^{-1}\in \calL^p$, and in (ii): $\{e^{-t\,\tilde{B}}\}_{t\geq 0}$ is a Gibbs semigroup, circumvent this potential obstruction.
\end{remark}
\bigskip

The obstacle encountered in Remark \ref{p<q} (b) can be overcome by using $weak$-$\calL^p$ spaces. To this aim,  we shall utilise the ideals $\calL^{p,\infty}$ (generated by the operator Diag\,$\{(k ^{-1/p}\}_{k \in \bbN}$) and the Lorentz ideal $\calM^{1,\infty}$ which is the dual of the Macaev ideal \cite{LSZ}:


\begin{definition}
\label{def-Zp}
Let $p>0\,$.
\begin{align*}
\calL^{p,\infty} &\vc \{\, A\in \calL^\infty \,\vert \, s_k(A) =\OO_\infty(k^{-1/p})\,\},\text{ equipped with } \norm{A}_{p,\infty} \vc \sup_{k\geq 1} \,k^{1/p} \,s_k(A)\,.\\
\end{align*}

\vspace{-1cm}

\noindent \text{We shall also use the spaces introduced in \cite{CRSS} and \cite{CGRS}}: for $p\geq 1$,
\begin{align*}
\calM^{p,\infty} & \vc \{\,A\in \calL^\infty \,\vert \,\sum_{k=1}^n s_k(A)^p =\OO_\infty(\ln(1+n))\, \}\, , \text{ with }  \norm{A}^p_{\calM^{p,\infty}} \vc \sup_{n\in \bbN} \,\frac{1}{\ln(1+n)}\,\sum_{k=1}^n s_k(A)^p\,,\\
\calZ_p & \vc \{A \in \calL^\infty \,\vert\,\,  \norm{A}_{\calZ_p} \vc
\limsup_{q \downarrow p} \,\big((q-p) \Tr \, \abs{A}^q \big)^{1/q} < \infty \}\,.
\end{align*}
\end{definition}
We evoke that, for  $p>0$, $\norm{\cdot}_{p,\infty}$  is a quasi-norm and
$\calL^{p,\infty}$ is a quasi-Banach space and for $p>1$, $\norm{\cdot}_{p,\infty}$ is equivalent to a (unitarily invariant) Banach norm.
The vector spaces $\calM^{p,\infty}$ and $\calZ_p$ coincide, see \cite[page 387]{CGRS} (but equipped with different norms) and since $\norm{\,A}_{\calZ_1} = (p\norm{\,A^p}_{\calZ_1})^{1/p}\,$, we get for $A>0$, $A\in \calZ_p$ if and only if $A^p\in \calZ_1$. Note the strict inclusions $\calL^p  \subsetneq \calL^{p,\infty} \subsetneq \calL^{p,\infty}  \subsetneq \calM^{p,\infty} = \calZ_p\,$. For details, see \cite{LSZ}.

For convenience, we now recall a useful result from \cite[Section 5]{CRSS} (see also \cite{CGRS}).
\begin{proposition}\label{prop-trace versus Zp}
Let $A$ be a  positive self-adjoint operator.

\emph{(a)} Given $p>0$, the following are equivalent:

\hspace{0.3cm} \emph{(i)}  $A^{-1} \in \calL^{p,\infty}\,$.

\hspace{0.3cm}\emph{(ii)} $\Tr \, e^{-tA} = \OO_0(t^{-p})\,$.

\emph{(b)} Let $p\geq 1$. If $A^{-1} \in \calZ_p\,$, then  $\{e^{-t A}\}_{t\geq 0}$ is a Gibbs semigroup with behaviour $\Tr e^{-tA} =\OO_0(t^{-\tfrac{p}{1-\varepsilon p}})$ for any $0<\varepsilon<1/p$.
\end{proposition}

\begin{proof}
(a) (i)$\implies$(ii) If $A^{-1}\in\calL^{p,\infty}\,$, then $\lambda_k^{-1}\vc\lambda_k(A)^{-1}=\lambda_k(A^{-1}) = \OO_\infty(k^{-1/p})$. Thus there exists $c>0$ and $K\in \bbN$ such that if $k>K$ then $\lambda_k^{-1} <c\,k^{-1/p}$. \\
As a consequence,
$\Tr\,e^{-tA}=\sum_{k=1}^\infty e^{-t\lambda_k} \leq \sum_{k=1}^K e^{-t\lambda_k} +
\sum_{k=1}^\infty e^{-tc^{-1} \,k^{1/p}} = \OO_0(t^{-p})\,$, as shown in \eqref{sum and integral}.

(a) (ii)$\implies$(i) If $\Tr \,e^{-tA}=\OO_0(t^{-p})$, Corollary \ref{thm-Karamata for semigroup} (iii) gives $\lambda_k(A)^{-p}=\OO_\infty (k^{-1})$, so $A^{-1}\in \calL^{p,\infty}$.

(b) Since $A^{-p}\in \calZ_1=\calM^{1,\infty}$, we have
\begin{align*}
 \frac{n\,s_n(A^{-p})}{\ln(1+n)} <\frac{1}{\ln(1+n)}\, \sum_{k=1}^n s_k(A^{-p}) \leq  \norm{A^{-p}}_{\calM^{1,\infty}} <\infty\,,
 \end{align*}
and for some $0<\varepsilon<1/p$, there exists $c_\varepsilon >0$ such that for any $n\in \bbN$ we have the inequalities (see Lemma \ref{ln estimate} for the second one; an alternative is to use \eqref{estimates by monomials})
\begin{equation*}
\label{inequality for Zp}
s_n(A^{-1}) \leq \big(\norm{A^{-p}}_{\calM^{1,\infty}} \,{n}^{-1}\, {\ln(1+n)}\big)^{1/p} \leq c_\varepsilon \,
n^{- \, (1/p-\varepsilon)} \ ,
\end{equation*}
and the estimate
\begin{equation}
\label{Tr-estim1}
\Tr \, e^{-t A} =\sum_{n \geq 1} \,e^{-t\, {\lambda_n(A)}} \leq
\sum_{n \geq 1}\,e^{-tc_\varepsilon^{-1}\, n^{a}} <\infty \, ,
\quad \text{ with }a:= \tfrac{1}{p}-\varepsilon \, ,
\end{equation}
 which proves that $\{e^{-t A}\}_{t\geq 0}$ is a Gibbs semigroup.
 \\
 As in \eqref{sum and integral},
 $\sum_{n \geq 1}\,e^{-tc_\varepsilon^{-1}\, n^{a}}  \underset{t \downarrow 0}{\sim} \int_0^\infty \dd x \,e^{-tc_\varepsilon^{-1}\, x^{a}} = c_\varepsilon^{1/a} \Gamma(1+1/a)\, t^{-1/a} =\OO_0( t^{-1/a})\,$, thus \eqref{Tr-estim1} gives $\Tr e^{-tA} =\OO_0(t^{-\tfrac{p}{1-\varepsilon p}})$.
\end{proof}

\begin{remark}
\label{rem-an estimate}
Since $\tfrac{p}{(1-\epsilon p)} > p > 0$, one gets $t^{-\tfrac{p}{1-\epsilon p}} > t^{-p}$,
for $0<t<1$ and $\OO_0(t^{-p})\subset \OO_0(t^{-\tfrac{p}{1-\epsilon p}})$.
\end{remark}

\begin{remark}
Let $A>0$. If $A^{-1} \in \calZ_1\,$, then, as we have seen above there exists a constant $c>0$ such that $s_n(A^{-1}) \leq\,c\,n^{-1}\, \ln(1+n)\,,\,\forall n\in \bbN\,$.
\\
 However, the converse is false: Let $f$ be the non-increasing mapping defined as $f (x) = x^{-1}\ln(1+x)$ for $x>0$. By changing $A$ to $cA$, we can assume that $c\leq 1$. \\
 Suppose $s_n(A^{-1})=n^{-1}\ln(1+n)=f(n)$ when $n\geq 1$. Since $f(k+1)\leq \int_k^{k+1} \dd x\,f(x)\leq f(k)$, we have $\sum_{k=1}^n \,s_k(A^{-1})-s_1(A^{-1}) \leq \int_1^n \dd x\,f(x)\leq\sum_{k=1}^{n-1}\,  s_k(A^{-1})\,.$
 Thus,
 \begin{align*}
 \frac{1}{\ln(1+n)} \,\sum_{k=1}^n \,s_k(A^{-1})\, &\underset{n\uparrow \infty}{\sim} \,\frac{1}{\ln(1+n)} \int_1^n \dd x\,\frac{\ln(1+x)}{x}=  -\frac{1}{\ln(1+n)}\,[\,\frac{\pi^2}{12} + \text{Li}_2(-n)\,]
 \end{align*}
 where the dilogarithm Li$_2$ satisfies the inversion formula (see \cite[equation (1.7)]{Lewin})
 $$
 \text{Li}_2(-n) + \text{Li}_2(-1/n)=-\tfrac{1}{2} \ln^2(n) -\tfrac{\pi^2}{6}.
 $$
 Since
 $\text{Li}_2(-1/n)= \sum_{k=1}^\infty (-1/n)^k/k^2 \,\sim_{n\uparrow \infty} \,- {1}/{n}\,$, we get $\text{Li}_2(-n) \sim_{n\uparrow \infty}-\tfrac{1}{2} \ln^2(n)\,$. This shows that $ \frac{1}{\ln(1+n)} \,\sum_{k=1}^n \,s_k(A^{-1}) \to +\infty$ when $n$ increases, so that $A^{-1}\notin \calZ_1\,$.
 \end{remark}

 In noncommutative geometry, a spectral triple $(\calA,\calH,\calD)$ is called {\it $\theta$-summable} (see \cite{Connes1988}, \cite[Chapter 4, section 8.$\alpha$]{ConnesNCG}) when $\Tr e^{-t\,\calD^2} <\infty$ for $t>0$,  which precisely means that $\{e^{-t\,\calD^2} \}_{t\geq0}$ is a Gibbs semigroup. The triple is called {\it $p$-summable }when $(\calD+i)^{-1} \in \calL^{p,\infty}\,$ for some $p>0$.

\begin{corollary}
\label{p-summability}
Let $(\calA,\calH,\calD)$ be a $\theta$-summable spectral triple and $p>0\,$.\\
Then it is $p$-summable if and only if $\norm{e^{-t\calD^2}}_1=\OO_0(t^{-p/2})\,$.
\end{corollary}

\begin{proof}
This is a direct consequence of previous proposition for $A = \calD^2$.
\end{proof}
Note that the $\theta$-summability hypothesis is quite restricting since it is easy to get (trivial) spectral triples with pathologies. For example, let $\calA=l^\infty(\bbN)$ seen as diagonal operators on $\calH=l^2(\bbN)$ and $\calD$ be a self-adjoint positive invertible  diagonal operator defined by its spectrum.  If $\sigma(\calD)=\{[\ln(n+1)]^{1/2}\,\vert \, n\in \bbN\}$, then $\Tr e^{-t\calD^2}= \sum_{n=1}^\infty (n+1)^{-t}$ is finite only for $t>1$ (thus $\{e^{-t\calD^2}\}_{t\geq 0}$ is not an immediately Gibbs semigroup) even though $\{e^{-t\calD^2}\}_{t\geq 0}$ is a compact semigroup (see \cite[Remark 4.4]{Zagrebnov2}).
\\
Similarly, if $\sigma(\calD)=[\ln^{\circ 2}(n+2)]^{1/2} \,\vert \, n\in \bbN\}$ (see Definition \ref{def-lnk}),
then $\Tr e^{-t\calD^2}= \sum_{n=1}^\infty [\ln(n+2)]^{-t}$ is not finite for
$t>0\,$.

The following result establishes an ''integrated counterpart'' to the $t$-behaviour of $\norm{e^{-tA}}_1$ in comparison with Proposition \ref{prop-trace versus Zp}.
\begin{theorem}\label{4.10}
Let $A$ be a self-adjoint positive invertible operator seen as generator of a semigroup. Then,

\hspace{0.5cm}(i) \,\, If $A^{-1} \in \calZ_1\,$, then the function $f: t\in (0,\infty) \mapsto \norm{\,A^{-1} e^{-tA}}_1\,$ is such that $-f$ is a primitive of the function$: t \mapsto \norm{e^{-tA}}_1\,$.
\begin{align}
\label{Z1}
(ii) \,\,\,\, A^{-1} \in \calZ_1 \iff
\norm{A^{-1} e^{-tA}}_1 = \OO_0(\ln t^{-1})\,.
\end{align}
\end{theorem}

\begin{proof}
(i) By Proposition \ref{prop-trace versus Zp} (b),  $\{e^{-tA}\}_{t\geq 0}$ is a Gibbs semigroup and  $f(t) \leq \norm{A^{-1}}\,\norm{e^{-tA}}_1<\infty$, so the function $f$ is well defined. Let us show that $f$ is derivable on $[0,\infty)$ with $f'(t)=-\Tr e^{-tA}\,$:
If $U(t)=e^{-tA}\,$, then s-$\lim_{s\to 0} \,A^{-1} U(t){(U(s)-\bbbone)}/{s}
=-U(t)\,$ for any $t>0\,$, thus by using the $\norm{\cdot}_1$-continuity of multiplication on $\calL^1$
\begin{align*}
\norm{\cdot}_1\text{-}\lim_{s\to 0} \,A^{-1}\frac{U(t+s) -U(t)}{s} = -U(t)\,.
\end{align*}

(ii) To begin with, we recall the following chararacterisation of positive elements in $\calZ_1\,$
(see \cite[Lemma 5.1]{CRSS} and \cite[page 388 and Corollary 3.5]{CGRS}):
\begin{align}
A^{-1} \in \calZ_1=\calM^{1,\infty} & \iff \sup_{q\downarrow 1}\, (q-1)\,\norm{\,A^{-1}}_q <\infty \quad \text{(see Definition \ref{def-Zp} and \cite[Theorem 2.1]{CGRS})} \nonumber\\
& \iff  \sup_{x>0} \,\frac{1}{\ln(1+x)} \,\int_0^x \dd s \,\,s^{-2}\,\Tr \,e^{-s^{-1} A}<\infty  \quad \text{(see \cite[Corollary 4.6]{Gay-Suk}}) \nonumber\\\
&  \iff \sup_{x>0} \,\frac{1}{\ln(1+x)} \,\int_{1/x}^\infty \dd t \,\Tr \,e^{-t A}<\infty  \nonumber\\\
& \iff  \sup_{x>0} \,\frac{1}{\ln(1+x)} \,\Tr\, A^{-1}\,e^{-x^{-1} A} <\infty  \nonumber\\\
& \iff  \sup_{t>0} \,\frac{1}{\ln(1+t^{-1})} \,\Tr\, A^{-1}\,e^{-t A} <\infty \label{integrated version}
\end{align}
where the penultimate equivalence follows from the second one since:
\begin{align*}
\int_0^x \dd s \,\,s^{-2}\,\Tr \,e^{-s^{-1} A} = \Tr\, \int_0^x \dd s \,\,s^{-2}\,e^{-s^{-1} A}= \Tr\, A^{-1}\,e^{-x^{-1} A}\,.
\end{align*}

Now, let $g(t) \vc f(t)/ \ln(1+t^{-1})\,$.
\\
($\Longrightarrow$): If $A^{-1} \in \calZ_1$,
the implication $\Longrightarrow$ in \eqref{integrated version} shows that $g(t)\,\underset{t\downarrow 0}{\sim}\, f(t)/\ln t^{-1}$ is bounded in a vicinity of $t=0^+$. Therefore,  $\norm{\,A^{-1} e^{-tA}}_1=\OO_0(\ln t^{-1})\,$.\\
($\Longleftarrow$): Since $g$ is continuous on $(0,\infty)$, by the implication $\Longleftarrow$ in \eqref{integrated version}, we only need to show that $g$ is bounded near zero and infinity. The hypothesis $\norm{\,A^{-1} e^{-tA}}_1 = \OO_0(\ln t^{-1})$ implies the boundedness of $g$
near $t=0^+$. Since $A$ generates a Gibbs semigroup, we have $f(t) \leq \norm{\,A^{-1}} \norm{e^{-tA}}_1\leq \norm{\,A^{-1}} \,c_1e^{-c_2t}$ for some $c_1,c_2 > 0$ (see \eqref{estim-T}). This shows that $g$ is bounded at infinity as
\begin{align*}
g(t) \leq \frac{c_1e^{-c_2t}}{\ln(1+t^{-1})} \to 0 \quad \text{when } t\to \infty\,.
\end{align*}
\end{proof}

Note that if $A$ is not invertible, we can define the symbol $A^{-1}$ used in previous theorem by its extension with $A^{-1} u \vc 0$ for $u\in \Ker A\,$.

In Theorem \ref{singularity-ter}, we shall see that the trace-norm asymptotics of a holomorphic Gibbs semigroup can be compatible with differentiation when the function driving the behaviour at $t=0^+$ is in $\calS\calR\,$. Here we do not differentiate such behaviour, but intead we integrate it! However, such integration is less stable since it does not apply to asymptotics:
\\
Assume $A^{-1} \in \calZ_1 \backslash \calL^{1,\infty}\,$ and $\norm{\,A^{-1} e^{-tA}}_1=\OO_0(\Ln t^{-1})\,$ (note that $f(t)= \Ln t$ is in $\SR_0$). Then
$(\norm{\,A^{-1}e^{-tA}}_1)'(t) =-\norm{e^{-tA}}_1$ and we have $-(\Ln t^{-1})'(t)=(t+t^2)^{-1} \sim_{t\uparrow 0}t^{-1}$. The behaviour $\norm{e^{-tA}}_1=\OO_0(t^{-1})$ would imply $A\in \calL^{1,\infty}$ by Proposition \ref{prop-trace versus Zp}, which leads to a contradiction.

We can generalise the previous Theorem \ref{4.10} under the same hypothesis on $A$, using the equivalence $A^{-1}\in \calZ_p \iff A^{-p} \in\calZ_1\,$:
\begin{align*}
A^{-1} \in \calZ_p &\iff \norm{\,A^{-p}e^{-tA^p}}_1=\OO_0(\ln t^{-1})\,,\quad p\geq 1\,,\\
&\iff \norm{\,A^{-1}e^{-tA^p}}_p=\OO_0(\ln^{1/p} t^{-1})\,,\quad p\geq 1\,.
\end{align*}

\begin{remark}
 Example of an operator $A\in \calZ_1=\calM^{1,\infty}$ such that $0<A\notin \calL^{1,\infty}$ inspired by \cite[Lemma 1.2.8]{LSZ}: Given a partition of $\bbR^+$ with segments of rapidly increasing length, the idea is to define eigenvalues of $A$ as constant on these segments with a control over their decay. \\More precisely, for $k_\ell\vc2^{\ell^2}$ with $\ell\in \bbN$ giving a partition of $[2,\infty)\,$, let $\{\lambda_k\}_{k\geq 2}$ be the non-increasing family defined by
 \begin{align*}
 \lambda_{k_\ell}\vc \frac{\ell}{k_{\ell+1}-k_\ell}\,, \quad \lambda_{k_\ell}= \lambda_{k_\ell+1}=\cdots =\lambda_{k_{\ell+1}-1}\quad \text{for any $\ell\in \bbN\,$.}
 \end{align*}
Within the sequence $\{\lambda_k\}_{k\geq 2}\,$, the different values $\lambda_{k_\ell}$ have multiplicity $m_\ell=k_{\ell+1}-k_\ell\,$.\\
Let now $A$ be a positive invertible operator with spectrum equals to $\{\lambda_k\}_{k\geq 2}\,$. \\
If $n=1+\sum_{\ell=1}^{\ell_0}m_\ell=2^{(\ell_0+1)^2} -1$ with $\ell_0\in \bbN$, we have
\begin{align*}
\sum_{k =2}^n \lambda_k(A)=\sum_{\ell=1}^{\ell_0}\, \ell =\frac{1}{2}\ell_0(\ell_0+1)<\frac{1}{2} (\ell_0+1)^2 = \frac{1}{2\ln 2} \ln (1+n)
\end{align*}
which proves that $\sum_{k =1}^n \lambda_k(A)=\OO_\infty(\ln (1+n))$ and hence $A\in \calM^{1,\infty}= \calZ_1\,$.\\
Moreover, $A \notin \calL^{1,\infty}$ since we now show that for every $c>1$, $\{\lambda_k\}_{k\geq 2} \leq c \{k^{-1}\}_{k \geq 2}$ fails: For any $c>1$, choose $m= \lceil c\rceil $. Then for $n =2^{(m+1)^2}-1\,$, we get $\lambda_n = m/(2^{(m+1)^2}-2^{m^2}) > m/(2^{(m+1)^2}-1) \geq c\,n^{-1}\,$.

We have
\begin{align*}
\Tr \, e^{-t\,A^{-1}} =\sum_{k \geq 2} \, e^{-t\,\lambda_k(A)^{-1}}=\sum_{\ell\geq 1}\,\ell \lambda_{k_\ell}^{-1}\,e^{-t\lambda_{k_\ell}^{-1}}
\end{align*}
and the behaviour at $t=+0$ is not easy to estimate even if we know the behaviour of eigenvalues: \\ Since $\ell=(\rm{log}_{2}\, k_\ell)^{1/2}$, where $\rm{log}_{2}(\cdot)$ is the base-2 logarithm and
\begin{align*}
\lambda_{k_\ell}= \frac{\ell}{2^{(\ell+1)^2}-2^{\ell^2}} = \frac{\ell}{2^{\ell^2}(2^{2\ell+1}-1)} \, \underset{\ell \uparrow \infty}{\sim}\,
 \frac{\ell}{2^{\ell^2}}2^{-(2\ell+1)}
\end{align*}
we obtain
\begin{align*}
\lambda_{k_\ell}^{-1}\underset{\ell \uparrow \infty}{\sim}\, \frac{2\,k_\ell}{\sqrt{\rm{log}_{2}\,k_\ell}}\,\,2^{2\,\sqrt{\rm{log}_{2}\,k_\ell}}=  \frac{a\,k_\ell}{\sqrt{\ln\,k_\ell}}\,\,e^{b\sqrt{\ln\,k_\ell}}\,.
\end{align*}
\end{remark}
Since we were unable to estimate the asymptotics of $\Tr \,e^{-tA^{-1}}$, an unsolved question remains:

\noindent
\underline{Open question:}\\
Given $0<A^{-1} \in \calZ_1$, does there exist a function $f$ such that $\Tr e^{-tA} =\OO_{0}(f(t))$ provides a sharper  estimate than the one in Proposition \ref{prop-trace versus Zp} (b) for $p=1$? The characterization of $\calZ_1$ given in Lemma \ref{corol-h-stronger} highlights the difficulty of finding an answer. \\
Furthermore, if possible, can we find an explicit function $f$ such that $\Tr e^{-tA} \underset{t\downarrow 0}{\sim} f(t)$?

\begin{remark}
\label{rem-prime zeta}

Let $\underline{p}\vc\{p_n\}_{n\in \bbN}$ be the strictly non-decreasing sequence of prime numbers, and let $A$ be the self-adjoint positive operator defined by $A=A_{\underline{p}}\vc\text{Diag}(\,\underline{p} )$ acting on $l^2(\bbN)\,$. \\
The Prime Number Theorem \cite[Chapter I, Theorem 12]{Ingham} states that
\begin{align*}
N_A(\lambda) \,\underset{\lambda \uparrow \infty}{\sim}\,\Pi(\lambda)\text{\, where \,\,}\Pi(\lambda)\vc \lambda\,(\ln \lambda)^{-1}\, \quad \text{(see definition of $N_A$ in \eqref{def-NA})}\,.
\end{align*}
\noindent
It can be verified that $\Pi \in \SR_1$ (see Definition \ref{def-SR}). Therefore, Theorem \ref{thm-Kamarata for series} (c) implies that
\begin{align*}
\Tr\,e^{-tA} =\sum_{n\geq1} e^{-tp_n}  \,\underset{t \downarrow 0}{\sim}\, \Gamma(2) \,\Pi(t^{-1})= - \frac{1}{t\,\ln t}\,.
\end{align*}
Note that Theorem \ref{singularity} $(ii)$ can be applied here. Thus we obtain
\begin{align*}
(-1)^n\Tr\,A^ne^{-tA} \,\underset{t \downarrow 0}{\sim}\, g^{(n)}(t)\,, \text{ where } g(t)=\Pi(t^{-1})\,, \text{ for any }  n\in \bbN\,,
\end{align*}
so that
\begin{align*}
\Tr \, A\,e^{-tA} \,\underset{t \downarrow 0}{\sim}\, \frac{1+\ln t}{(t\ln t)^2}\,, \quad \Tr \, A^2\,e^{-tA} \,\underset{t \downarrow 0}{\sim}\, -\frac{2 +  [3 + 2 \ln t]\ln t}{(t\ln t)^3} \,, \quad\text{etc}.
\end{align*}

Note that $A_{\underline{p}}^{-1} \notin \calL^1$ since $\sum_{p_k\leq n} p_k^{-1} \,\underset{n \uparrow \infty}{\sim} \,\ln(\ln n)\,$ \cite[Theorem 7 Chapter I]{Ingham}, but $A_{\underline{p}}^{-1} \calL^{1,\infty}$ since $p_n  \underset{n\uparrow \infty}{\sim} n\ln n$ as shown in \cite[Theorem 13, Chapter I]{Ingham}, and hence
\begin{align*}
s_k(A^{-1})=p_k^{-1} \underset{k\uparrow \infty}{\sim} (k\ln k)^{-1}=\OO_\infty(k^{-1})\,.
\end{align*}
Note that we could also have used \cite[Corollary 4.8]{Gay-Suk}: $A_{\underline{p}}^{-1}$ is in the Lorentz space $\calM_\psi$ given by $\psi(t)=\ln(1+\ln(1+t))\,$, while $\calM_\psi=\calM^{1,\infty}$ for $\psi(t)=\ln(1+t)\,$.
\end{remark}

\begin{lemma}
\label{ln estimate}
For any $\varepsilon \in (0,1]\,$, there exists a constant $c_\varepsilon>0$ such that
$\ln(1+x)\leq c_\varepsilon\,x^\varepsilon$ for $x\in [0,\infty)$.
\end{lemma}
\begin{proof}
Let  $F_\epsilon(x) \vc x^\varepsilon /\ln(1+x)$ be defined on $\bbR_{0}^{+} =
[0,\infty)$.
Then, on $\bbR^{+}$, one has
\begin{align*}
\partial_x F_\epsilon (x)=0 &\iff \varepsilon x^{\varepsilon-1} \ln(1+x) =
{x^\varepsilon}/{(1+x)}  \iff h(x)\vc x^{-1}{(1+x)\ln(1+x)}=\varepsilon^{-1} \geq 1 .
\end{align*}
Hence, we obtain that $\lim_{x\downarrow 0}h(x)=1$, and the function $h$ is
increasing on $[0,\infty)$, where by definition $h(0):=1$. Therefore, the equation $h(x)=\varepsilon^{-1}$ has
 a unique solution $x_\varepsilon \in [0,\infty)$ for $\varepsilon \in (0,1]\,$.
\\
Note that the function $x \mapsto F_\epsilon(x)$ is continuous and strictly positive on
$\bbR^{+}$, and for $\varepsilon \in (0,1]$ we have $\lim_{x\downarrow 0} F_\epsilon(x)
= \lim_{x\downarrow 0} \ {x^{\varepsilon}}/ {\ln(1+x)} =\infty \,$
and $\lim_{x\uparrow \infty} F_\epsilon(x) = \infty$.\\
Therefore, the solution $x_\varepsilon$ minimises $F_\epsilon(x)$, that is,
$F_\epsilon(x)\geq F_\epsilon(x_\varepsilon) >0 $, and hence we obtain the desired inequality
$\ln(1+x) \leq \, x^\varepsilon / F_\epsilon(x_\epsilon) = c_\varepsilon \,x^\varepsilon$ for all $x\in [0,\infty)\,$ where $c_\varepsilon \vc F_\varepsilon(x_\varepsilon)^{-1}\,$.
\end{proof}
To determine the bounds of $c_\varepsilon = 1/F_\varepsilon(x_\varepsilon)$ we calculate the derivative of the function: $\varepsilon \mapsto F_\varepsilon(x_\varepsilon)$.
Since $x_\varepsilon$ is a minimiser of $F_\epsilon(x)$, we have
\begin{equation}\label{deriv-F}
\frac{d}{d\varepsilon} F_\varepsilon(x_\varepsilon) =
\partial_{x_\varepsilon} F_{\varepsilon}(x_\varepsilon) \ \partial_{\varepsilon} {x_\varepsilon} + (\partial_{\varepsilon} F_\varepsilon) (x_\varepsilon) =
F_\varepsilon (x_\varepsilon) \, \ln x_\varepsilon \, .
\end{equation}
Now we claim that $\varepsilon \mapsto x_\varepsilon \, $  is a monotonically decreasing function. Indeed, if $0<\eta<\varepsilon<1$, then $h(x_\varepsilon)=\varepsilon^{-1}<\eta^{-1}=h(x_\eta)$ and since $h$ is increasing, we infer $x_\varepsilon<x_\eta$.
Since $h: \bbR_{0}^{+} \rightarrow [1, +\infty)$, there is a unique
$\varepsilon^* \in (0,1)$
such that $x_{\varepsilon^*} =1$, while $x_\varepsilon \,
\vert_{\, \varepsilon <\varepsilon^*} > 1 $ and $x_\varepsilon \,
\vert_{\, \varepsilon^*< \varepsilon } < 1$.
Then according to (\ref{deriv-F}), we obtain
\begin{equation}\label{deriv-F1}
\frac{d}{d\varepsilon} F_\varepsilon(x_\varepsilon) \,
\vert_{\, \varepsilon <\varepsilon^*} \
> 0
\quad \text{and} \quad
\frac{d}{d\varepsilon} F_\varepsilon(x_\varepsilon) \,
\vert_{\, \varepsilon^*< \varepsilon }
< 0 \, .
\end{equation}
As a consequence of (\ref{deriv-F1}), the \textit{extremum} $F_{\varepsilon^*} (x_{\varepsilon^*})\ $
of the function $\varepsilon \mapsto F_\varepsilon (x_\varepsilon)$, is actually a \textit{maximum}:
$F_\varepsilon (x_\varepsilon) < F_{\varepsilon^*} (x_{\varepsilon^*})
\simeq 1, 44\, $. In addition, by explicit calculations we also find that
$\lim_{\varepsilon \rightarrow 0}\ F_\varepsilon (x_\varepsilon) = 0$, while
$\lim_{\varepsilon \rightarrow 1}\ F_\varepsilon (x_\varepsilon) = 1$, which
yields that the coefficient $c_\varepsilon > 0$ exists for any $\varepsilon \in (0,1]$.
\\
Remark that $\lim_{\varepsilon \rightarrow 0}\, c_\varepsilon = +\infty$ bears to keep
inequality $\ln(1+x)\leq c_\varepsilon\,x^\varepsilon$ ($x \geq 0$) for small
$\varepsilon$.
On the other hand, $\lim_{\varepsilon \rightarrow 1}\, c_\varepsilon = 1$ provides
the known inequality $\ln(1+x)\leq \, x$ for $x \in \bbR_{0}^{+}$.

\bigskip

While two functions may be asymptotically equivalent, this does not guarantee their derivatives share a comparable relationship. For example, consider the functions $g(t)=t^{-1}+\sin(t^{-2})$ and $h(t)=t^{-1}$, where $g(t)\sim_{t\downarrow 0} h(t)\,$. However, we have neither $g'(t)=\OO_0(h'(t))$ nor $h'(t)=\OO_0(g'(t))\,$.

Nevertheless, the following theorem establishes that for Gibbs semigroups with positive generators, an asymptotic behaviour of the trace, such as $\norm{e^{-tX}}_1=\OO_0(f(t^{-1}))\,$, remains valid under differentiation, at least for smooth functions $f$ exhibiting regular variation at infinity ($f \in \SR$ as defined in Definition \ref{def-SR}).

We start by presenting the following preliminary result, which will be utilised in the next theorem.

\begin{lemma}\label{prop-derivative of trace-norm}
Let $A$ be a positive self-adjoint generator of a Gibbs semigroup. \\Then, the function
$F(t)=\norm{e^{-tA}}_1\,$ for $t>0\,$ is infinitely differentiable and its $n$-th derivative $F^{(n)}$ satisfies
\begin{align}
\label{derivative positive case}
(-1)^n\,F^{(n)}(t)=\Tr \,A^n e^{-tA} = \norm{\, A^n e^{-tA}}_1\,.
\end{align}
Moreover, there exists constants $c_n>0$ and $a\in (0,1[$ such that
\begin{align}
\abs{F^{(n)}(t)}  \leq c_n \, t^{-n} F(a t) \,\text{ for any } n\in \bbN\,, t>0\,. \label{estimate derivative of F}
\end{align}
\end{lemma}
\begin{proof}
Note that by \cite[Corollary 4.32]{Zagrebnov2}, cf. Theorem \ref{corol-main-1},
the self-adjoint Gibbs semigroup $\{e^{-tA}\}_{t\geq 0}$ has a trace-norm holomorphic extension into an open sector in the right half-plane $\bbC_{+}$.
As a consequence,  $z \mapsto F(z):= \Tr e^{-zA}$ is holomorphic in
$\bbC_{+}$ and $F^{(n)}(t) = (-1)^n\,\Tr \,A^n e^{-tA}$ for any $n\in \bbN_0\,$.

Moreover, applying the Cauchy formula for the derivatives of $F(z)$ at $z = t > 0$, we obtain for the function $F_n(t)\vc (-1)^n\,F^{(n)}(t) >0$ that
\begin{equation*}
F_n(t)=\frac{(-1)^nn!}{2\pi i}\int_{C_\rho} \,\dd \zeta \ \frac{F(\zeta)}{(\zeta-t)^{n+1}} =\frac{(-1)^nn!}{2 \pi\rho^n} \int_0^{2 \pi} \dd \varphi \, e^{- i n\varphi} \,
F(t + \rho \, e^{i \varphi}) \, , \quad n \in \bbN \, ,
\end{equation*}
where the circle $C_\rho$ of radius $\rho  < t$ is centered at $t>0$. This equation  yields the estimates
\begin{equation}\label{cauchy3}
 F_n(t) \leq  \frac{n!}{2 \pi\rho^n} \int_0^{2 \pi} \dd \varphi \
\vert F(t + \rho \, e^{i \varphi})\vert \, ,  \quad n \in \bbN \, .
\end{equation}
Since, by the hypothesis of the lemma, the operator $A^{-1}$ is compact, the spectrum of $A$ consists of eigenvalues $\{\lambda_k=\lambda_k(A) > 0\}_{k\geq 1}\,$, and we obtain an upper bound for the integrand in \eqref{cauchy3}:
\begin{align}
\label{cauchy4}
\vert F(t + \rho \, e^{i \varphi})\vert &=
\vert\Tr e^{-(t + \rho \, e^{i \varphi})\, X}\vert \leq
\vert \sum_{k=1}^{\infty} e^{-(t + \rho \, e^{i \varphi})\, \lambda_k}\vert
\leq \sum_{k=1}^{\infty} e^{-(t + \rho \, \cos \varphi)\, \lambda_k}  \nonumber\\
&\leq \sum_{k=1}^{\infty} e^{-(t - \rho)\, \lambda_k} =
\Tr\ e^{-(t-\rho)X}=F(t-\rho)\ .
\end{align}
If we choose $\rho \vc t  \,\sin \theta$ in (\ref{cauchy4}), where $0<\theta < \pi/2$, then by (\ref{cauchy3}) we get the desired estimate:
\begin{align*}
F_{n}(t) \leq c_n t^{-n} F(a t)\quad \text{ with $c_n=n!\, (\sin \theta)^{-n}$ and
$a = 1-\sin \theta \in (0,1)$}.
\end{align*}
\end{proof}

\begin{theorem} \label{singularity}
Let $A$ be a positive self-adjoint generator of a Gibbs semigroup, and consider the functions
$F(t)\vc\norm{e^{-tA}}_1\,$,  $f\in \SR$ and $g(t)\vc f(t^{-1})\,$ for $t>0\,$. Then, for any $n\in \bbN$, we have:

\emph{(i)} If $F(t)=\OO_0(g(t))\,$, then $F^{(n)}(t)=\OO_0(g^{(n)}(t))\,$.

\emph{(ii)} If $F(t) \,\underset{t \downarrow 0}{\sim}\, g(t)\,$, then  $F^{(n)}(t) \,\underset{t \downarrow 0}{\sim}\,g^{(n)}(t)\,$.
\end{theorem}

\vspace{-0.3cm}
We restrict ourselves to the case $f \in \SR$ with $p=\text{ind}_f>0$ since the behaviour
$\Tr \,e^{-tA}=\OO_0(t^{-p})$ for $p\leq 0$ is impossible as previously mentioned in Proposition \ref{prop-behaviour at 0}.

\begin{proof}
Since $f \in \SR$, let $p=\text{ind}_f>0$ (see definition of $\text{ind}_f$ in Appendix).

(i)
Note that due to (\ref{estimate derivative of F}), we obtain in particular
\begin{align*}
\big{\vert} \frac{F^{(n)}(t)}{g^{(n)}(t)} \big{\vert} \leq c_n \,\frac{F(a t)}{t^n \abs{g^{(n)}(t)}} = c_n \,\cdot\,\frac{g(t)}{t^n \abs{g^{(n)}(t)}}\,\,\cdot\, \frac{F(a t)}{f((a t)^{-1})} \,\cdot\,\frac{f(a^{-1}t^{-1})}{g(t)}\,.
\end{align*}
Therefore, the proof of the assertion follows from the next lemma since we have
\begin{align*}
\underset{t\downarrow 0}{\limsup}\, \big{\vert}\frac{F^{(n)}(t)}{g^{(n)}(t)}\big{\vert} \leq c_n \,\cdot\,\frac{\Gamma(p)}{\Gamma(p+n)}\,\cdot\,\underset{t\downarrow 0}{\limsup}\, \frac{F(a t)}{f((a t)^{-1})} \,\cdot\,a^{-p} <\infty \, .
\end{align*}

(ii)  Define the $\sigma$-finite measures on $[0,\infty)$, $\nu_n \vc \sum_{k\geq 1} \lambda_k^n \,\delta_{\lambda_k}$ for each $n\in\bbN_0\,$. This choice gives
\begin{align*}
F(t) =L_{\nu_0}(t)\,,\quad   (-1)^nF^{(n)}(t) = L_{\nu_n}(t)\, \quad \text{and}\,\quad \dd \nu_n(x)=x^n\,\dd \nu_0(x)\,.
\end{align*}
The hypothesis $L_{\nu_0}(t) \underset{t\downarrow 0}{\sim} f(t^{-1})$ implies, using Theorem \ref{appendix:Thm-Karamata} (c):
\begin{align*}
\nu_0([0,\lambda)) \underset{t\downarrow 0}{\sim} \Gamma(I+p)^{-1}\,f(\lambda)\,.
\end{align*}
Note that the function $\lambda\geq 0 \mapsto \nu_0([0,\lambda))$ has the same index at infinity as $f$ and being increasing, is locally of bounded variation on $[0,\infty)$. Applying \cite[Theorem 1.6.4]{RegVar}, we obtain
\begin{align*}
\nu_n([0,\lambda))=\int_{[0,\lambda)}\! \dd \nu_0(x)\,x^n  &\underset{\lambda \uparrow \infty}{\sim} \frac{p}{p+n}\,\lambda^n\,\nu_0([0,\lambda))
\\
&\underset{\lambda \uparrow \infty}{\sim} \frac{p\,}{(p+n)\,\Gamma(1+p)}\,\lambda^n\,f(\lambda) = \frac{1}{(p+n)\,\Gamma(p)}\,\lambda^n\,f(\lambda)\,.
\end{align*}
Since the index of the map: $\lambda \mapsto \lambda^n\,f(\lambda)$ is $p+n$, a new application of Theorem \ref{appendix:Thm-Karamata} (c) to $\nu_n$ gives the following asymptotics for $L_{\nu_n}(t)$:
\begin{align}
\label{eq-derivatives of F}
(-1)^n\,F^{(n)}(t)=\Tr \,A^n\,e^{-tA}=L_{\nu_n}(t) \,\underset{t\downarrow 0}{\sim}\, \frac{\Gamma(1+p+n)}{(p+n)\,\Gamma(p)}\,t^{-n}\,f(t^{-1}) = \frac{\Gamma(p+n)}{\Gamma(p)}\,t^{-n}\,f(t^{-1})\,.
\end{align}
Finally, from \eqref{eq-derivatives of F} and the next lemma, we deduce the desired asymptotics:
\begin{align*}
\frac{F^{(n)}(t)}{g^{(n)}(t)} =\frac{(-1)^n\,F^{(n)}(t)}{ t^{-n}\,g(t)}\cdot \frac{g(t)}{(-1)^n \,t^n\,g^{(n)}(t)}
\,\underset{t\downarrow 0}{\sim}\,\frac{\Gamma(p+n)}{\Gamma(p)}\,\cdot\, \frac{\Gamma(p)}{\Gamma(p+n)}
= 1\,.
\end{align*}
\end{proof}
\begin{lemma}\label{next-lemma}
Let $f\in \SR$ and $g(t)\vc f(t^{-1})\,$. Then for $n\in \bbN_0\,$, we have
\begin{align}
\label{derivative of g}
\frac{(-1)^n\,t^n\,g^{(n)}(t)}{g(t)}\, \underset{t\downarrow 0}{\to} \, \frac{\Gamma(\text{ind}_f+n)}{\Gamma(\text{ind}_f)}\,,
\end{align}
where $\Gamma$ is the usual Gamma-function.
\end{lemma}
\begin{proof}
The \textit{Faà di Bruno's} formula for $g(t)=f(t^{-1})$ (see \cite[equation 0.431, page 23]{Gradshteyn}) gives
\begin{align*}
g^{(n)}(t)= \sum_{m=1}^n \frac{n!}{m!} \binom{n-1}{m-1} (-1)^nt^{-(n+m)}\,f^{(m)}(t^{-1})\,, \quad n\in \bbN\,.
\end{align*}
Since \eqref{derivatives for SR} holds true, we obtain with $p\vc \text{ind}_f>0$
\begin{align*}
\lim_{t\downarrow 0}\frac{(-1)^n\,t^n\,g^{(n)}(t)}{g(t)}& =  \sum_{m=1}^n \frac{n!}{m!} \binom{n-1}{m-1}\, \lim_{t\downarrow 0}\,\frac{t^{-m}\,f^{(m)}(t^{-1})}{f(t^{-1})}
\\ &
= \sum_{m=1}^n \frac{n!}{m!} \binom{n-1}{m-1}\,p(p-1)\cdots (p-m+1)
\\&
=  p(p+1)\cdots (p+n-1)\,,
\end{align*}
where the last equality which is a  rewriting of \eqref{derivative of g}, can also be checked directly by recurrence: \\
Let $C_n := p(p+1)\cdots (p+n-1)$ and $G_{n}(t) :=  (-1)^n\,g^{(n)}(t)/g(t)$ and suppose that $\lim_{t\downarrow 0}t^k \ G_{k}(t) = C_k$, for $k = 1,2, \ldots, n$. Then we obtain
\begin{equation}\label{Phi-C1}
t \, \partial_t (t^n \ G_{n}(t)) = n \, t^n \ G_{n}(t) -
t^{n+1} \ G_{n+1}(t) + t^n \ G_{n}(t) \ t \, G_{1}(t) \, .
\end{equation}
To calculate $\lim_{t\downarrow 0} t^{n+1} \,G_{n+1}(t) = C_{n+1}$
from (\ref{Phi-C1}) thanks to
\begin{equation}\label{Phi-C2}
C_{n+1} = C_{n} (C_{1} +n )  - \lim_{t\downarrow 0}
t \, \partial_t (t^n \ G_{n}(t)) \, ,
\end{equation}
one only has to prove that
\begin{equation}\label{Phi-C3}
\lim_{t\downarrow 0} t \, \partial_t (t^n \ G_{n}(t)) =0 \, .
\end{equation}
To this aim we note that
$\int_{0}^{t} \dd s  \ \partial_s (s^n \ G_{n}(s)) = t^n \ G_{n}(t) -
C_{n} \,$.
Then by virtue of the hypothesis we obtain
\begin{equation*}
\lim_{t\downarrow 0}  t^n \ G_{n}(t) - C_{n} =
\lim_{t\downarrow 0} \ \int_{0}^{t} \dd s \ \partial_{s} (s^n \ G_{n}(s)) = \lim_{t\downarrow 0} t \, \partial_{\tau} (\tau^n \ G_{n}(\tau)) = 0 \, ,
\quad 0 < \tau < t \, ,
\end{equation*}
which proves the claim  (\ref{Phi-C3}) and hence the formula
$C_{n+1} = C_{n} (C_{1} + n) $ due to (\ref{Phi-C2}).
\end{proof}

\section{Certain overall cases}\label{sec3}

An immediate generalisation of Proposition \ref{prop-trace versus Zp} is the following statement:
\begin{proposition}
Let $A$ be a normal operator generating a  compact $C_0$-semigroup
and let $p>0\,$. The following are equivalent:

\hspace{0.3cm} \emph{(i)}  $R_{\Re A}(z)\in \calL^{p,\infty}\,\, \text {for } z\in \rho(\Re A)\,$.

\hspace{0.3cm}\emph{(ii)} $\norm{e^{-tA}}_1= \OO_0(t^{-p})\,$.
\end{proposition}

\begin{proof}
There exists $\omega_0\in \bbR$ such that $X=\Re A + \omega_0 \bbbone>0\,$ as in Lemma \ref{lem-ReA et omega0}. Thus,
\begin{align*}
R_{\Re A}(z)\in \calL^{p,\infty}\,\, \text {for some $z\in \rho(\Re A)$}
&\iff X^{-1}=R_{\Re A}(-\omega_0) \in \calL^{p,\infty} \quad \text{by resolvent identity,}\\
&\iff \norm{e^{-tX}}_1=\OO_0(t^{-p}) \quad \text{ by Proposition \ref{prop-trace versus Zp},}\\
& \iff \norm{e^{-tA}}_1=\norm{e^{-t\Re A}}_1=e^{\omega_0 t}\,\norm{e^{-tX}}_1=\OO_0(t^{-p})\,.
\end{align*}
\end{proof}

Similarly, our goal now is to generalise the previous Lemma \ref{lem-behaviour for positive operator} to a \textit{non-self-adjoint} generator $A$, strengthened by a specific \textit{invertibility} condition,
see Remark \ref{rem:2.5}. 
Since the link between resolvent and semigroup involves the Bochner integral, see \eqref{resolvent-B},
we begin by collecting the following preliminaries.

First, if $A\in Q(M,\omega_0)$, then
we can deduce that $\norm{e^{-t(A - z\bbbone)}}_1$ decreases exponentially as $t\to\infty$ for $\Re(z) < -\omega_0\,$. In fact, for any $T > 0$, we have:
\begin{align} \label{estim-T}
\norm{e^{- t\,(A - z\bbbone)}}_1 \leq \norm{\,e^{- T \,(A - z\bbbone)}}_1  \ M \
e^{(t -T) \,(\omega_0 + \re z)} \leq M_{z,T} \, e^{-t\abs{\omega_0 + \re z}} \ , \quad
\re z < -\omega_0 \, , \ t > T\, .
\end{align}
where $M_{z,T}$ is a constant. Hence, from (\ref{estim-T}), we obtain $\norm{e^{-t(A - z\bbbone)}}_1 = \OO_\infty(e^{-t|\omega_0 + \Re(z)|})$.

Second, let us recall a characterisation of the Bochner integral in a complex Banach space $\cal{X}$ equipped with a norm $\norm{\cdot}_{\cal{X}}$. We take this characterisation for our definition:

\begin{definition} {\rm (see e.g., \cite[Theorem 3.7.4]{HillePhillips})}
Let $I \subset \bbR$ be an interval (bounded or unbounded) in $\bbR$.
A function $f: I \rightarrow \cal{X}$ is Bochner integrable if the function
$I \ni t \mapsto f(t)$ is strongly measurable  \emph{(}which is in particular the case when $f$ is strongly continuous\emph{)}, and the function $I \ni t \mapsto \norm{f(t)}_{\cal{X}}$ is Lebesgue integrable.
\end{definition}

\begin{proposition}
\label{Bochner}
If $f$ is $\norm{\cdot}_{\cal{X}}$-Bochner integrable then,
\begin{equation}
\label{B-integral}
\big{\Vert}\int_{I} \!\mathrm{d}t\, f(t) \, \big{\Vert}_{\cal{X}} \, \leq \,
\int_{I} \!\mathrm{d}t\, \norm{f(t)}_{\cal{X}} < \infty\, .
\end{equation}
\end{proposition}

Now we can provide a characterisation of $C_0$-semigroups that ensures they are Gibbs semigroups with the asymptotic property $\norm{e^{-t\,A}}_1= \OO_0(t^{-p})$ for some $p\geq 1$.
\begin{theorem} \label{main-1B}
Let $\{e^{- t\,A}\}_{t\geq0}$ be a compact $C_0$-semigroup on a Hilbert space $\calH$ with an invertible generator $A\in Q(M,\omega_0)$ (that is,  $\Ker (A )= \{0\}$).
If $p\geq 1$, then the following statements are equivalent\emph{:}

\emph{(a)} $\{U_A(t) = e^{- t\,A}\}_{t\geq0}$ is a Gibbs semigroup
with the asymptotics $\norm{e^{-t\,A}}_1= \OO_0(t^{-p})\, $.

\emph{(b)} Let $q > p$\emph{:}

\hspace{1cm} \emph{(i)}  The mapping: $t \mapsto e^{-tA}\,$, is $\norm{\cdot}_q$-continuous at $t \in \bbR^+$ for some $q > p$.

\hspace{1cm} \emph{(ii)} The resolvent family $\{R_A(z)\, \}_{z \in \rho(A)}$
is defined using the Laplace transform of $\{e^{-tA}\}_{t\geq0}$ by means of a $\norm{\cdot}_q$-Bochner integral and $(R_A(z))^{\,q} \in \calL^{1}$.
\end{theorem}

\begin{remark}
\label{main-rem}
The assumption (i) in (b) does not entail that $U_A(t) \in \calL^{q}$ or that the semigroup is {holomorphic}. It is important to note that, because of  (\ref{3.13b}) and (\ref{3.14a}), we cannot relax the condition $\Ker(A) = \{0\}$ in our proof. \
However, as a corollary, we can extend our arguments below from $q \in \bbN$ to real values of $q$ such that $q > p \geq 1$. In doing so, we will establish the existence of a \textit{fractional power} of the resolvent $(R_A(z))^{\,q}$.
\end{remark}

\begin{proof}
(a)$\implies$(b)
Suppose that the $C_0$-semigroup $\{e^{-tA}\}_{t\geq 0}$ is a Gibbs semigroup.
Then it is trace-norm continuous for $t \in \bbR^+$ and consequently,
is continuous in the topology of the ideal $\calL^{q} \supset \calL^{1}$ for $q > p \geq 1$; that proves (i).

To establish (ii), we observe that the Gibbs semigroup $\{U_A(t) = e^{-tA}\}_{t\geq 0}$ is both operator-norm continuous and compact for $t \in \bbR^+$, since $\calL^{1} \subset \calL^{\infty}$.
Since $A\in Q(M,\omega_0)$, we get $(-\infty, -\omega_0) \subset \rho(A)$.
Then choosing $\lambda > \omega_0$ (i.e., $-\lambda \in \rho(A)$), we have
\begin{align}
\label{norm ineq}
\norm{e^{-\lambda t} \, e^{-t\,A}} \leq M \, e^{(\omega_0 - \lambda) \, t}.
\end{align}
Now, by virtue of compactness and operator-norm continuity of the family
$\{e^{-tA}\}_{t>0} \,$, the resolvent of generator $A$ at $z = -\lambda$ can be represented as the Bochner integral (Laplace transform)
\begin{equation}
\label{resolvent-B}
R_{A}(-\lambda) = \int_0^\infty \!\mathrm{d}t\, e^{-\lambda t} \,e^{- t\,A} \, ,
\quad  \omega_0 < \lambda \, ,
\end{equation}
which is operator-norm convergent, due to the good behaviour at infinity deduced from \eqref{norm  ineq}.

Therefore, by compactness of the integrand, the operator-norm convergent integral (\ref{resolvent-B}) defines in the left-hand side
a compact operator $R_{A}(-\lambda)$. Furthermore, by the resolvent identity, $R_{A}(-\lambda) \in \calL^{\infty}$ for some $-\lambda \in \rho(A)$ implies that resolvent $R_{A}(z)$ is compact for any $z \in \rho(A)$.

Now, using the representation (\ref{resolvent-B}) and applying Fubini's theorem, we can deduce the following expression for the powers $q \in \bbN$ of the resolvent:
\begin{equation}\label{resolvent-q}
(R_{A}(z))^{\, q} = \Gamma(q)^{-1} \int_0^\infty \!\mathrm{d}t\, t^{q-1} \,e^{- t\,(A - z\bbbone)} \ , \quad \omega_0 < \re(- z) \, ,
\end{equation}
where the integral is interpreted as an operator-norm convergent Bochner integral, as in (\ref{resolvent-B}).\\
Note that a Gibbs semigroup is $\norm{\cdot}_1$-continuous (hence $\norm{\cdot}_1$-measurable) on $\bbR^+$, and the $t$-function $\norm{t^{q-1} \,e^{- t\,(A - z\bbbone)}}_1$ is integrable near zero due to the condition $\norm{e^{-t\,A}}_1= \OO_0(t^{-p})\ $ for $q>p$, and it is integrable at infinity due to \eqref{estim-T}.
Hence, the right-hand side of \eqref{resolvent-q} is a $\norm{\cdot}_1$-convergent Bochner integral, and
applying the inequality \eqref{B-integral}, we obtain the following estimate:
\begin{equation}\label{resolvent-q-norm}
\norm{(R_{A}(z))^{\, q} }_1 \leq \Gamma(q)^{-1} \int_0^\infty \!\mathrm{d}t\, t^{q-1} \, \norm{e^{-t\,A}}_1 \, e^{t \re(z)}<\infty \, , \quad \text{when \ }
\omega_0 < \re(- z) \, .
\end{equation}
This shows that $(R_{A}(z))^{\, q} \in \calL^{1}$ for $z \in \rho(A)$, thereby establishing assertion (ii).

(b) $\implies$ (a)
The converse implication is less straightforward since it requires to construct a Gibbs semigroup, which is the integrand in equation (\ref{resolvent-q}). To this aim we primarily use the condition (ii) for the resolvent.

Under the hypotheses of the theorem, the \textit{family} $\{U_A(t)\vc e^{-tA}\}_{t\geq0}$ is a compact $C_0$-semigroup. Hence, there exists a value $\omega_0$ such that $(-\infty,-\omega_0) \subset \rho(A)$.
To proceed, we fix $\lambda > \omega_0$ and define a family of bounded operators using the operator-norm convergent Bochner integral:
\begin{equation}\label{3.11b}
F(t) \vc \int_0^t \! \mathrm{d}\tau\, e^{-\lambda\tau} \ U_A(\tau) \, ,
\quad t \geq 0 \, .
\end{equation}

Next, integrating the equation:
\begin{equation*}
\partial_t (e^{-\lambda t}U_A(t) \ u) = - \, e^{-\lambda t}
U_A(t) \ (\lambda \bbbone + A) \ u \, , \quad u \in \Dom A \, ,
\end{equation*}
we obtain
\begin{equation*}
(\bbbone - e^{-\lambda t} U_A(t)) \ u = \int_0^t
\!\mathrm{d}\tau \, e^{-\lambda\tau}U_A(\tau) \ (\lambda \bbbone + A) \ u\,,
\end{equation*}
or, for $w \vc (\lambda \bbbone + A)\, u\,$, the equation
\begin{equation}\label{3.12b}
(\bbbone - e^{-\lambda t} U_A(t)) \ (A + \lambda \bbbone)^{-1} \, w = \int_0^t
\!\mathrm{d}\tau \, e^{-\lambda\tau} U_A(\tau) \ w \, .
\end{equation}
Since $-\lambda \in \rho(A)\,$, the range $\Ran(\lambda \bbbone + A) = \calH$. Consequently, the equation (\ref{3.12b}) implies that the family of operators (\ref{3.11b}) belongs to the ideal $\calL^q$:
\begin{equation} \label{3.13b}
F(t) = (\bbbone - e^{-\lambda t}U_A(t)) \ (A + \lambda \bbbone)^{-1} \in
\calL^q \, , \quad t \geq 0 \, ,
\end{equation}
since, according to condition (ii), the resolvent $R_{-\lambda}(A) \in \calL^q$, and $F(0)=0\cdot \bbbone$.
As a result,
\begin{equation} \label{3.13bb}
F(t+\delta) - F(t) = \int_t^{t+\delta}\!\!\mathrm{d}\tau \,
e^{-\lambda \tau} U_A(\tau) \in \calL^q \, , \quad 
\delta > 0 \, ,
\end{equation}
for any $t > 0$. 

On account of condition (i) the $\Vert\cdot\Vert_q$-continuity of the family 
$\{U_A (\tau)\}_{\tau > 0}$ in the neighbourhood of the corresponding $t>0$ yields the estimate
\begin{align}\label{3.14a}
\big{\Vert} \frac{1}{\delta} \int_t^{t+\delta}\! \!\!\mathrm{d}\tau \ U_A(\tau) -
U_A(t) \big{\Vert}_q \leq \frac{1}{\delta} \int_t^{t+\delta} \!\!\mathrm{d}\tau \
\norm{U_A(\tau) - U_A(t)}_q \leq \varepsilon_t(\delta) \, , \quad \delta > 0
 \, ,
\end{align}
where $\varepsilon_t(\delta) := \sup_{\tau \in [t, t+\delta ]} \ \norm{U_A(\tau) -
U_A(t)}_q$ and $\lim_{\delta \downarrow0} \ \varepsilon_t(\delta) = 0$.
Then, because of (\ref{3.13bb}) the estimate (\ref{3.14a}) yields
\begin{align}\label{3.14b}
  U_A(t) = \ \Vert\cdot\Vert_q \text{-}\lim_{\delta \downarrow0}
     \frac{1}{\delta} \int_t^{t+\delta} \!\!\! \mathrm{d}\tau \  U_A(\tau) \, ,
\end{align}
for any $t > 0$. 

Note that, by the invertibility of the generator $A$ and
by virtue of (\ref{3.13b}) and (\ref{3.13bb}) for $\lambda = 0$, the integral in
(\ref{3.14b}) coincides with the operator $F(t+\delta) - F(t)$, which belongs to the
ideal $\calL^q$. Because the limit (\ref{3.14b}) exists and also holds in the $\Vert\cdot\Vert_q$-norm
for the $\Vert\cdot\Vert_q$-valued integrals, the operator $U_A(t)$ in the left-hand side of
(\ref{3.14b}) belongs to the $\calL^q$-class for any $t > 0$.
The semigroup property now implies that $U_A(t) \in \calL^1$ for $t > 0$, and consequently, the $C_0$-semigroup $\{e^{-tA} \}_{t \geq 0}$  is a Gibbs semigroup.

Seeing that by (\ref{3.14b}), the family $\{U_A (t)\}_{t > 0}$ belongs to $\calL^q$
and that it is a $\calL^q$-continuous function by condition (i), the integrand in (\ref{3.11b}) is $\calL^q$-continuous. Then, for $\omega_0 < \lambda $, we can construct the representation of the resolvent
$R_{A}(z= -\lambda)$ as a $\calL^q$-Laplace transform. To this end, it is sufficient to take in (\ref{3.11b}) and (\ref{3.13b}) the $\norm{ \cdot }_q$-$\lim_{t \to \infty} \, F(t) \,$, which can then be extended to $z \in \rho(A)$. As a result, we obtain
\begin{equation}\label{resolvent-q-n}
\norm{R_{A}(z)}_q = \big{\Vert} \int_0^\infty \!\!\mathrm{d}t \, e^{t \, {z}} \,e^{- t\,A} \ \big{\Vert}_q \, ,
\quad {\re }\,z<-\omega_0 \, ,
\end{equation}
where the left-hand side of (\ref{resolvent-q-n}) is finite by condition (ii). By the same condition, similarly to (\ref{resolvent-q}) and (\ref{resolvent-q-norm}), we also get a $\calL^1$-Bochner integral for a Laplace transform of the power $q$:
\begin{equation}\label{resolvent-1-n}
\norm{(R_{A}(z))^{\, q} }_1 = \big{\Vert}\Gamma(q)^{-1} \int_0^\infty \!\!\mathrm{d}t \, t^{q-1} \,
e^{-t\,A} \ e^{t \,z}\big{\Vert}_1 \, <\infty \, , \quad \text{when } {\re }\,z<-\omega_0 \, .
\end{equation}
Since by Proposition \ref{Bochner}, the integrand in (\ref{resolvent-1-n}) is $\norm{\cdot }_1$-integrable, we obtain for any $q>p$:
\begin{equation}\label{resolvent-1-n-bound}
\Gamma(q)^{-1} \int_0^\infty \! \!\mathrm{d}t \, t^{q-1} \,
\norm{ e^{-t\,A}}_1 \ e^{t {\re } \,z} \, <\infty \, , \quad \text{for \ }
{\re }\,z<-\omega_0 \, ,
\end{equation}
which implies the asymptotics $\norm{ e^{-t\,A}}_1 = \OO_0(t^{-p})\,$.
\end{proof}
It is worth noting that (\ref{resolvent-q-n}) and
(\ref{resolvent-1-n-bound}) yield the asymptotics $\norm{ e^{-t\,A}}_q = \OO_0(t^{-p/q})$ for the $\norm{\cdot }_q$-norm of the semigroup, where $q > p \in \bbN$.
\begin{remark} \label{non-entiers}
Regarding (a)$\implies$(b): If a normal operator $A$ satisfies the assumptions of Corollary \ref{thm-Karamata for semigroup}, then it is known that hypothesis (a) implies $\lambda_k(\Re A)=\OO_\infty(k^{-1/p})$ not only for $p\geq 1$, but also under the weaker assumption $p>0$. Consequently, we have $s_k(\Re A)=\abs{\lambda_k(\Re A)}=\OO_\infty(k^{-1/p})\,$. Moreover, $\norm{A^{-q}}_1<\infty$ for any $q>p$, which implies that $R_A(z)^{\,q} \in \calL^1$ for any $z\in \rho(A)\,$. We have thus verified condition (b) independently of the resolvent representation \eqref{resolvent-B}.

Besides that, in the proof of (a)$\implies$(b), we \textit{formally} end up with $\norm{R_A(z)^{p+\varepsilon}}_1 < \infty$ for any $\varepsilon >0$. It is  known that
\textit{a priori} one cannot expect to get $\norm{R_A(z)^p}_1 <\infty$, cf.  Remark \ref{p<q}. Then it is interesting to control the behaviour of $\norm{R_A(z)^{p+\varepsilon}}_1$ as a function of $\varepsilon$. To this aim, we are going to prove below (as indicated in Remark \ref{main-rem}) that there exists a \textit{fractional power} of the resolvent $(R_A(z))^{\,q}$ for $q > p \geq 1$.
\end{remark}

\begin{corollary} \label{non-entiers-2}
The assertion of Theorem \emph{\ref{main-1B}} holds for any real number $q > p$, where
$p \geq 1$.
\end{corollary}
\begin{proof}
The key point is to show that the fractional power of the resolvent can be defined by the right-hand side of the representation \eqref{resolvent-q} for any real $q > p$. First, we introduce the function
\begin{equation}
\label{function-q}
F_z : \ (p,\infty) \ni q \ \mapsto \ \Gamma(q)^{-1} \int_0^\infty \!\!\mathrm{d}t\, t^{q -1} \,
e^{- t\,(A - z\bbbone)} \ , \quad {\re }\,z < - \omega_0 \ ,
\end{equation}
where $t \mapsto t^{q -1} : = e^{(q -1) \, \ln  t}$ is the first branch of the fractional power function corresponding to the branch $t \mapsto \ln  t$ of the
logarithm on $\bbR^+$.
Based on (\ref{resolvent-1-n}) and (\ref{resolvent-1-n-bound}), the function \eqref{function-q} is well-defined and $\norm{\cdot}_1$-continuous on $(p, + \infty)\,$.

For an integer $q$ with $q > p$, definitions \eqref{resolvent-q} and \eqref{function-q} yield
$R_A(z)^{\,q} = F_z (q)$ with the composition law for integers $q_1,q_2: \,R_A(z)^{\,q_1} R_A(z)^{\,q_2} =
R_A(z)^{\,q_1 + q_2}$.
To ensure that the representation \eqref{function-q} defines the \textit{fractional powers} of the resolvent, one has to show that $F_z$ satisfies the \textit{composition law}:
\begin{equation}
\label{resolvent-q-prod}
F_z (q_1) \ F_z (q_2) = F_z (q_1 + q_2) \, ,  \quad  {\re }\, z < - \omega_0 \ ,\text{ for any real $q_i > p$.}
\end{equation}
 After a change of variables: $t_1 = x^2$, $t_2 = y^2\,$, and then $x = r \cos \theta$, $y = r \sin \theta$, we obtain:
\begin{align}
\label{resolvent-zeta-prod2}
F_z (q_1) \ F_z (q_2) &= \Gamma(q_1 )^{-1} \Gamma(q_2 )^{-1}
\int_0^\infty \!\!\mathrm{d}t_1\, \int_0^\infty \!\! \mathrm{d}t_2 \ t_1^{q_1 -1} \,
t_2^{q_2 -1} \, e^{- (t_1 + t_2) \,(A - z\bbbone)} \nonumber \\
& =\frac{2 }{\Gamma(q_1 ) \Gamma(q_2 )}
\int_0^\infty\! \!\!\mathrm{d}r^2 \,  \ r^{2 \, (q_1 + q_2) -2} \,
e^{- \, r^2 \, (A - z\bbbone)} \int_0^{\pi/2} \!\! \!\mathrm{d}\theta \
(\cos \theta)^{2 \, q_1 -1} \, (\sin \theta)^{2 \, q_2 -1} \, .
\end{align}
The $\theta$-integral is equal to $B(q_1, q_2)/2$, with the Beta-function $B(q_1, q_2) \vc
\Gamma(q_1 ) \, \Gamma(q_2 ) / \Gamma(q_1 + q_2)\,$. Consequently, the relation
\eqref{resolvent-zeta-prod2} and the representation \eqref{function-q} prove
\eqref{resolvent-q-prod}.
\end{proof}

\begin{remark}\label{rem-main-1B}
See \cite[Example 2.11]{EckIoc}.
In Theorem \ref{main-1B}, we assumed that the generator $A$ is \textit{invertible}, i.e., $\Ker \,(A)= \{0\}$.
Note that if it is \textit{not} the case, the \textit{Mellin transform}
$\cal M$ of $f(t) = \Tr \,e^{-tA}$ that we used previously, cf. (\ref{resolvent-1-n-bound}), namely:
\begin{align*}
{\cal M}[f]:\,z \in \bbC \to \,
\Gamma(z)^{-1} \int_0^\infty \dd t \, t^{z-1} \, f(t)
\end{align*}
does not exist even for a \textit{positive} generator $A$. Indeed, considering that for a
finite dimensional $\Ker \,(A)$, we have $\lim_{t\to \infty} \Tr \,e^{-tA} =
\dim \Ker \,(A)\, $, the integrand with $f(t)=\Tr \,e^{-tA}$ converges at infinity only for $\Re\,z<0$. On the other hand, if $\Tr  e^{-tA} = \OO_0(t^{-p})$, the convergence at zero requires $\Re(z) > p$.
\\
Another way to see this is as follows: Let $A'\vc A + P$, where $P$ is a projection on the (\textit{finite-dimensional}) kernel of $A$. Then the operator $A'$ is \textit{invertible}, and we obtain
\begin{align}
\label{A invertible}
\Tr \,e^{-tA}=\Tr \,e^{-t(A + P)}-(e^{-t}-1)\dim \Ker (A)\, .
\end{align}
Consequently, $\Tr\,e^{-tA}=\OO_0(t^{-p})$ if and only if $\Tr\,e^{-tA'}=\OO_0(t^{-p})$
for $p\geq 0$. Looking at the decomposition on the right-hand side of \eqref{A invertible}, we infer that
$\Tr \,e^{-tA'}$ has a \textit{Mellin} transform for $\Re\,z>p$, while
${\cal M}[t\to e^{-t}-1](z)$ exists only for $\Re \,z\in (-1,0)$ (and in this case it is
equal to $\Gamma(z)$, see \cite[page 13]{Flajolet}). This also shows that
${\cal M}[t>0\to \Tr\,e^{-tA}]$ exists \textit{nowhere} in $\bbC$.
\end{remark}

Another benefit of Theorem \ref{main-1B} is that that the assertion (b) (i) is a sufficient condition to characterise the Gibbs semigroups.
Recall that in Definition \ref{def:GibbsSG}, cf. \cite[Definition 4.1]{Zagrebnov2},
the Gibbs semigroups are specified as $C_0$-semigroups with \textit{values} in the trace-class $\calL^1$.
On the other hand, one can also construct a Gibbs semigroup from the \textit{resolvent}
$R_{A}$ of its generator $A$ under the hypothesis that operator $A$
is a $p$-generator, see \cite[Definition 4.26 and Proposition 4.27]{Zagrebnov2}.
The advantage of the characterisation of Gibbs semigroups provided by Theorem \ref{main-1B} is that it relaxes the condition regarding the $p$-generator. Thus, in addition to Definition \ref{def:GibbsSG} and equation (\ref{Riesz-Dunford}), one obtains yet another characterisation of Gibbs semigroups that does not involve the Cauchy representation for $p$-generators and, consequently, avoids reference to analyticity.
\begin{corollary} \label{III.15}
The $C_0$-semigroup $\{e^{-tA}\}_{t\geq0}$ is a Gibbs semigroup
if it is $\norm{\cdot}_q$-continuous on $(0, + \infty)\,$ for some $q \geq 1$ and the resolvent $R_A(z) \in {\calL^q}$ for $z\in\rho(A)$ such that also $0 \in \rho(A)\,$.
\end{corollary}

We note that a similar claim is known for the \textit{compact} $C_0$-semigroups, that is, for $q = \infty$, see for example \cite[ Chapter II, 4.29 Theorem]{Engel-Nagel}.

By using a hypothesis that is stronger than a simple requirement for asymptotics, such as $\norm{e^{-tA}}_1=\OO_0(t^{-p})$, we can determine the behaviour of $\norm{R_A(z)^{p+\varepsilon}}_1$ as $\varepsilon$ approaches zero.

\begin{proposition}
\label{h-stronger}
Assume that $A$ generates a Gibbs semigroup
and that $\norm{e^{-t \,A}}_1$ has an asymptotic expansion in powers of $t^{-1}$ with the leading term $Ct^{-p}$ (i.e., $\norm{e^{-t\,A}}_1 =\OO_0(t^{-p})\,$). Then for
$\varepsilon > 0\,$, we have
\begin{align}
\label{eq-limitexists}
\norm{R_A(z)^{\,p+\varepsilon}}_1 =\OO_0(\varepsilon^{-1}) \,\,\text{ for $z\in \rho(A)\,$.}
\end{align}
\end{proposition}
\begin{proof}
We follow the proof of \cite[Proposition 5.3]{CRSS}. \\
By Theorem \ref{main-1B} and its corollary, we already know that  $\norm{ \,R_A(z)^{\,p+\varepsilon} }_1 <\infty$ for any
$\varepsilon>0$. Therefore, it remains to prove the asymptotics \eqref{eq-limitexists}. To do so, we define two integrals:
\begin{eqnarray*}
 &\zeta_1(s) \vc \Gamma(s)^{-1} \int_0^1 \! \dd t\, t^{s-1} \ \norm{e^{-t(A-z\bbbone)}}_1
\quad \text{ for } s>p \, , \\
&\zeta_2(s) \vc  \Gamma(s)^{-1} \int_1^\infty \!\!\!\dd t\,t^{s-1} \
\norm{e^{-t(A-z\bbbone)}}_1 \quad \text{for } s>0 \, .
\end{eqnarray*}
These integrals are convergent since the integrands are continuous and have an integrable behaviour as $t \to 0$ due to the asymptotic property $\norm{e^{- t A}}_1 =\OO_0(t^{-p})$, and also because of the exponential decay of $\norm{e^{-t(A-z\bbbone)}}_1$ for $z\in \rho(A)$ as $t\to \infty$, as stated in (\ref{estim-T}). Note that $\zeta_2$ is holomorphic in a complex neighborhood of $s=p$, and $\norm{R_A(z)^{\,s}}_1\leq \zeta_1(s) +\zeta_2(s)$ exists when $\Re(s)>p$.

Using $\Gamma(s)^{-1}\! \int_0^1\dd t\, t^{s-1} t^{-p} = [\Gamma(s)\, (s-p)]^{- 1}$, we obtain $\norm{R_A(z)^{\,s}}1 \leq {C} \ [\Gamma(s)\, (s-p)]^{- 1} + f(s)$, where $f$ is a function that is holomorphic around $s=p$. It should be mentioned that lower-order terms in the asymptotic expansion do contribute to the function $f(s)$, but these contributions are holomorphic around $s=p$. Therefore, we obtain the asymptotics \eqref{eq-limitexists} because $\lim_{s\downarrow p} (s-p) \,f(s)=0$.
\end{proof}

In the following result, which is due to \cite[Proposition 5.3]{CRSS}, $\Tr_\omega$ denotes a \textit{Dixmier trace} associated with a state $\omega$ on $l_\infty$ that is invariant under dilation (see \cite[Definition 3.8]{CRSS} or \cite[Theorem 10.1.2 (a)]{LSZ}, and \cite[Theorem 2.20]{Gof-Usa} for the property of dilation invariance). Specifically, we have
\begin{align*}
 \Tr_\omega\,A^{-p} \vc \omega\,{\Big(}{\big\{}\frac{1}{\ln(n+2)}\,
\sum_{k=0}^n \lambda_{k+1}(A^{-1})^{\,p}{\big \}}_{n=0}^\infty\,{\Big )}\ .
\end{align*}

\begin{lemma}\label{corol-h-stronger}
Under the hypothesis of previous proposition, assume additionally that $A$ is positive and invertible. Then
\begin{align*}
 \lim_{\varepsilon \downarrow 0} \,\varepsilon \Tr \,A^{-(p+\varepsilon)} = p\Tr_\omega A^{-p} \,.
\end{align*}
Moreover, the existence of this limit is equivalent to
$A^{-1} \in \calZ_p \supsetneq \calL^{p,\infty}$, or $A^{-p}\in \calM^{1,\infty}$.
\end{lemma}

Now, we focus on the case of holomorphic Gibbs semigroups.
To this end, we establish conditions on the generator $A$ that ensure a bounded $C_0$-semigroup $\{U_A(t)\}_{t \geq 0}$ admits a holomorphic extension $\{U_A(z)\}_{z\in S_{\theta}\cup {0}}$ into a sector $S_{\theta} = \{ z \in \bbC \, : \, {\Re }(z) > 0  \text{ and } \abs{\arg(z)} < \theta \}$ with semi-angle $\theta \in (0, \pi/2)$. Then, under additional conditions on the resolvent of $A\,$, we derive the following generalisation of Lemma \ref{lem-behaviour for positive operator} (i) and (ii), extending the results to \textit{non-self-adjoint} holomorphic Gibbs semigroups.

\begin{theorem} \label{corol-main-1}
Given a bounded $C_0$-semigroup $\{e^{- t\,A}\}_{t\geq0}$ and $p\geq 1$, the following assertions are equivalent:

\emph{(i)} The operator $A$ is $m$-sectorial with spectrum $\sigma(A) \subset S_{\pi/2 -\theta}\,$, and there exist $z \in \bbC \backslash S_{\pi/2 - \theta} \subset \rho(A)$ such that $R_{A}(z) \in \calL^p$ for some $p \geq 1$.

\emph{(ii)} The semigroup extends to a bounded holomorphic Gibbs semigroup $\{e^{- z\,A}\}_{z\in S_\theta \, \cup \, \{0\}}$.
\\
{Under these hypotheses,} this semigroup exhibits the asymptotic behaviour $\norm{e^{-z A}}_1 = \OO_{0}(\abs{z}^{-1})\,$ for $z\in \overline{S}_{\theta'}\,$, that is in a smaller closed sector where $\theta'<\theta\,$.

\end{theorem}

\begin{proof}

(i) $\Longrightarrow$ (ii): If $A$ is an $m$-sectorial operator with spectrum $\sigma(A) \subset S_{\pi/2 -\theta}\,$, one can construct a holomorphic semigroup with generator $A$ using the \textit{Riesz--Dunford} formula:
\begin{align}
\label{Riesz-Dunford}
U_{A}(z) =\frac{1}{2\pi i} \int_\Gamma \dd \zeta \,e^{-\zeta z}\, (\zeta \bbbone- A)^{-1} \ , \quad z\in S_{\theta} \ .
\end{align}
Here, the integrand, which is (up to sign) the resolvent of an $m$-sectorial operator,
satisfies, for any $\varepsilon \in (0,\theta)$, with \( M_\varepsilon \) independent of \( \zeta \), the following conditions:
\begin{align}
\label{Riesz-Dunford-cond}
\norm{(\zeta \bbbone - A)^{-1}}  \leq \frac{M_\varepsilon}{\abs{\zeta}}\ , \quad
\zeta \in D_{\theta -\varepsilon}:=
\bbC \backslash S_{\pi/2 - \theta + \varepsilon} \ \subset \rho(A)\ .
\end{align}
Then, the integral (\ref{Riesz-Dunford}) is absolutely convergent for $t > 0$ in the operator-norm topology if $\Gamma \subset D_{\theta -\varepsilon}$ is a positively oriented contour around $S_{\pi/2 -\theta + \varepsilon}$ and define a \textit{bounded holomorphic} semigroup
$\{U_{A}(z)\}_{z\in S_\theta \, \cup \, \{0\}}$  (\ref{Riesz-Dunford}), which is uniformly bounded and strongly continuous within the smaller sector $S_{\theta-\varepsilon}$, see e.g.,
\cite[Chapter IX \S1.6]{Kato}.

Furthermore, since the resolvent condition $R_{A}(\zeta) \in \mathcal{L}^p$ holds for some $ p \geq 1$ and $\zeta \in D_{\theta -\varepsilon}$, the holomorphic semigroup defined by the (Bochner) integral (\ref{Riesz-Dunford}) also belongs to the class $\calL^p$.
(In this case, $A$ is referred to as a $p$-\textit{generator}.) \\
Moreover, the semigroup property of $\{U_{A}(z)\}_{z\in S_\theta \, \cup \, \{0\}} \,$ implies that $U_{A}(z) \in \calL^1$ for $z\neq 0$. Consequently, it is a Gibbs semigroup.

(ii) $\Longrightarrow$ (i):
If $\{U_{A}(z)\}_{z\in S_\theta \, \cup \, \{0\}} \,$ is a holomorphic Gibbs semigroup, then
the family $\{U_A (t)\}_{t > 0}$ belongs to $\calL^1$ and it is $\calL^1$-continuous. Hence, we can express the resolvent $R_A(z)$ via the Laplace transform as the $\calL^1$-Bochner integral (\ref{resolvent-B}). This representation yields that
$\bbC_{-}\vc \{z \in \bbC : \Re (z) < 0\} \subset \rho(A)$ and that, in this region,
$R_{A}(z) \in \calL^1$. Consequently, it follows that $R_{A}(z) \in \calL^p$ for any $p \geq 1$ and $z \in \bbC_{-}\,$.\\
Since the semigroup $\{U_{A}(z)\}_{z\in S_\theta \, \cup \, \{0\}} \,$ is holomorphic,
by varying in (\ref{resolvent-B}) the integration variable $\lambda$ within the sector $t\in S_\theta$, and adjusting $\lambda \in \bbC$ in such a way that
$\lambda t \in [0, + \infty)$, one checks that the resolvent set is larger than the half-plane $\bbC_{-}\,$, and in fact, $\rho(A) \subset \bbC \backslash S_{\pi/2 - \theta}$. This proves that the generator $A$ is $m$-sectorial with its spectrum confined to the sector $S_{\pi/2 - \theta}\,$.


To analyse the claimed asymptotics, it is convenient to introduce the auxiliary $m$-sectorial generator $A_\lambda := A + \lambda \bbbone$ with a parameter $\lambda > 0$ to ensure invertibility (i.e., the existence of the resolvent): by virtue of (\ref{Riesz-Dunford-cond}), $(- \lambda) \in \rho(A)$. Thus, the resolvent condition $ R_{A}(- \lambda) \in \calL^1$ given by hypothesis (ii) yields the estimate
\begin{align}
\label{estimate-p}
\norm{U_{A}(z)}_1 \leq \norm{A_{\lambda} U_{A}(z)} \ \norm{R_{A}(- \lambda)}_1\,.
\end{align}
Since $-\lambda \in D_{\theta -\varepsilon}\,$, the formula (\ref{Riesz-Dunford}) and the inequality (\ref{Riesz-Dunford-cond}) remain valid for $A_\lambda$. Then changing the integration variable to $\zeta' = \zeta z$, we obtain
\begin{align}
\label{Riesz-Dunford-new}
U_{A_\lambda}(z) =\frac{1}{2\pi i z} \int_\Gamma \dd \zeta' \,e^{-\zeta'}\,
({\zeta'}/{z} \ \bbbone- A_\lambda)^{-1} \ , \quad z \in S_{\theta - \varepsilon} \,.
\end{align}
As the left-hand side of (\ref{Riesz-Dunford-new}) is a bounded holomorphic semigroup, we get for any $n\in \bbN$, the estimates for operator-norm derivatives:
\begin{align}
\label{Riesz-Dunford-deriv}
\norm{\partial_{z}^{n}U_{A_\lambda}(z)}
= \norm{A_{\lambda}^n \, U_{A}(z)\, e^{-\lambda\,z}} &=
\big{\Vert}\frac{1}{2\pi i z^{n+1}} \!\int_\Gamma \dd \zeta' \ (\zeta')^n \,e^{-\zeta'}\, ({\zeta'}/{z} \ \bbbone- A_\lambda)^{-1} \big{\Vert}
\nonumber \\
&\leq  M_{\varepsilon, n}\,\abs{z}^{-n}\,,
\,\,\, z \in S_{\theta - \varepsilon}.
\end{align}
As a result, the inequalities (\ref{estimate-p}) and (\ref{Riesz-Dunford-deriv}) for $n=1$ yield
an upper bound for the asymptotic behaviour: $\norm{e^{-z A}}_1 = \OO_{0}(\abs{z}^{-1})$ for $z\in \overline{S}_{\theta' = \theta - \varepsilon}\,$.
\end{proof}

In the case of a quasi-bounded holomorphic semigroup with generator $A\in Q(M,\omega_0)$, the proof of the assertion in Theorem \ref{corol-main-1} follows \textit{mutatis mutandis} from the arguments presented above replacing $A$ with the generator $A +\omega_0 \bbbone$, see for instance \cite[Chapter IX \S1.6]{Kato} and \cite[Proposition 4.27]{Zagrebnov2} for details.\\
{It is worth noting that, in the preceding theorem, we refine the asymptotic behaviour of $\norm{e^{-z A}}_1$ established in \cite[Theorem 4.1 and Corollary 4.2]{Blunck} within the framework of bounded holomorphic semigroups acting on Banach spaces $X$, with, in their notation, $\calA(X)=\calL^p(\calH)$. Specifically, we show that $R_A(z) \in \calL^1$ is equivalent to $R_A(z)\in \calL^p$ for any $p>1\,$.}

{Now, let us extend Theorem \ref{singularity} since the positivity of the generator is not necessary.
\begin{theorem}\label{corol-main-2}
Given a quasi-bounded self-adjoint Gibbs semigroup, we still have assertion \emph{(i)} of Theorem \emph{\ref{singularity}}, as well as for \emph{(ii)} we obtain
\begin{align}\label{selfadjoint case}
F^{(n)}(t) \underset{t \downarrow 0}{\sim} \norm{\,A^n\,e^{-tA}}_1\, .
\end{align}
\end{theorem}
}

\begin{proof}
By hypothesis, the function  $F(t)\vc \norm{e^{-tA}}_1=\Tr e^{-tA}=\sum_k e^{-t\lambda_k}$ is infinitely differentiable, as shown in the proof of Lemma \ref{prop-derivative of trace-norm}. However, when $A$ is not positive, \eqref{derivative positive case} is modified to:
\begin{align*}
& (-1)^n F^{(n)}(t)=\Tr A^n e^{-tA}=\sum_k \lambda_k^n \,e^{-t\lambda_k}  \leq \sum_k \abs{\lambda_k}^n \,e^{-t\lambda_k} =\norm{\,A^n e^{-tA}}_1 \, ,
\end{align*}
with a strict inequality when $n$ is odd.

Despite of this, the assertion (i) of Theorem \ref{singularity} holds true. For a $C_0$-semigroup (Lemma \ref{lem-ReA et omega0}), there exists $\omega_0\in \bbR$ such that $B=A +\omega_0\bbbone$ is a positive self-adjoint operator, and $F(t) =e^{t\omega_0}\norm{e^{-tB}}_1\sim_{t \downarrow 0} \norm{e^{-tB}}_1\,$. Thus we may apply Theorem  \ref{singularity} (i) to $B\,$.
\\
Now, let us assume the hypothesis of Theorem \ref{singularity} (ii) but again, only for a semibounded from below self-adjoint generator $A\,$.
The assertion (ii) is \textit{a priori} modified as
$\norm{(A+\omega_0 \bbbone)^n\,e^{-tA}}_1 \sim_{t\downarrow 0}  g^{(n)}(t)\,$ since $\norm{e^{-tB}}_1 \sim_{t\downarrow 0} g(t)\,$ and $B$ is positive. Let us prove that we still obtain for $n\in \bbN_0$:
\begin{align}
\label{derivative-case selfadjoint}
F^{(n)}(t)=(-1)^n\,\Tr A^n\,e^{-tA} \underset{t \downarrow 0}{\sim} g^{(n)}(t)
\end{align}
Indeed, using \eqref{derivative of g}, we have, for $p=\text{ind}_f$ and $k\in \bbN_0$ with $k\leq n$,
\begin{align}
\label{quotient of derivatives}
\frac{g^{(k)}(t)}{g^{(n)}(t)} =\frac{g^{(k)}(t)}{g(t)}\cdot \frac{g(t)}{g^{(n)}(t)} \, \underset{t\downarrow 0}{\sim} \, \frac{\Gamma(p+k)}{\Gamma(p)}\,t^{-k}\frac{\Gamma(p)}{\Gamma(p+n)} \,t^n =\frac{\Gamma(p+k)}{\Gamma(p+n)}\,t^{n-k} \,\underset{t \downarrow 0}{\sim} \,\delta_{n,k} \, t^0\,.
\end{align}
This means $\lim_{t \downarrow 0} \frac{g^{(k)}(t)}{g^{(n)}(t)}=\pm \delta_{n,k} $ since $g$ is real valued.
Thus, given $G(t)\vc \norm{e^{-tB}}_1$, we have by applying Theorem \ref{singularity} (ii)
\begin{align}
\label{a}
G(t) \underset{t \downarrow 0}{\sim} g(t) \,\text{ and }\,G^{(k)}(t) \underset{t \downarrow 0}{\sim} g^{(k)}(t)\,.
\end{align}
Since $F^{(n)}(t) = \sum_{k=0}^n$ {\small $\binom{n}{k}$}  $(e^{t\omega_0})^{(n-k)}(t)\,G^{(k)}(t)\,$,
we obtain by virtue of \eqref{quotient of derivatives} and \eqref{a}
\begin{align*}
 \frac{F^{(n)}(t)}{g^{(n)}(t)}
= \sum_{k=0}^n\,\binom{n}{k}
 \,\omega_0^{n-k}\,e^{t\omega_0} \,  \frac{G^{(k)}(t)}{g^{k)}(t)}\,\frac{g^{(k)}(t)}{g^{(n)}(t)} \, \underset{t\downarrow 0}{\to}\, \pm 1\,.
\end{align*}
\\ \indent
In summary, we have
\begin{align*}
F^{(n)}(t)\,\underset{t\downarrow 0}{\sim} \, g^{(n)}(t)\,\underset{t\downarrow 0}{\sim} \,\norm{(A+\omega_0 \bbbone)^n\,e^{-tA}}_1\,.
\end{align*}

Note that $F^{2n}(t)\underset{t\downarrow 0}{\sim} \norm{\,A^{2n}\,e^{-tA}}_1\,$, so to prove \eqref{selfadjoint case}, it is sufficient to consider that $n$ is odd. \\
Then $0 \leq \norm{\,A^n\,e^{-tA}}_1 - \Tr\,A^n\,e^{-tA} = 2\sum_{\lambda_k <0} \,\abs{\lambda_k}^n\, e^{-t\lambda_k} \,$.
Since this sum has only a finite number of terms and approaches $2\sum_{\lambda_k <0} \abs{\lambda_k}$ when $t\to 0\,$, while $\norm{\,A^n\,e^{-tA}}_1$ goes to infinity using Proposition \ref{prop-behaviour at 0} as $\norm{e^{-tA}}_1 \leq \norm{A^{-n}}\,\norm{\,A^n\,e^{-tA}}_1$, we conclude that $\norm{\,A^n\,e^{-tA}}_1 $ and $\Tr\,A^n\,e^{-tA}$ have the same asymptotic behaviour.
\end{proof}
As a side result, we obtained that $\norm{\,A^n\,e^{-tA}}_1 \underset{t\downarrow 0}{\sim} \,\norm{(A+\omega_0 \bbbone)^n\,e^{-tA}}_1\,$.

\begin{remark}
\label{normal-case}
Let us now relax the requirement of self-adjointness and assume that the generator $A$ of a Gibbs semigroup is a \textit{normal} operator. In this case, we note that $\norm{e^{-tA}}_1= \norm{e^{-t\Re A}}_1\,$, as seen in \eqref{normal trace-norm}. This leads us to examine the associated self-adjoint Gibbs semigroup $\{e^{-t\Re A}\}_{t\geq 0}\,$, which, as mentioned earlier, forms a trace-norm holomorphic semigroup in the right half of the complex plane. As a result, the function $F(t)\vc \norm{e^{-t\Re A}}_1 = \Tr \, e^{-t\Re A}$ becomes infinitely differentiable. We can then establish $(-1)^n\,F^{(n)}(t) = \Tr \,(\Re A)^n e^{-t\Re A}\,$. Therefore, the assertion (i) of Theorem \ref{singularity} remains valid, while assertion (ii) takes on a modified form: $\norm{(\Re A+\omega_0\bbbone)^n\,e^{-t\Re A}}_1 \sim_{t\downarrow 0} g^{(n)}(t)\,$.
\\
On the contrary, in the general case, we cannot assert that $\norm{(\Re A)^n \, e^{-t\Re A}}_1 = \norm{\,A^n\,e^{-tA}}_1$, meaning that the equivalence established for the self-adjoint case is \textit{lost}.
In Remark \ref{rem-non-selfadjoint case} we shall provide an example of a \textit{normal}
operator $A$, generating a Gibbs semigroup such that
$F'(t) \sim_{t \downarrow 0} t^{-2}$ whereas $\norm{\,A \,e^{-tA}}_1  \sim_{t \downarrow 0} ct^{-3}\,$.
\end{remark}

From Remark \ref{normal-case}, we infer that for non-self-adjoint but \textit{normal} generators, a direct connection between $(\norm{e^{-tA}}_1)^{(n)}(t)$ and $\norm{\,A^n e^{-tA}}_1$ may be lost. Note that the semigroup in the example of Remark \ref{rem-non-selfadjoint case} is \textit{not} holomorphic. This gives a hint on how one can strengthen the result by involving a \textit{holomorphic} Gibbs semigroup $\{e^{-z A}\}_{z \in S_\theta \cup \{0\}}$ with a semi-angle $\theta$. A corresponding condition applies to the generator $A$ of a holomorphic semigroup,
as stated in Proposition \ref{prop-Karamata-hol}. It is noteworthy that the following theorem restores the equivalence.

Furthermore, it is also important to mention the existence of \textit{normal} infinitely trace-norm-differen-tiable Gibbs semigroups with \textit{no} holomorphic extension, as we will demonstrate in Remark \ref{rem-non-selfadjoint case}.
\begin{theorem} \label{singularity-ter}
Let $A$ be the generator of a quasi-bounded holomorphic Gibbs semigroup $\{e^{-z\, A}\}_{z\in S_{\theta}}$ in the sector $S_{\theta}$ with semi-angle $\theta <  \pi/2\,$. Let $F(t)\vc \norm{e^{-t \, A}}_1\,$, $f\in \SR$ and $g(t)\vc f(t^{-1})\,$. \\
Then, for any $n \in \bbN_0\,$, $\norm{\,A^n \, e^{- t \, A}}_1 =\OO_0(g^{(n)}(t))\,$ ,
whenever $F(t)=\OO_0(g(t))\,$.
\end{theorem}
\begin{proof}
Using the Cauchy formula for derivatives of the function $z \mapsto e^{-z\, A}$ at $z = t > 0$, we obtain the expression:
\begin{equation}\label{cauchy5}
A^n \, e^{- t \, A} =\frac{(-1)^n n!}{2\pi i}\int_{C_\rho} \,\dd \zeta \
\frac{e^{- \zeta \, A}}{(\zeta-t)^{n+1}} =\frac{(-1)^nn!}{2 \pi\rho^n} \int_0^{2 \pi}
\dd \varphi \, e^{- i n\varphi} \,
e^{- (t + \rho \, e^{i \varphi}) \, A} \, , \quad n \in \bbN_0 \, ,
\end{equation}
where the circle $C_\rho \subset S_{\theta - \varepsilon}$ of radius
$\rho = t \, \sin (\theta - \varepsilon)\, $ is centered at $t>0$ and
$\theta <  \pi/2$.
Then, the equation (\ref{cauchy5}) yields the operator-norm estimates:
\begin{equation}\label{cauchy6}
\norm{\,A^n \, e^{- t \, A}} \leq  \frac{n!}{2 \pi\rho^n} \int_0^{2 \pi}
\dd \varphi \ \norm{ e^{- (t + \rho \, e^{i \varphi}) \, A}} =
\frac{n! \ M_{\varepsilon}}{(t \, \sin (\theta - \varepsilon))^n} =:
M_{n, \, \varepsilon,\theta}\,t^{-n}\ ,
\quad t > 0  \, , \quad n \in \bbN_0 \, .
\end{equation}
Now, let $0< \delta < t$. Utilising the estimates (\ref{cauchy6}), we deduce
\begin{equation}\label{estim1}
\norm{\,A^n \, e^{- t \, A}}_1  \leq \norm{A^n \, e^{- (t - \delta) \, A}} \,
\norm{e^{- \delta \, A}}_1 \leq
M_{n, \, \varepsilon,\theta} \,(t - \delta)^{-n} \
\norm{e^{- \delta A}}_1 \,,\,\,  t > 0  \, , \,\, \theta <  \pi/2 \, ,
\,\, n \in \bbN_0 \, .
\end{equation}
To summarise, we have shown the following: given $\delta = at$ with $0 < a < 1$ and
$F_n(t)\vc\norm{\,A^n \, e^{- t \, A}}_1\,$ for $ n\in \bbN_0\,$, so $F_0(t)$
coincides with $F(t)$, we obtained
\begin{align*}
F_n(t) \leq c_n\, t^{-n} F_0(at)\,, \quad \text{where } c_n \vc M_{n, \, \varepsilon,\theta}\,(1-a)^{-1}\,,\,\,n\geq 1\,.
\end{align*}
By applying the same argument as in the proof of Theorem \ref{singularity}, we find:
\begin{align*}
\frac{F_n(t)}{\abs{g^{(n)}(t)}} \leq c_n\,\cdot\, \frac{F_0(at)}{f((at)^{-1})}\,\cdot\,\frac{f((at)^{-1})}{g(t)}\,\cdot \,\frac{g(t)}{t^n\,\abs{g^{(n)}(t)}} \,.
\end{align*}
By hypothesis, $F_0(t) = \OO_0(g(t))$. Combining this with \eqref{derivative of g}, we obtain:
\begin{align*}
\limsup_{t\downarrow 0}\frac{F_n(t)}{\abs{g^{(n)}(t)}} \leq c_n\,\cdot\, \limsup_{t\downarrow 0}\frac{F_0(at)}{g(at)}\,\cdot\,a^{-\text{ind}_f}\,\cdot \,\frac{\Gamma(\text{ind}_f)}{\Gamma(\text{ind}_f+n)} <\infty\,,
\end{align*}
and the proof is complete.
\end{proof}

\begin{remark}
\label{rem-non-selfadjoint case}
In Theorem \ref{singularity-ter}, we derived certain properties related to the
function $F(t)=\norm{e^{-tA}}_1$, and one may wonder if a weaker hypothesis, namely that
$\{e^{- tA}\}_{t\geq 0}$ is bounded and (infinitely) ${\norm{\cdot}}_1$-differentiable, is sufficient to reach
the conclusion. However, this is not the case, as demonstrated by the counterexample of a generator $A$ for an infinitely differentiable Gibbs semigroup that lacks a holomorphic extension: \\
Let $A = \sum_k \lambda_k P_k$, where $\{\lambda_k=k+ik^2\}_{k\geq 1}$ are the eigenvalues and $P_k$ are the corresponding projections on orthonormal basis $\{e_k\}_{k\geq 1}$ of $\calH = \ell^2(\bbN)$. Then the operator $A$ is \textit{normal} with $\Re A>0\,$, and it generates a Gibbs semigroup such that
\begin{align}\label{exGibbs0}
e^{-t A} =  \sum_{k = 1}^{\infty}\, e^{-t \lambda_k} \, P_k \, ,
\quad t \geq 0 \, .
\end{align}
where the series in the right-hand side of \eqref{exGibbs1} is trace-norm convergent \textit{uniformly} in $t>0$. Given that
$t \mapsto e^{-t \lambda_k} \, P_k$ is ${\norm{\cdot}}_1$-continuous for $t > 0$,
the series in (\ref{exGibbs0}) and thus $t \mapsto e^{- tA}$ is also
${\norm{\cdot}}_1$-continuous for $t > 0$. Similarly to (\ref{exGibbs0}), we obtain the representation
\begin{align}\label{exGibbs1}
A\, e^{-t A} =  \sum_{k = 1}^{\infty}\, \lambda_k \, e^{-t \lambda_k} \, P_k  =
\sum_{k = 1}^{N}\, \lambda_k \, e^{-t \lambda_k} \, P_k  +
\sum_{k = N + 1}^{\infty}\, \lambda_k \, e^{-t \lambda_k} \, P_k \, ,
\quad t > 0 \, .
\end{align}
By virtue of the limit
\begin{align}\label{exGibbs10}
\big{\Vert}\sum_{k = N + 1}^{\infty}\, \lambda_k \,
e^{-t \lambda_k} \, P_k\big{\Vert}_1 \leq
\sum_{k = N + 1}^{\infty}\, \vert \lambda_k \, e^{-t \lambda_k}\vert
 \, \,\norm{P_k}_1 =
\sum_{k= N + 1}^\infty \sqrt{k^2+k^4} \ e^{-t k} \underset{N\to \infty}{\to} 0 \, ,
\quad t > 0 \, .
\end{align}
we infer that series in (\ref{exGibbs1}) is  ${\norm{\cdot}}_1$-convergent \textit{uniformly} in $t>0$. As a consequence the Gibbs semigroup
$\{e^{- tA}\}_{t\geq 0}$ is ${\norm{\cdot}}_1$-differentiable for $t>0$
and the derivative $\partial_t(\cdot)$ commutes with the sum in (\ref{exGibbs1}).
\\ \indent
Since, in turn, the series (\ref{exGibbs1}) is term-wise
${\norm{\cdot}}_1$-differentiable and an estimate similar to (\ref{exGibbs10})
holds uniformly for $t>0\,$, the second ${\norm{\cdot}}_1$-derivative of semigroup
$\{e^{- tA}\}_{t\geq 0}$ exists. Again, due to the commutativity of
${\norm{\cdot}}_1$-$\partial_t (\cdot)$ with the sum, the second derivative can be expressed in the same manner as in (\ref{exGibbs1}), but with coefficients
$\{\lambda_k^2 \, e^{-t \lambda_k}\}_{k\geq 1}\,$.

The corresponding estimate in (\ref{exGibbs10}) allows for the possibility of iterating this procedure for any ${\norm{\cdot}}_1$-$\partial_{t}^n (\cdot)\,$, where $n \in \mathbb{N}$.
\\ \indent
Notwithstanding, the Gibbs semigroup $\{e^{- tA}\}_{t\geq 0}$ has no
analytic extension $t\in S_{\theta }$ in sector $S_{\theta }$ for any $\theta < \pi/2$.
By the semigroup composition law, it is sufficient to check that
$\{e^{- tA}\}_{t\geq 0}$ has \textit{no analytic extension} from $\bbR^+$ in the operator-norm topology.
The latter follows from the estimate:
\begin{align}\label{estim-t2}
\norm{\,A \,e^{-tA}}= \sup_{k\geq 1} \sqrt{k^2+k^4} \ e^{-t k} \geq
\frac{4 \, e^{-2}}{t^2}  \, , \quad t > 0 \, ,
\end{align}
which means that a fundamental inequality for an analytic extension, namely $\norm{\,A \,e^{-tA}} \leq M/t$, fails, see (\ref{Riesz-Dunford-deriv}).
The reason is that for any $\theta < \pi/2$ the spectrum $\sigma(A)$ is \textit{not contained} in the closed sector $\overline{S}_{\pi/2-\theta}$.

For this example, it can be also verified that for the given generator $A$, one has
\begin{align*}
F(t)\vc \norm{e^{-tA}}_1=\norm{e^{-t\Re A}}_1=\sum_{k\geq 1}\, e^{-tk} =\frac{e^{-t}}
{1-e^{-t}}\,\underset{t\downarrow 0}{\sim}\,t^{-1} \, ,
\end{align*}
see \eqref{normal trace-norm}, as well as that
\begin{align*}
F'(t)= \frac{- \, e^{t}}{(e^t-1)^2} \,\underset{t\downarrow 0}{\sim}\, t^{-2} \quad\text{whereas \,\,\, } \norm{\,A\, e^{-tA}}_1=\Tr \,\abs{A} e^{-t\Re A}=\sum_{k=1}^\infty \sqrt{k^2+k^4} \,e^{-t k} \,\underset{t\downarrow 0}{\sim}\,ct^{-3}
\end{align*}
since
\begin{align*}
 2\,t^{-3} \underset{t\downarrow 0}{\sim} \frac{e^t+e^{2t}}{(e^t-1)^3}=\sum_{k=1}^\infty k^2e^{-tk}<\sum_{k=1}^\infty \sqrt{k^2+k^4} \,e^{-t k} < \sqrt{2}\sum_{k=1}^\infty k^2e^{-tk} \underset{t\downarrow 0}{\sim}\,2\sqrt{2}\,t^{-3} \,.
\end{align*}
Similarly one obtains
\begin{align*}
& F''(t)=\frac{e^t+e^{2t}}{(e^t-1)^3} \,\underset{t\downarrow 0}{\sim}\, 2\,t^{-3}&\\
& \norm{\,A^2 e^{-tA}}_1=\Tr \,AA^* e^{-t\Re A}=\sum_{k=1}^\infty (k^2+k^4) \,e^{-t k}=\frac{2e^t+10e^{2t}+10 e^{3t}+2e^{4t}}{(e^t-1)^5}\,\underset{t\downarrow 0}{\sim}\, 24\,t^{-5} \, ,
\end{align*}
so the conclusion of Theorem \ref{singularity-ter} (ii) fails for $n=1,\,2, \, $ etc.,
since the normal operator $A$ is not the generator of a holomorphic Gibbs semigroup.
\end{remark}

\begin{remark} \label{rem-III.15}
(a) Let a \textit{self-adjoint} generator $A$ be such that $\norm{e^{-t A}}_1 < \infty$ for $t>0$. The self-adjointness implies that $C_0$-semigroup
$\{e^{-t \, A}\}_{t\geq 0}$ is \textit{infinitly} strongly (and in operator-norm) differentiable for $t>0$. As a consequence, it is an (immediately) trace-norm holomorphic Gibbs semigroup in sector $S_{\theta < \pi/2}\ $.
\\
\indent (b) In contrast to the self-adjoint case, for \textit{non-self-adjoint} generators, it is possible to construct examples of Gibbs semigroups that, although being \textit{infinitely} trace-norm differentiable for $t> 0$, are \textit{not} holomorphic. For certain examples (see one in Remark \ref{rem-non-selfadjoint case}), we obtain for any $n\in \bbN$ that, $\partial_{t}^{n} F_{A}(t) = \Tr \, A^n \, e^{-t A}$, where $F_{A}(t) := \Tr\, e^{-t A}\,$, but $\norm {\partial_{t} e^{-t A}} \geq M t^{-\, s}$ for \(s > 1\) (as \eqref{estim-t2}),
in contrast to the fundamental condition for an analytic extension: $\norm{\,A \,e^{-tA}} \leq M/t$, which we mentioned in (\ref{Riesz-Dunford-deriv}).
\end{remark}


We conclude this section with a complementary result that involves the concept of the
\textit{integrated semigroups.} First, we recall the basic definition, see e.g. \cite{Arendt}.
\begin{definition} \label{def-IntSG}
Let $\{U_A (t)\}_{t \geq 0}$ be a $C_0$-semigroup on $\calH$.
The family $\{S_A (t)\}_{t \geq 0}\,$, which is given by
\begin{align}
\label{IntSG}
S_A(t) \vc \int_0^{t} \!\mathrm{d}\tau\, U_A(\tau)  \, ,
\end{align}
is referred to as the (one-time) integrated semigroup generated by the operator $A\,$.
\end{definition}
\begin{theorem}
\label{main-2}
Given a $C_0$-semigroup $\{ U_A(t)=e^{-t\, A}\}_{t \geq 0}\,$, the set of resolvent operators $\{R_A(z)\}_{z \in \rho(A)}$ belongs to the ideal $\calL^{p}$ if and only if the integrated semigroup
$\{S_A (t)\}_{t \geq 0}$ belongs to the ideal $\calL^{p}$ for $p > 0$ and $t > 0\,$.
\end{theorem}

Note that Theorem \ref{main-1B} elucidates a \textit{direct} relation between the {resolvent} $\{R_A(z)\}_{z \in \rho(A)}$ and the $C_0$-{semigroup}
$\{U_A (t)\}_{t \geq 0} $. This assertion is \textit{stronger} than Theorem \ref{main-2}.
Specifically, Theorem \ref{main-2} provides a connection between the
\textit{resolvent} and the \textit{integral} (\ref{IntSG}), whereas
Theorem \ref{main-1B} establishes a relation between the
\textit{resolvent} and the \textit{integrand} in (\ref{IntSG}), which requires additional conditions.

\begin{proof}
\textit{Necessity.} Suppose that $\{R_A(z)\}_{z \in \rho(A)} \subset \calL^{p}\,$. For $A\in Q(M,\omega_0)\,$, we get $(-\infty,-\omega_0)\subset \rho(A)$.
\\ To proceed, we introduce $A_{\lambda} := A + \lambda \bbbone$ for $\lambda > \omega_0\,$,
and define the associated integrated semigroup
\begin{equation}
\label{IntSG-lambda}
S_{A_{\lambda}}(t) := \int_0^{t} \!\mathrm{d}\tau\, U_{A_{\lambda}}(\tau)  \, ,
\end{equation}
where $\{U_{A_{\lambda}} (t)\}_{t \geq 0}$ is a bounded semigroup:
$\Vert U_{A_{\lambda}} (t) \Vert \leq M$ for $t\geq 0$.
Using a similar argument as in (\ref{3.12b}), we deduce from (\ref{IntSG-lambda}) the representation
\begin{equation}
\label{3.12-N}
 (\bbbone - U_{A_{\lambda}}(t)) \ (A + \lambda \bbbone)^{-1}  =
 \int_0^t  \!\mathrm{d}\tau \, U_{A_{\lambda}}(\tau) = S_{A_{\lambda}}(t) \, .
\end{equation}
Therefore, by virtue of $R_A(-\lambda) \in \calL^{p}$ and (\ref{3.12-N}), we conclude that
 $\{S_{A_{\lambda}}(t)\}_{t \geq 0} \subset \calL^{p}$.
\\
The same conclusion also holds for the integrated semigroup ${S_{A}(t)}_{t \geq 0}$.

\textit{Sufficiency.}
Now we assume that $\{S_{A}(t)\}_{t \geq 0} \subset \calL^{p}$. First, we note that there exists an \textit{equivalent norm} in $\calH$ such that the \textit{bounded} semigroup $\{U_{A_{\lambda}} (t)\}_{t \geq 0}$ becomes a \textit{contraction} semigroup in this norm, see e.g. \cite[ Lemma 3.10]{Engel-Nagel}. Therefore, we keep the same
notations and infer that $\Vert U_{A_{\lambda}}(t) \Vert  < 1$ for any $ t> 0$.
By iterating the equation
\begin{equation*}
\int_0^{\infty}\!\mathrm{d}\tau \, U_{A_{\lambda}}(\tau) = S_{A_{\lambda}}(t) +
\int_t^{\infty}\!\mathrm{d}\tau \, U_{A_{\lambda}}(\tau) = S_{A_{\lambda}}(t) +
U_{A_{\lambda}}(t)  \int_0^{\infty}\!\mathrm{d}\tau \, U_{A_{\lambda}}(\tau)\, ,
\end{equation*}
we obtain a representation of the resolvent for the generator $A$ at $z = - \lambda$:
\begin{equation*}
R_A(-\lambda) = \int_0^{\infty}\!\mathrm{d}\tau \, U_{A_{\lambda}}(\tau) =
\sum_{m =0}^{\infty} U_{A_{\lambda}}(m t)\int_0^{t} \! \mathrm{d}\tau\,U_{A_{\lambda}}(\tau) = \sum_{m =0}^{\infty} \,U_{A_{\lambda}}^m (t) \ S_{A_{\lambda}}(t) \, , \quad t>0 \, .
\end{equation*}
Since $\Vert U_{A_{\lambda}}(t>0) \Vert  < 1$,
the series $\sum_{m = 0}^{\infty} U_{A_{\lambda}}^m (t)$ is norm-convergent and since $\{S_{A}(t)\}_{t > 0} \subset \calL^{p}$, we conclude that $R_A(-\lambda) \in \calL^{p}$. Furthermore,
one can also extend this result to $\{R_A(z)\}_{z\in \rho(A)} \subset \calL^{p}\,$.
\end{proof}

{This result should be compared with the analogous theorem in \cite{Blunck}, which was established in the more general setting of Banach spaces $X$. Adapting their notation to our context, let $\calA(X) = \calL^p(\cal{H})$ denote a closed operator ideal equipped with the norm $\norm{\cdot}_p\,$, requiring $p \geq 1$. Our Theorem~\ref{main-2} thus extends \cite[Theorem 3.2]{Blunck} to the broader range $p > 0\,$.
}

\section{On stability under perturbations of trace-norm asymptotics}
\label{perturbation}

Let $A$ be the generator of a Gibbs semigroup, and let $F_A(t)\vc \norm{e^{-tA}}_1$. We consider a perturbation $B$ such that $H=A+B$ (with definition of the sum precised later on) also generates a Gibbs semigroup.

In this section, we focus on elucidating the relationships between the asymptotic behaviours of $F_A$ and $F_{H}$ and investigate two eventual assertions:

\quad $(a)$ \,If $F_A(t)=\OO_0(f(t))\,$, then $F_{H}(t)=\OO_0(f(t))\,$,

\quad $(b)$ \,$F_{H}(t)\,\underset{t\downarrow 0}{\sim} \, F_A(t)\,$.

We begin with the easy case. Let perturbation $B$ be such that inequality
\begin{align}
\label{inequality in perturbation}
\norm{e^{-tH}}_1 \leq c_1 \norm{e^{-c_2 tA}}_1\,,\quad  \, \text{for any }t>0\,,
\end{align}
holds true for some  $c_1,c_2>0\,$. Then one can check that assertion $(a)$ holds true.

To this end, in the following proposition and subsequent remark, we present conditions
ensuring that inequality \eqref{inequality in perturbation} is satisfied.
\begin{proposition}\label{III.35}
Let $A$ be the generator of a self-adjoint Gibbs semigroup $\{U_A(t)\}_{t \geq 0}$ on
${\cal H}$ and $B$ be a symmetric operator that is Kato-small with respect to $A\,$, meaning that
\begin{align}
\label{def-relatively Kato-small}
\norm{Bu}\leq a \norm{u} +b\norm{Au},\quad u\in \Dom(A)\subset \Dom(B),\quad
a\geq 0 , \  0< b <1 \, .
\end{align}
Then the sum $A+B$ defines on $\dom(A)$ a self-adjoint operator, which is the generator
$H := A + B$ of a quasi-bounded self-adjoint Gibbs semigroup $\{U_H(t)\}_{t \geq 0}$ such that \eqref{inequality in perturbation} holds true.
\end{proposition}
\begin{proof} Proof of the first part of the statement is the Kato theorem about
stability of self-adjointness of operator $A$ relative to $\calP_{b<1}$-perturbations $B$ satisfying (\ref{def-relatively Kato-small}) \cite[Chapter V \S 4.1, Theorem 4.3]{Kato}. The second part of the proof can be found in \cite[Proposition 4.45]{Zagrebnov2}.
\end{proof}

\begin{remark}\label{III.37}
The condition that the symmetric perturbation $B$ in Proposition \ref{III.35}
is $A$-small with relative bound $b <1$ (element of the class of {Kato-small}
$\calP_{b<1}$-perturbations), can be relaxed under the following assumptions:
let the symmetric (non necessarily Kato-small) perturbation be a non-negative operator $B \geq 0$ such that the  domain $D := \dom (A) \cap \dom (B)$ is dense in ${\cal H}$.

To illustrate this proposal, let us assume that the generator $A$ of a self-adjoint
Gibbs semigroup $\{U_A(t)\}_{t \geq 0}$ satisfies $A \geq \alpha \bbbone$ for
$\alpha > 0$. Since the densely
defined symmetric operator $H_0 : = A + B$ for  $\dom (H_0) = D$, is
semibounded from below by $\alpha \bbbone$, it admits a self-adjoint (Friedrichs)
extension $\widetilde H_0 =: H \geq \alpha \bbbone\,$. Moreover, since $B \geq 0$,
we have $H \geq A$, which means that $H$ is bounded from below by $\alpha \bbbone\,$, $\alpha > 0$.
\\
Since $\alpha > 0$, the operators $H^{-1}$ and $A^{-1}$ exist and satisfy $0 \leq H^{-1} \leq A^{-1}$. Then, $A^{-1}$ is compact as inverse of a Gibbs semigroup generator, implying that $H^{-1}$ is also compact. Furthermore,
for any eigenvalue $\lambda_n(H)$ of $H$, we have: $\lambda_n(H) \geq \lambda_n(A)$ and  $n \geq 1$. This implies that \eqref{inequality in perturbation} is satisfied for
$c_1 = c_2 = 1$. As a result, we have shown that $\{U_{H}(t)\}_{t \geq 0}$ is a
self-adjoint Gibbs semigroup, which can be extended to a holomorphic semigroup for
$z \in S_{\pi/2}\,$.
\end{remark}


Next we consider another class of perturbations introduced by Hille--Phillips \cite[Definition 13.3.5]{HillePhillips} (see also \cite[Chapter 11.4] {Davies}):
\\
Given a generator $A$ of a $C_0$-semigroup, a closed operator $B$ is in the class of perturbations $\calP(A)$ if
\begin{align}
& \text{(i) $\Dom(B) \supset  \bigcup_{t>0} U_A(t) \calH\,$,} \nonumber
\\
&\text{(ii)  $\int_0^1 \dd t\, \norm{BU_A(t)} <\infty\,$.} \label{condition on BUA}
\end{align}
The perturbations $\calP(A)$ belong to the class of \textit{infinitesimally} small \textit{unbounded} perturbations
$\calP_{b=0^+}$, which is between \textit{bounded} perturbations $\calP_{b=0}$ and
the \textit{unbounded} {Kato-small} $\calP_{b<1}$-perturbations, cf. \cite[Corollary 4.58]{Zagrebnov2}.
Using the Dyson--Phillips series, we obtain the following proposition:

\begin{proposition}
\label{prop-perturbation of A sectorial}
Let $A$ be an $m$-sectorial operator such that $e^{-t \re A}\in \calL^1$ for $t>0\,$, and let $B\in \calP(A)$. Then the operator $H=A+B$ with $\Dom ( H)=\Dom (A)\,$ generates a (holomorphic) Gibbs semigroup
$\{e^{-tH}\}_{t\geq 0}\,$, which satisfies assertion (a).
\end{proposition}

\begin{proof}
Since $A$ is $m$-sectorial, it generates a contraction semigroup, meaning $A\in Q(M=1,0)$.
Moreover, by Proposition \ref{prop-Karamata-hol}, the real part of $A$ is a self-adjoint, positive operator such that $\{e^{-t\re A}\}_{t\geq 0}$ is a Gibbs semigroup, satisfying the estimate $\norm{e^{-tA}}_1\leq \norm{e^{-t\re A}}_1=\Tr \,e^{-t\re A}\,$.

Now applying the trace-norm convergent Dyson--Phillips series for the perturbation $B\in \calP(A)$ :
\begin{align*}
& e^{-tH} = \sum_{n=0}^\infty S_n(t)\,,\quad \text{where \ }S_0(t)\vc e^{-tA}\ ,\quad S_n(t) \vc \int_0^t \dd s\, e^{-(t-s) A} (-B) \,S_{n-1}(s)\, \text{ for } n\geq 1\, ,
\end{align*}
we need only to prove the inequality \eqref{inequality in perturbation}.
\\ \indent
Given $\varepsilon>0$, the estimate $\norm{S_n(t)}_1\leq (2\gamma_\varepsilon)^n \,\norm{e^{-t\re A/8}}_1$ (provided in \cite[(4.99)] {Zagrebnov2} and valid for $t\in (0,2\varepsilon]$) implies, for a choice of $\varepsilon $ such that $c_1=\sum_{n=0}^\infty\, (2\gamma_\varepsilon)^n <\infty$, that $\norm{e^{-tH}}_1 \leq c_1 \norm{e^{-t\re A/8}}_1\,$. Then for $t$ small enough this inequality proves an estimate equivalent to \eqref{inequality in perturbation}, and thus the assertion (a).
\end{proof}
\begin{remark}
In \cite{Boulton} (see also \cite{Dimoudis}), condition (ii) in \eqref{condition on BUA} was modified as follows:
\begin{align*}
\text{(ii')} \,\int_0^1 \dd t\, \norm{B\,U_A(t)}_{p} <\infty\,,\quad \text{for some $p\geq 1\,$.}
\end{align*}
Given the generator $A$ of a Gibbs semigroup, it is proved in \cite[Lemma 1]{Boulton} that for any closed operator $B$ satisfying (i) and (ii'), the operator $H=A+B$ with domain $\dom\,(A)$ also generates a Gibbs semigroup such that $\norm{e^{-t(A+B)}}_p=\OO_0(\norm{e^{-tA}}_p)\,$. Consequently, the inequality
\eqref{inequality in perturbation} holds true for $p=1$.

The trade-off for relaxing the hypothesis that $A$ is $m$-sectorial (cf. Proposition \ref{prop-perturbation of A sectorial}, where the holomorphy of semigroup was used to control the trace-norm topology) is a direct introduction of this topology into condition (ii').
\end{remark}

Let us now turn to the study of assertion (b).\\
Motivated by the trace-norm convergence of the Dyson--Phillips series for perturbed Gibbs semigroups, the following class of perturbations was introduced in \cite{Boulton-Dimoudis} with preliminary results in \cite{Boulton} and \cite{Dimoudis}:
\\
\indent
Given the generator $A$ of a $C_0$-semigroup $\{U_A(t)\}_{t\geq 0}\,$, the operator $B$ belongs to the class $\calB_p(A)$ if

(i) $B$ with $\dom (B)=\dom (A)$ is relatively bounded with respect to $A\,$
for some $b>0$, see (\ref{def-relatively Kato-small}).   \\
This condition is equivalent to $B\,R_\lambda(A)$ is bounded for $\lambda\in \rho(A)\,$, thus $\Dom (B) = \Dom (A) \subset \Dom (\tilde{B})\,$, where $\tilde{B}$ is the unique extension of $B$ defined $u\in \Dom (\tilde{B})$ if and only if
$\lim_{\lambda\to -\infty}\abs{\lambda}\, B\, R_A(\lambda) \,u$ exits.

(ii) $B\,U_A(t)$ is bounded on $\Dom (A)$ for $t>0\,$,

(iii) $\int_0^1 \dd s\,\norm{\tilde{B}\,U_A(s)} <\infty\,$.

\noindent Moreover, in the context of Gibbs semigroups, the following conditions are required for some $p\geq 1$:

(iv) $\tilde{B}\,U_A(t)\in \calL^p\,$,

(v) $\int_0^1 \dd s\,\norm{\tilde{B}\,U_A(s)}_p <\infty\,$.

\noindent
As it was proved in \cite[Corollary 3.1]{Boulton-Dimoudis} that, given a semigroup generator $A$ and $B\in \calB_p$,  the operator $H=A+B$ with $\Dom (H)=\Dom (A)$ also generates a Gibbs semigroup.
\\
\indent It is worth mentioning that the class $\calB_p$ determines an equivalence relation: If $A_1$ and $A_2$ are generators of immediately norm continuous semigroups then  $A_1 \overset{p}{\approx} A_2$ if and only if $A_2=A_1 +B$ for $B\in \calB_p(A_1)\,$.

\begin{theorem}
Let $A$ be the generator of a Gibbs semigroup and $B\in \calB_1(A)\,$. Then,
\begin{align*}
\norm{e^{-t(A+B)}}_1 \,\underset{t\downarrow 0}{\sim} \,\norm{e^{-tA}}_1\,.
\end{align*}
\end{theorem}
\begin{proof}
Since by condition $B\in \calB_1(A)\,$, the (Bochner) integral in Duhamel's formula: $e^{-t(A+B)}-e^{-tA}= \int_0^t \dd s \,e^{-(t-s)(A+B)}\,\tilde{B}\,e^{-sA}\,$ for $t>0\,$,
converges in the $\norm{\cdot}_1$-norm, it follows that for $t\leq 1\,$, we obtain
\begin{align}
\label{estim 1}
\big{\Vert} \int_0^t \dd s\,e^{-(t-s)(A+B)}\,\tilde{B}\,e^{-sA}\,\big{\Vert}_1 \leq \sup_{r\in [0,1]} \,\norm{e^{-r(A+B)}} \int_0^1 \dd s\, \norm{\tilde{B}\,e^{-sA}}_1 \cv M_{A,B}\,.
\end{align}
Then the triangle inequality yields the estimate: $\norm{e^{-t(A+B)}}_1 \leq \norm{e^{-tA}}_1 +M_{A,B}\,, \, t\leq 1\,$.
Since the quoted equivalence relation implies $(-B)\in \calB_1(A+B)$, an application of the previous estimate provides
\begin{align}
\label{estim 2}
\norm{e^{-tA}}_1 \leq \norm{e^{-t(A+B)}}_1 +M_{A+B,-B} \,, \quad t\leq 1\,.
\end{align}
Then on account of Proposition \ref{prop-behaviour at 0} (i), the inequalities \eqref{estim 1} and \eqref{estim 2} prove the claim.
\end{proof}

Several examples of perturbation of differential Schr\"odinger operators on domains
in $\bbR^d$, where the hypotheses of the previous theorem hold, are considered in \cite{Boulton-Dimoudis}. In this context, we note that within the framework of relation \eqref{an expansion with log 2}, we have
$\Tr e^{-tA} \sim_{t\downarrow 0} \,a_0 \,t^{-d/m}$, where $a_0$ depends only on the principal symbol of $A\,$. Therefore, if the perturbation $B$ is also a pseudodifferential operator such that $A+B$ has the same symbol as $A\,$, then $\Tr e^{-t(A+B)} $ and $\Tr e^{-tA}$ exhibit the same asymptotic behaviour.

\clearpage

\appendix
\renewcommand{\thetheorem}{A.\arabic{theorem}} 
\renewcommand{\theequation}{A.\arabic{equation}} 
\setcounter{theorem}{0}

\section*{Appendix: Regularly varying functions}\label{Appendix}

Recall that a (Lebesgue) measurable function $L\!:\,[a,\infty) \mapsto (0,\infty)$ with $a> 0$, is considered {\it slowly varying at infinity} if, for all $x>0$, $\lim_{\alpha \to \infty} L(\alpha x)/L(\alpha) =1$.
\\ If $a\neq 0$, we can extend $L$ to $(0,\infty)$ by defining $L_{ext}(x)=L(a)$ for $x\in (0,a[$ and for  $x\geq a$, $L_{ext}(x)=L(x)$ since it does not alter the previous behaviour at infinity.

In a more general setting, let $f$ be a measurable strictly positive function defined on $[a, \infty)$, where $a > 0$ (and extended on $(0, \infty)$ as described above.) Then $f$ is said to be {\it regularly varying} if, for all $x > 0$, $\lim_{0<\alpha \to \infty} f(\alpha x)/f(\alpha) \in (0, \infty)\,$.

The {\it Karamata characterisation theorem} (see \cite[Theorem 1.4.1]{RegVar}) states that a measurable function $f: (0, \infty) \rightarrow (0, \infty)$ is regularly varying if and only if there exists $p \in \mathbb{R}$ and a function $L$ that varies slowly at infinity such that $f(x) = x^p L(x)\,$.
\\ \indent
The number $p$ is called  the {\it  index} of $f$ and it is denoted  by $ind_f\,$.
It is worth noting here that:

- If the functions $L_i$ vary slowly, then $\sum_i c_i\,L_i^{r_i}$ varies slowly for any $c_i>0$ and $r_i\in \bbR$.

- We have the following estimates (see \cite[Proposition 1.3.6 (v)]{RegVar} and \cite[Theorem 2, Chapter XIII.5]{Feller}): If $L$ varies slowly, for any $\varepsilon > 0$,
\begin{align}
&\quad \text{$L(x)=\oo_\infty(x^\varepsilon)\,$,} \label{L and small o}\\
& \quad x^{-\varepsilon} < L(x) < x^\epsilon\, \quad\text{for any $x$ sufficiently large.}
\label{estimates by monomials}
\end{align}
When $f$ varies regularly, the following property holds:
\begin{align}
\text{ If $\text{ind}_f > 0$, then $f(x) \underset{x\to \infty}{\longrightarrow} \infty$ (see \cite[Proposition 1.5.1]{RegVar}) } \label{behaviour of f} .
\end{align}

\begin{definition}
\label{def-RV}
Let $\RV_p$ denote the set of measurable functions $f\!:\,(0,\infty) \mapsto (0,\infty)$ that vary regularly with ind$_f=p$. Define $\RV$ as the union $\RV \vc \underset{p\geq 0}{\cup} \RV_p\,$.
\end{definition}
Note that if $f\in \RV$ then, applying \eqref{L and small o}, we have:
\begin{align}
\label{behaviour of f(t-1)}
\OO_0(f(t^{-1}))=\oo_0(t^{-(\text{ind}_f + \varepsilon)}) \,, \quad \text{for any } \varepsilon >0\,.
\end{align}

Any positive measurable function (or its extension $f_{ext}$) with a positive limit at infinity belongs to $\RV$. However, there are non-trivial examples like $f(x) = x^p \ln^{\circ k}(x)^r$, where $\ln^{\circ (k+1)} = \ln \circ \ln^{\circ k}$ and $\ln^{\circ 0} (x) = x$.

In this context, it is worth noting that $\ln x > 0$ for $x \geq a > 1$, and we need to choose an extension $\ln_{ext}$ for the natural logarithm. If we are interested in both the behaviours at zero and at infinity, it is better to avoid an extension of the function $f$. Therefore, we define for $k\in \bbN$,
\begin{align}
\label{def-Lnk}
\Ln_{k+1}(x) \vc \ln(1 + \Ln_k(x)) \text{ with }\Ln_0 (x)= x\,.
\end{align}
All these functions $\Ln_k$ define slowly varying functions on
$(0,\infty)$.\\
For non-logarithmic examples we refer to \cite[page 16]{RegVar}.

The following result on the Abelian/Tauberian theorem, originally due to Karamata and presented in \cite[Section 5]{Ambrosio} (see also  \cite[Theorem 1.7.1]{RegVar}, \cite[Theorem 2 in Chapter XIII.5]{Feller}, \cite[Theorem 10.3]{Simonbook}, \cite[Theorem 1.1]{Simon83} and \cite{Metafune}), is reformulated here with slight refinements extendind its applicability to the broader space $\RV$ and incorporating the limit inferior in point (b) of the theorem.

\begin{theorem}
\label{appendix:Thm-Karamata}
Let $\mu$ be a non-negative $\sigma$-finite Borel measure on $[0,\infty)$ and $f\in \RV$. Then, for $L_\mu$ defined in \eqref{def-lmu},

\emph{(a)} The following are equivalent:

\hspace{0.3cm} \emph{(i)} $\mu[0,\lambda] =\OO_\infty(f(\lambda))\,$.

\hspace{0.3cm} \emph{(ii)} $L_\mu(t) = \OO_0(f(t^{-1}))\,$.

\emph{(b)}
 \!\emph{(iii)} If  $\underset{\lambda \uparrow \infty} {\liminf} \,\,f(\lambda)^{-1}\,\mu([0,\lambda]) \geq c\,$, then \,\,$\underset{t \downarrow 0} {\liminf} \,\,f(t^{-1})^{-1} \, L_\mu(t) \geq c \,\Gamma(1+\textup{ind}_f)\,$.

\hspace{0.3cm} \emph{(iv)} If $\underset{t \downarrow 0} {\liminf}\, \,f(t^{-1})^{-1}\,L_\mu(t)>0$ \,\, {and}  \,\,$L_\mu(t)  =\OO_0(f(t^{-1}))\,$, then \,\,
$\underset{\lambda \uparrow \infty} {\liminf} \,\,f(\lambda)^{-1}\,\mu([0,\lambda]) >0\,$.

\emph{(c)} For any $C \in [0,\infty)\,$, the following are equivalent

\hspace{0.3cm} \emph{(v)} $\mu([0,\lambda)) \underset{\lambda \uparrow \infty}{\sim}\, C\,f(\lambda)\,$.

\hspace{0.3cm} \emph{(vi)} $L_\mu(t)  \underset{t \downarrow 0}{\sim}\, \Gamma(1+\textup{ind}_f)\, C\,f(t^{-1})\,$.
\\
When $C=0$, \emph{(v)} is interpreted as $\mu([0,\lambda)) =\OO_\infty(f(\lambda))$ and  \emph{(vi)} as $L_\mu(t)=\OO_0 (f(t^{-1}))\,$, an equivalence found in \emph{(a)}.

\emph{(d)} Moreover in this theorem, if $f(x) =x^p\, (\Ln_k x)^r$ for some $p\geq 0,\,r\in \bbR,\, k\geq 1\,$,
then $f(x)$ can be replaced by $x^p\,(\ln^{\circ \,k} x)^r$ with
\begin{align}
\label{def-lnk}
\ln^{\circ k}(\cdot)\vc \underbrace{\ln\circ\dots\circ \ln}_k\,(\cdot)\,.
\end{align}
\end{theorem}
	
\begin{proof}
(a) (Abelian result) (i) $\implies$ (ii)
For $\lambda>0$ and $t>0$,
\begin{align}
\label{Lmu < series}
\int_{[0,\infty)} \dd \mu(x) \, e^{-tx}
& = \int_{[0,\lambda]}\dd \mu(x)\,e^{-tx} +\sum_{k=1}^\infty \,\int_{]k\lambda,(k+1)\lambda]} \dd \mu(x)\, e^{-tx} \nonumber
\\
& \leq \mu([0,\lambda]) +\sum_{k=1}^\infty e^{-t k\lambda } \,\mu[0,(k+1)\lambda])=\sum_{k=0}^\infty e^{-t k\lambda} \,\mu[0,(k+1)\lambda])\,.
\end{align}
The hypothesis means that $c_{ls} := \limsup_{\lambda \uparrow \infty} {\mu([0,\lambda])}/{f(\lambda)}$ is finite, so for any $\epsilon>0$, there exists $\lambda_0$ such that ${\mu([0,\lambda])}/{f(\lambda)} \leq  c_{ls}+\epsilon$  for any $\lambda>\lambda_0 $.
Thus, for $\lambda=t^{-1}\geq \lambda_0$, we have:
\begin{align}
\label{f Lmu < series}
 f(t^{-1})^{-1}\int_{[0,\infty)} \dd \mu(x) \, e^{-tx}\leq (c_{ls} +\epsilon)\,\sum_{k=0}^\infty e^{-k} \, {f((k+1)t^{-1})}/{f(t^{-1})}\,.
\end{align}
By Potter's theorem, see \cite[Theorem 1.5.6 (iii)]{RegVar}, for any $a>1$ and $\varepsilon>0$, there exists a constant $c := c_{a,\,\varepsilon}$ such that $f(y)/f(x)\leq a\max\{(y/x)^{p+\varepsilon},\,(y/x)^{p-\varepsilon}\}$
when $x\geq c,\,y\geq c$. As a consequence,
for $t^{-1} > \max\{\lambda_0,\,c \}$, ${f((k+1)t^{-1})}/{f(t^{-1})} \leq a \max\{(k+1)^{p+\varepsilon},\,(k+1)^{p-\varepsilon}\} \leq a (k+1)^{p+\varepsilon}$ where $p$ is the index of $f$. \\
Since the series $\sum_{k=0}^\infty e^{-k} (k+1)^{p+\varepsilon}$ converges, we have proved (ii).

\medskip

(a) (Tauberian result) (ii) $\implies$ (i) For any $\lambda>0$ and $t>0$,
\begin{align*}
\mu([0,\lambda]) \leq e^{t\lambda} \int_0^\lambda \dd \mu(x) \,e^{-tx} \leq e^{t \lambda } \int_0^\infty \dd \mu(x) \,e^{-tx}.
\end{align*}
By hypothesis $C_{ls} \vc \underset{t\downarrow 0}{\limsup}\, f(t^{-1})^{-1}\int_{[0,\infty)}\dd \mu(x)\,e^{-tx} <\infty\,$.  So for any $\epsilon>0$, there exists $t_\epsilon$ such that
\begin{align*}
\int_0^\infty \dd \mu(x) \,e^{-tx} \leq (C_{ls}+\epsilon) \,f(t^{-1}) ,\quad t<t_\epsilon\,.
\end{align*}
Thus $\mu([0,\lambda]) \leq e^{t\lambda} (C_{ls}+\epsilon) \,f(t^{-1})$ for any $\lambda>0$ and $t<t_\epsilon$. \\
Taking $\lambda=t^{-1}$, and letting $t\downarrow 0$ achieves the proof as we obtain:
$\limsup_{\lambda \uparrow \infty} {\mu([0,\lambda])}/ {f(\lambda)}
\leq e \, C_{ls}\,$.

 \medskip

(b) (Abelian result) (iii) We apply Fatou's lemma to estimate
\begin{align*}
\underset{t \downarrow 0} {\liminf} \,\,f(t^{-1})^{-1} \int_{[0,\infty)} \dd \mu(x)\, e^{-tx}
& \geq
\int_{[0,\infty)} \dd \mu(x)\ e^{-x} \, \underset{t \downarrow 0} {\liminf} \,
{\mu([0,xt^{-1}])}/{f(t^{-1})} \\
&= \int_{[0,\infty)} \dd \mu(x)\,x^{\textup{ind}_f}\, e^{-x} \,
\underset{t \downarrow 0} {\liminf} \,\, {\mu([0,xt^{-1}])}/{f(xt^{-1})}\\
&\geq c \,\int_{[0,\infty)} \dd x \,x^{\textup{ind}_f} \,e^{-x} =
c \ \Gamma(1+\textup{ind}_f) \,.
\end{align*}

\begin{remark}
\label{NB}
To establish the inequality: $\liminf_{t\downarrow 0}\,
{\mu([0,xt^{-1}])}/{f(xt^{-1})} \geq  \liminf_{\lambda \uparrow \infty}\,
{\mu([0,\lambda])}/{f(\lambda)} $ used in the last line of the proof,
let $c_{li}:=\liminf_{\lambda \uparrow \infty}\, {\mu([0,\lambda])}/{f(\lambda)} $.
This means that for any $\varepsilon>0$, there exists $\lambda_0$ such that
\begin{align*}
\lambda > \lambda_0 \implies {\mu([0,\lambda])}/{f(\lambda)} \geq c_{li} -\varepsilon.
\end{align*}
Let $x$ be given and choose $t_0=x\lambda_0^{-1}$. For each $\lambda>0$ define $t=x\lambda^{-1}$. Then
$\lambda>\lambda_0 \iff t<t_0$ and consequently for each $t<t_0$ we have
\begin{align*}
 {\mu([0,xt^{-1}])}/{f(xt^{-1})} \geq c_{li}- \varepsilon & \implies
\liminf_{t \downarrow 0} {\mu([0,xt^{-1}])}/{f(xt^{-1})} \geq c_{li}- \varepsilon\\
& \implies  \liminf_{t \downarrow 0} {\mu([0,xt^{-1}])}/{f(xt^{-1})} \geq c_{li}\,,
\end{align*}
proving the inequality. By the same reasoning, we can also show the converse, which leads to the conclusion:
$\liminf_{t\downarrow 0}\, {\mu([0,xt^{-1}])}/{f(xt^{-1})} =
\liminf_{\lambda \uparrow \infty}\,{\mu([0,\lambda])}/{f(\lambda)} \,$.
\end{remark}

(b) (Tauberian result) (iv)  By the hypothesis, we have $0<C_{li} \vc \underset{t \downarrow 0} {\liminf} \,\,f(t^{-1})^{-1} \!\int_{[0,\infty)} \dd \mu(x)\, e^{-tx}$ and $C_{ls} = \underset{t \downarrow 0} {\limsup} \,f(t^{-1})^{-1} \int_{[0,\infty)} \dd \mu(x)\, e^{-tx} <\infty$.
\\
Taking again $\lambda = t^{-1}$ in \eqref{Lmu < series}, we know by hypothesis that there exists $t_0>0$ such that for $t\in ]0,t_0[\,$,
\begin{align}
\label{ineq-1}
0<\tfrac{C_{li}}{2} &\leq f(t^{-1})^{-1} \int_{[0,\infty)} \dd \mu(x)\, e^{-tx} \leq f(t^{-1})^{-1} \sum_{k=0}^\infty \,e^{-k} \,\mu([0,(k+1)t^{-1}]) \nonumber\\
& = f(t^{-1})^{-1} \sum_{k=0}^{n-1}\, e^{-k} \,\mu([0,(k+1)t^{-1}]) + R_n(t)\,.
\end{align}
As shown in the proof of part (a), for $t < \min\{\lambda_0^{-1}, c_{a,\,\epsilon}^{-1}\}\,$, we have:
\begin{align*}
R_n(t)&=f(t^{-1})^{-1} \sum_{k=n}^\infty e^{-k} \,\mu([0,(k+1)t^{-1}]) \leq (c_{ls}+\epsilon)\sum_{k=n}^\infty e^{-k} \, {f((k+1)t^{-1})}/{f(t^{-1})} \\
& \leq (c_{ls}+\epsilon)  \sum_{k=n}^\infty e^{-k} (k+1)^{p+\varepsilon}
\end{align*}
where the last series is the tail of, in turn, a convergent series.
Therefore, for $n$ sufficiently large, we may choose $R_n(t) < {C_{ii}}/{2}\,$. Then,  for $t$ small enough, the estimate \eqref{ineq-1} yields to
\begin{align*}
0<f(t^{-1})^{-1} \sum_{k=0}^{n-1} e^{-k} \,\mu([0,(k+1)t^{-1}]) \leq f(t^{-1})^{-1} \sum_{k=0}^{n-1} \mu([0,(k+1)t^{-1}])\,.
\end{align*}
Taking the $\liminf_{t\downarrow 0}$ on both sides yields the claim (iv).

 \medskip

(c) This is well known: see for instance \cite[Theorem 1.7.1]{RegVar} or \cite[Theorem 2, Chapter XIII.5]{Feller}.

(d) The last assertion is justified by noting that the functions $(\ln(1+x))^r$ and $(\ln x)^r$ exhibit identical behaviours at infinity.
\end{proof}

\begin{remark}
\label{mu finite}
Under the assumptions of the previous theorem, the following inequalities hold for any $\lambda > 0$:
\begin{align*}
f(t^{-1})^{-1}\,L_\mu(t) \geq f(t^{-1})^{-1}\int_{[0,\lambda)} \dd \mu(x)\, e^{-tx} \geq e^{-t\lambda}\, {\mu([0,\lambda))}/{f(t^{-1})}\,.
\end{align*}
Thus, if condition $(vi)$ holds true, we obtain $C\Gamma(1+\textup{ind}_f) \geq \mu([0,\lambda))\liminf_{t\downarrow 0} \, {e^{-t\lambda}}/ {f(t^{-1})}\,$. When ind$_f>0$ and $t$ goes to zero, last inequality becomes trivial since, as mentioned in \eqref{behaviour of f}, the right limit is zero.
Besides, when ind$_f=0$ and  the function $f$ has a limit at infinity, the previous inequality implies that $\mu$ is finite, a point already considered in a remark following Theorem \ref{thm-Kamarata for series}.
\end{remark}	

For the study of asymptotic differentiability, we recall the following notion from \cite[Section 1.8]{RegVar}: A positive smooth function $f$ defined on $[a,\infty)$ for some $a > 0$ is said to {\it vary smoothly at infinity with index $p \in \mathbb{R}$}, denoted as $f \in \SR_p$, if the function
$h(x) \vc \ln f(e^x)$ is smooth and satisfies $\lim_{x \to \infty} h^{(1)}(x) = p$ and $\lim_{x \to \infty} h^{(n)}(x) = 0$ for all integers $n \geq 2$. These two conditions are equivalent to
\begin{align}
\label{derivatives for SR}
\frac{x^n f^{(n)}(x)}{f(x)} \underset{x \to \infty}{\longrightarrow} \,p(p-1)\cdots(p-n+1)\,, \quad n \geq 1.
\end{align}
This follows from $xf'(x)/f(x) = h'(\ln(x)) \underset{x \to \infty}{\longrightarrow}\, p\,$.

Note that $\SR_p \subset \RV_p\,$ and if $f\in \RV_p$, there exists $g\in \SR_p$ with $g \underset{x \uparrow \infty}{\sim} f$, see \cite[Theorem 1.8.2]{RegVar}.
\begin{definition}
\label{def-SR}
We denote by $\SR$ the union $\SR \vc \cup_{p>0}\, \SR_p\,$.
\end{definition}

\subsubsection*{Acknowledgements}

We would like to express our gratitude to Luigi Ambrosio for his insightful correspondence regarding the Appendix, as well as to Victor Gayral, Fedor Sukochev, and Dmitriy Zanin for providing clarifications on noncommutative Lorentz spaces and C\'edric Arhancet for discussions about related topics. We are also deeply grateful to the referee for their numerous helpful remarks and constructive suggestions, which were invaluable in correcting and improving the text of the original manuscript.

\bibliographystyle{plainnat}
\bibliography{notes-biblio.bib}

\end{document}